\documentclass[12pt,twoside,a4paper]{report}
\usepackage[dvips]{epsfig}
\usepackage{amssymb}
\begin{document}

\newcommand{\A}{$^{\hspace{1mm}*}$}
\newcommand{\lessim}{\stackrel{<}{\sim}}
\newcommand{\B}{$^{\hspace{1mm}**}$}
\newcommand{\STS}{\vspace{1mm}}
\newcommand{\GS}{\vspace{5mm}}
\newcommand{\C}{$^{\hspace{1mm}*/}$}
\newcommand{\D}{$^{\hspace{1mm}**/}$}
\newcommand{\ee}{$e^+e^-$\ }
\newcommand{\emem}{$e^-e^-$\ }
\newcommand{\LUM}{{\cal{L}}}
\newcommand{\NT}{\cdot }
\newcommand{\demi}{1\! /\! 2}
\newcommand{\ttb}{t\overline{t}}
\newcommand{\W}{{\it W} }
\newcommand{\WW}{{\it WW} }
\newcommand{\SM}{Standard Model }
\newcommand{\VA}{V \!\! + \!\! A}
\newcommand{\ra}{\to }
\newcommand{\epem}{e^+e^- }
\newcommand{\cs}{cross section }
\newcommand{\css}{cross sections }
\newcommand{\Hs}{Higgs-strahlung }
\newcommand{\p}{particle }
\newcommand{\ps}{particles }
\newcommand{\cp}{coupling }
\newcommand{\cps}{couplings }
\newcommand{\sce}{scenario }
\newcommand{\sces}{scenarios }
\newcommand{\ssy}{supersymmetric }
\newcommand{\tb}{\tan \beta }
\newcommand{\notE}{/\hspace{-2.85mm}E}
\renewcommand{\thechapter}{}
\renewcommand{\thesection}{\arabic{section}}
\renewcommand{\thesubsection}{\thesection.\arabic{subsection}}
\renewcommand{\thefigure}{\arabic{figure}}
\renewcommand{\thetable}{\arabic{table}}
\renewcommand{\theequation}{\arabic{equation}}

\catcode`@=11
\def\citer{\@ifnextchar [{\@tempswatrue\@citexr}
          {\@tempswafalse\@citexr[]}}

%

\def\@citexr[#1]#2{\if@filesw\immediate
\write\@auxout{\string\citation{#2}}\fi
  \def\@citea{}\@cite{\@for\@citeb:=#2\do
    {\@citea\def\@citea{--\penalty\@m}\@ifundefined
       {b@\@citeb}{{\bf ?}\@warning
       {Citation `\@citeb' on page \thepage \space undefined}}%
\hbox{\csname b@\@citeb\endcsname}}}{#1}}
\catcode`@=12

\thispagestyle{empty}

\begin{center}
\rule{10cm}{0mm}\\
\begin{flushright}
DESY 97--100\\
\end{flushright}
\vfill
\vspace{2cm}
{\LARGE PHYSICS WITH $e^+e^-$ LINEAR COLLIDERS}\\

\vfill
\end{center}


\noindent

\vspace*{5mm}

\noindent
E.~Accomando $^{44}$,  
A.~Andreazza $^{34}$, 
H.~Anlauf $^{14}$, 
A.~Ballestrero $^{44}$,
T.~Barklow $^{42}$, 
J.~Bartels $^{24}$, 
A.~Bartl    $^{49}$, 
M.~Battaglia   $^{20}$, 
W.~Beenakker $^{29}$,   
G.~B\'{e}langer $^{2}$
% ...
\newpage

\newpage
\thispagestyle{empty}
\par \noindent
{\footnotesize $^1$ RWTH Aachen, Physikzentrum, D--52074 Aachen; $^2$
  ENSLAPP, F--74941 Annecy--le--Vieux Cedex;
% ...
$^{10}$ Department of Physics, Northeastern
  University, Boston MA 02115;
% ...
$^{14}$ Institut f\"ur Kernphysik,
  Technische Hochschule Darmstadt, D--64289 Darmstadt;
% ...
$^{20}$
  CERN, CH-1211 Gen\`{e}ve 23;
% ...
$^{24}$ II. Institut f\"ur Theoretische Physik,
  Universit\"at Hamburg, D--22761 Hamburg;
% ...
$^{29}$ Lorentz Institute
  for Theoretical Physics, Rijksuniversiteit Leiden, NL--2300 RA
  Leiden;
% ...
$^{34}$ Dipartimento di Fisica, Universit\`{a} degli Studi
  di Milano and INFN, I--20133 Milano;
% ...
$^{42}$ SLAC, Stanford University,
  Stanford CA 94309;
% ...
$^{44}$ INFN and
  Dipartimento Fisica Teorica, Universit\`{a} degli Studi di Torino,
  I--10125 Torino;
% ...
$^{49}$ Institut f\"ur Theoretische
  Physik, Universit\"at Wien, A--1090 Wien;
% ...
  Hochenergiephysik, ETH, CH--8093 Z\"urich.}



\setcounter{page}{1}

\begin{abstract}
  
  We describe the physics potential of $e^+e^-$ linear colliders in
  this report. These machines are planned to operate in the first
  phase at a center--of--mass energy of 500~GeV, before being scaled
  up to about 1~TeV. In the second phase of the operation, a final
  energy of about 2~TeV is expected. The machines will allow us to
  perform precision tests of the heavy particles in the Standard
  Model, the top quark and the electroweak bosons. They are ideal
  facilities for exploring the properties of Higgs particles, in
  particular in the intermediate mass range. New vector bosons and
  novel matter particles in extended gauge theories can be searched
  for and studied thoroughly. The machines provide unique
  opportunities for the discovery of particles in supersymmetric
  extensions of the Standard Model, the spectrum of Higgs particles,
  the supersymmetric partners of the electroweak gauge and Higgs
  bosons, and of the matter particles.  High precision analyses of
  their properties and interactions will allow for extrapolations to
  energy scales close to the Planck scale where gravity becomes
  significant. In alternative scenarios, like compositeness models,
  novel matter particles and interactions can be discovered and
  investigated in the energy range above the existing colliders up to
  the TeV scale. Whatever scenario is realized in Nature, the
  discovery potential of $e^+e^-$ linear colliders and the
  high--precision with which the properties of particles and their
  interactions can be analysed, define an exciting physics programme
  complementary to hadron machines.

\end{abstract}

\tableofcontents

\newpage
\section[Synopsis]{Synopsis}

High-energy \ee colliders have been essential instruments to search
for the fundamental constituents of matter and their interactions.
Merged with the experimental observations at hadron accelerators, a
coherent picture of the structure of matter has evolved, that is
adequately described by the Standard Model.  The matter particles,
leptons and quarks, can be classified in three families with identical
symmetries.  The electroweak and strong forces are described by gauge
field theories, based on the symmetry group ${\rm SU}(3)_C \times {\rm
  SU}(2)_L \times {\rm U}(1)_Y$ \cite{N1,N1A}.  The third component of
the Standard Model, still hypothetical, is the Higgs mechanism
\cite{N2} through which the masses of the fundamental fermions and
gauge bosons are generated.

\STS
The Standard Model has been tremendously successful in predicting
the properties of new particles and the structure of the
basic interactions. 
In many of its facets it has been
tested at an accuracy significantly better
than 1 percent. 
The Higgs mechanism however has not been established
experimentally so far.

\STS Despite the success in describing leptons, quarks and their
interactions, the Standard Model cannot be considered as the {\it
  ultima ratio} of Nature.  Neither the fundamental parameters, masses
and couplings, nor the symmetry pattern are accounted for; these
elements are merely built into the model.  Moreover, gravity, with a
nature quite different from the electroweak and strong forces, is not
incorporated in the theory.

\STS First steps which could lead us to solutions of these problems
are associated with the unification of the electroweak and strong
interactions \cite{N3}, and with a possible supersymmetric extension
of the model \cite{N4}.  Supersymmetry provides a bridge from the
presently explored energy scales to the scale of grand unified
theories, which is close to the Planck scale where gravity becomes
important.  No such path is known, at the present time, for
alternative compositeness scenarios which may include several new
layers of matter between the low energy scale and the Planck scale.

\GS Two strategies can be followed in future experiments to explore
the area beyond the Standard Model and to reveal the signals of new
physical phenomena.  First, the properties of the particles and forces
in the Standard Model may be affected by new energy scales.  Precision
studies of the top quark and the electroweak gauge bosons can thus
reveal clues to the physics beyond the Standard Model.  Second, if the
machine energies are high enough to cross the relevant thresholds, new
phenomena can be searched for directly and studied thoroughly.  This
is of course the prime {\it raison d'$ \hat{e}$tre} for any new
accelerator.  While the presently operating collider facilities, the
\ee collider LEP2, the $ep$ collider HERA and the $p\overline{p}$
collider Tevatron, cover the energy range up to a scale of 200 to 300
GeV, the $pp$ collider LHC and \ee linear colliders will enable us to
explore the energy range up to the TeV scale.

\GS On the basis of this dual approach, a variety of fundamental
problems can be investigated that are so far unresolved within the
Standard Model, and that demand experiments at energies beyond the
range of the existing accelerators.

\STS {$(i)$} The mass of the {\it top quark} is much larger than the
masses of all the other quarks and leptons, and even of the
electroweak gauge bosons. Understanding the r$\hat{\rm o}$le of this
particle in Nature is therefore a key element of future experiments.
The experimental analysis of the $\ttb$ threshold region in \ee
collisions will allow the measurement of the top quark mass to an
accuracy less than 200 MeV, improving the accuracy of about 2~GeV at
the LHC significantly.  This is a highly desirable goal since future
theories of flavor dynamics should provide relations among the lepton
masses, quark masses and mixing angles in which the heavy top quark is
expected to play a key role.  In addition, stringent tests of the
electroweak sector in the Standard Model can be carried out at the
quantum level when the top mass is known accurately.  Analyses of the
($\gamma /Zt\bar{t}$) production vertices and of the $(tbW)$ decay
vertex will determine the magnetic dipole moments of the top quark and
the chirality of the decay current.  Bounds on the ${\cal{CP}}$
violating electric dipole moments of the $t$ quark can be set in a
similar way.

\STS {$(ii)$} The experimental study of the dynamics of the {\it
  electroweak gauge bosons} is an equally important task at high
energy \ee colliders.  The form and the strength of the triple and
quartic couplings of these particles are uniquely prescribed by the
non-abelian gauge symmetry of the Standard Model.  The triple gauge
boson couplings define the electroweak charges, the magnetic dipole
moments and the electric quadrupole moments of the $W^\pm$ bosons.
Any small deviation from the values of these parameters predicted in
the Standard Model, will destroy the unitarity cancellations of the
gauge theories.  Their effect will therefore be magnified by
increasing the energy, and the bounds will tighten considerably
with rising energy.

\STS {$(iii)$} While the LHC can cover the canonical mass range, \ee
colliders with an energy between 300 and 500 GeV are ideal instruments
to search for {\it Higgs particles} throughout the mass range
characterized by the scale of electroweak symmetry breaking.  The mass
of the Higgs \p is not determined by existing theory, but the
intermediate mass range below $\sim 200$ GeV is theoretically a most
attractive region for Higgs masses.  In this scenario, Higgs particles
remain weakly interacting up to the scale of grand unification, thus
providing a path for the renormalization of the electroweak mixing
angle $\sin^2 \theta_w$ from the symmetry value 3/8 in grand unified
theories down to $\sim 0.2$ which is close to the experimentally
observed value 0.23.  Once the Higgs particle is found, its properties
can be studied thoroughly, i.e. the external quantum numbers ${\cal
  J}^{\cal{PC}}$ and the Higgs couplings, including the self-couplings
of the particle.  These are fundamental tests to establish the nature
of the Higgs mechanism experimentally.

\GS Even though many aspects of the Standard Model are experimentally
supported to a very high accuracy, the embedding of the model into a
more comprehensive theory is to be expected.  The argument is based on
the mechanism of the electroweak symmetry breaking.  If the Higgs
boson is light, the Standard Model can naturally be embedded in a
grand unified theory.  The large gap which exists between the low
electroweak scale and the high grand unification scale in this
scenario, can be stabilized by supersymmetry.  If the Higgs boson is
very heavy, or if no fundamental Higgs boson exists, new strong
interactions between the massive electroweak gauge bosons are
predicted by unitarity at the TeV scale.  Thus, the next generation of
accelerators which will operate in the TeV energy range, can uncover
the structure of physics beyond the Standard Model.

\GS
The following two disgruent theories are the opposite endpoints in the 
arch of possible physics scenarios. They will be considered in 
detail:

\GS
{$(i)$} The electroweak and strong forces are traced back to a
common origin in {\it Grand Unified Theories}. 
This idea can be
realized in various scenarios some of which predict new
vector bosons and a plethora of new fermions. 
The mass scales of these novel particles could be 
as low as a few hundred GeV.

\STS Intimately related to the grand unification of the gauge
symmetries is {\it Supersymmetry}.  This symmetry unifies matter and
forces by pairing the associated fermionic and bosonic particles in
multiplets.  Several arguments strongly support the hypothesis that
this symmetry is realized in Nature.  (a) As argued before,
supersymmetry stabilizes light masses of Higgs particles in the
context of high energy scales as realized in grand unified theories.
(b) The Higgs mechanism itself can be generated in supersymmetric
theories as a quantum effect.  The breaking of the electroweak gauge
symmetry ${\rm SU}(2)_L \times {\rm U}(1)_Y$ can be induced
radiatively while leaving the electromagnetic gauge symmetry ${\rm
  U}(1)_{EM}$ and the color gauge symmetry ${\rm SU}(3)_C$ unbroken
for a top quark mass between 150 and 200 GeV.  (c) This symmetry
concept is strongly supported by the successful prediction of the
electroweak mixing angle in the minimal version of the theory.  The
particle spectrum in this theory drives the evolution of the
electroweak mixing angle from the GUT value 3/8 down to
$\sin^2\theta_w$ = 0.2336 $\pm$ 0.0017; this prediction coincides with
the experimentally measured value $\sin^2 \theta_w^{exp}$ = 0.2315
$\pm$ 0.0003, within the theoretical uncertainty of less than 2
permille.

\STS A spectrum of several neutral and charged Higgs bosons is
predicted in supersymmetric theories.  In nearly all scenarios, the
mass of the lightest Higgs boson is less than $\sim 150$ GeV while the
heavy Higgs particles have masses of the order of the electroweak
symmetry breaking scale.  Many other novel particles are predicted in
supersymmetric theories.  The scalar partners of the leptons could
have masses in the range of $\sim 200$ GeV whereas squarks are
expected to be considerably heavier.  The lightest supersymmetric
states are likely to be non-colored gaugino/higgsino states with
masses possibly in the 100 GeV range.  Searching for these
supersymmetric particles will be one of the most important tasks at
the LHC and at future \ee colliders.  Moreover, the high accuracy
which can be achieved in measurements of masses and couplings, will
allow the reconstruction of the key elements of the underlying grand
unified theories, which may be generated within the supersymmetric
extension of gravity.

\GS {$(ii)$} In the alternative scenario of heavy or no fundamental
Higgs bosons, {\it strong} {\it interactions between electroweak
  bosons} would be observed in the elastic scattering of these
particles at TeV energies.  New resonances would be formed, the
properties of which would uncover the nature of the underlying
microscopic interactions.

\STS Not only would the properties and interactions of the electroweak
bosons be affected but also those of the fundamental fermions, leptons
and quarks in such a scenario. In the most dramatic departure from the
Standard Model, these particles would be built up by new
subconstituents corresponding to a new layer in the structure of
matter. This alternative scenario would manifest itself in non-zero
radii of quarks and leptons and the existence of novel bound states
such as leptoquarks \cite{f5A}.

\GS While new vector bosons and particles carrying color quantum
numbers can be searched for very efficiently at the hadron collider
LHC, \ee colliders provide in many ways unique opportunities to
discover and explore the non-colored particles.  This is most obvious
in supersymmetric theories.  Combining LEP2 analyses with future
searches at the LHC, the individual light and heavy Higgs bosons can
be found only in part of the supersymmetry parameter space; even if
all channels are combined, the coverage of the entire parameter space
is guaranteed only if non-supersymmetric decay modes of the Higgs
bosons prevail.  Squarks and gluinos can be searched for very
efficiently at the LHC.  Yet, the detailed experimental study of their
properties is very difficult at this machine.  Likewise, cascade
decays proceeding in several steps, will allow the search for other,
non-colored supersymmetric particles, yet a general model--independent
analysis of gauginos/higgsinos and scalar sleptons can only be carried
out at \ee colliders with well-defined kinematics at the level of the
individual subprocesses.  They will allow to perform high-precision
studies which are impossible or very difficult to carry out at hadron
colliders.  Only the detailed knowledge of all the properties of the
colored and non-colored supersymmetric states, gathered both at the
LHC and \ee experiments, will finally enable us to reveal the
structure of the underlying theory.

\GS The physics programme of $e^+e^-$ linear colliders \citer{N5,N7},
summarized briefly in Table~\ref{t0}, is in many aspects complementary
to the programme of the proton collider LHC.  The properties of the
top quark, the electroweak gauge bosons, the Higgs particles, the
supersymmetric or other novel particles can be explored with high
accuracy in a universal way, independent of favorable circumstances.
These analyses will enable us to cover the energy range above the
existing machines up to the TeV region in a conclusive form.  This
will provide essential information for elucidating the structure of
matter at a much more basic level than accessible today -- in
particular, if grand unified theories are true, we will gain insight
into the most fundamental levels of all.

\begin{table}[ht]
\begin{center}
{\scriptsize THE ENERGY--PHYSICS MATRIX \hfill}\\
\vspace{5mm}
{\scriptsize
\begin{tabular}{|l||c|c|c||c||ccc|} 
\hline
\rule[-3mm]{-2mm}{9mm}  
& Top                                  & Gauge Bosons 
& Higgs                                & SUSY 
& & Compositeness &  \\
\hline \hline
\rule[-3mm]{-2mm}{9mm}                   LC350 
& mass                                 & $W$ mass 
& intermediate                         & light Higgs/profile 
& & radii of &  \\
\rule[-3mm]{-2mm}{7mm}       
& decays                               & $\sin^2\theta_w$ 
& Higgs/profile                        & light $\tilde{\chi},\tilde{l}$ 
& & particles; &  \\
\cline{1-5} 
\rule[-3mm]{-2mm}{9mm}                   LC500 
& static elw.                          & self-\cps 
& intermediate                         & light Higgs/profile 
& & excited &  \\
\rule[-3mm]{-2mm}{7mm}
& parameters                           & new bosons/fermions 
& Higgs/profile                        & light $\tilde{\chi},\tilde{l}$ 
& & states; &  \\
\cline{1-5} 
\rule[-3mm]{-2mm}{9mm}                   LC1000 
& stat. param.                         & self-couplgs. refined 
& heavy Higgs/                 & heavy Higgs, $\tilde{\chi},\tilde{l} $ 
& & novel &  \\
\rule[-3mm]{-2mm}{7mm}   
& refined                              & new bosons/fermions 
& profile                              & light $\tilde{t}, \tilde{b}$ 
& & particles: &  \\
\cline{1-5} 
\rule[-3mm]{-2mm}{9mm}                   LC2000 
& stat. param.                         & new bosons/fermions 
& heavy Higgs                          & spectrum {\it in toto}: 
& & leptoquarks &  \\
\rule[-3mm]{-2mm}{7mm} 
& refined                              & strong $WW$ interact. 
& Higgs potential   & $H, \tilde{\chi},\tilde{l}, \tilde{q}, \tilde{g}$ 
& & dileptons etc. &  \\
\hline
\end{tabular}
}
\caption[]{{\it The physics spectrum that can be explored
    in experiments at \ee linear colliders with energies extending
    from LEP up to 2~TeV. \label{t0} }}
\end{center}
\end{table}

\vspace*{-5mm}
\GS 
The discussion in this report will focus on the physics with \ee
colliders in the first phase, corresponding to center-of-mass energies
above LEP2 up to $\sqrt{s} = 500$ GeV; it will be assumed that the
energy will be upgraded adiabatically up to about 800~GeV.  Where
necessary, we will refer to the second high-energy phase of the
machine, anticipating an energy of about 1.6~TeV. To cover some
physical scenarios it is necessary to extend the energy of the two
phases up to 1 and 2 TeV, respectively. The integrated luminosity for
energies at and below 500 GeV will in general be taken as $\int \LUM
=$ 50 fb$^{-1}$, corresponding to operating the machine at these
energies over 1 to 2 years.  Above 500 GeV, the required integrated
luminosity will be assumed to increase with the square of the c.m.
energy.  This implies about $\int \LUM$ = 125 fb$^{-1}$ at 800~GeV and
500 fb$^{-1}$ at 1.6~TeV.

\STS The polarization of the electron and positron beams is a powerful
tool in \ee colliders.  At a technical level, the polarization of the
beams can be used to enhance signals and to suppress backgrounds;
quite often, polarized electron beams are sufficient for this purpose.
At a deeper level, the polarization is of great advantage in
performing the microscopic diagnosis of the properties of the
fundamental particles, their interactions and the underlying symmetry
concepts.

\GS For some specific problems, operating linear colliders in the
$e^-e^-$ and the $e \gamma$ or $\gamma \gamma$ satellite modes will be
very useful. The high energy photons can be generated by Compton
back-scattering of laser light on the high energy electron (and
positron) bunches of the collider. The luminosities in these modes
will be slightly reduced compared with the \ee collisions; in contrast
to electron and positron bunches, electron--photon and photon--photon
bunches do not attract each other while electron--electron bunches
even repel each other.  Longitudinal and transverse photon
polarizations can be generated in Compton colliders by choosing the
appropriate polarizations of the initial electron/positron and laser
$\gamma$ beams.

\STS The problems which can be tackled in the \emem \cite{N8} and the
$\gamma$ \cite{N9} collider modes of the machine, will be addressed in
the appropriate physics context. Initial states with exotic lepton
quantum numbers are generated in \emem collisions, which are the
proper basis for studies of dilepton states, doubly charged Higgs
bosons and other particles, in particular Majorana neutrinos. $e
\gamma$ and $\gamma \gamma$ collisions provide one of the most complex
test grounds for QCD at high energies. Moreover, important aspects of
Higgs physics and other areas in the electroweak sector can only be
studied in $\gamma \gamma$ collisions.

\GS In many examples the physics potential of \ee linear colliders will 
be compared with the results which are expected at the
high-energy hadron-collider LHC.  Any such comparison cannot be
complete since only a selected set of processes has been simulated
experimentally in detail so far. However, this set includes most of the
problems associated with electroweak symmetry breaking and the Higgs
mechanism, and essential elements  of supersymmetry analyses at the LHC.
The LHC comments are based primarily on the material presented in
the ATLAS and CMS Technical Proposals \cite{801}, analyses of the DPF
studies Ref.\cite{803}, and results presented at the LHCC Workshop on
Supersymmetry \cite{804}.

\section[Basic Standard Processes]{Basic Standard Processes}
\STS The study of Standard Model processes at high energy colliders
serves several purposes.  On the one hand, high-precision analyses of
these classical processes can be exploited to determine the properties
of the particles in the Standard Model very accurately, and to detect
or set limits on anomalous properties, such as anomalous multipole
moments or potentially non-pointlike structures of the particles.  On
the other hand, Standard Model reactions are often unwanted background
processes, which mask novel reactions predicted in the physical
scenarios beyond the Standard Model\nopagebreak\ and which should
therefore be suppressed as much as possible.

\begin{figure}[t]
\begin{center}
\vspace*{-1cm}
\epsfig{file= 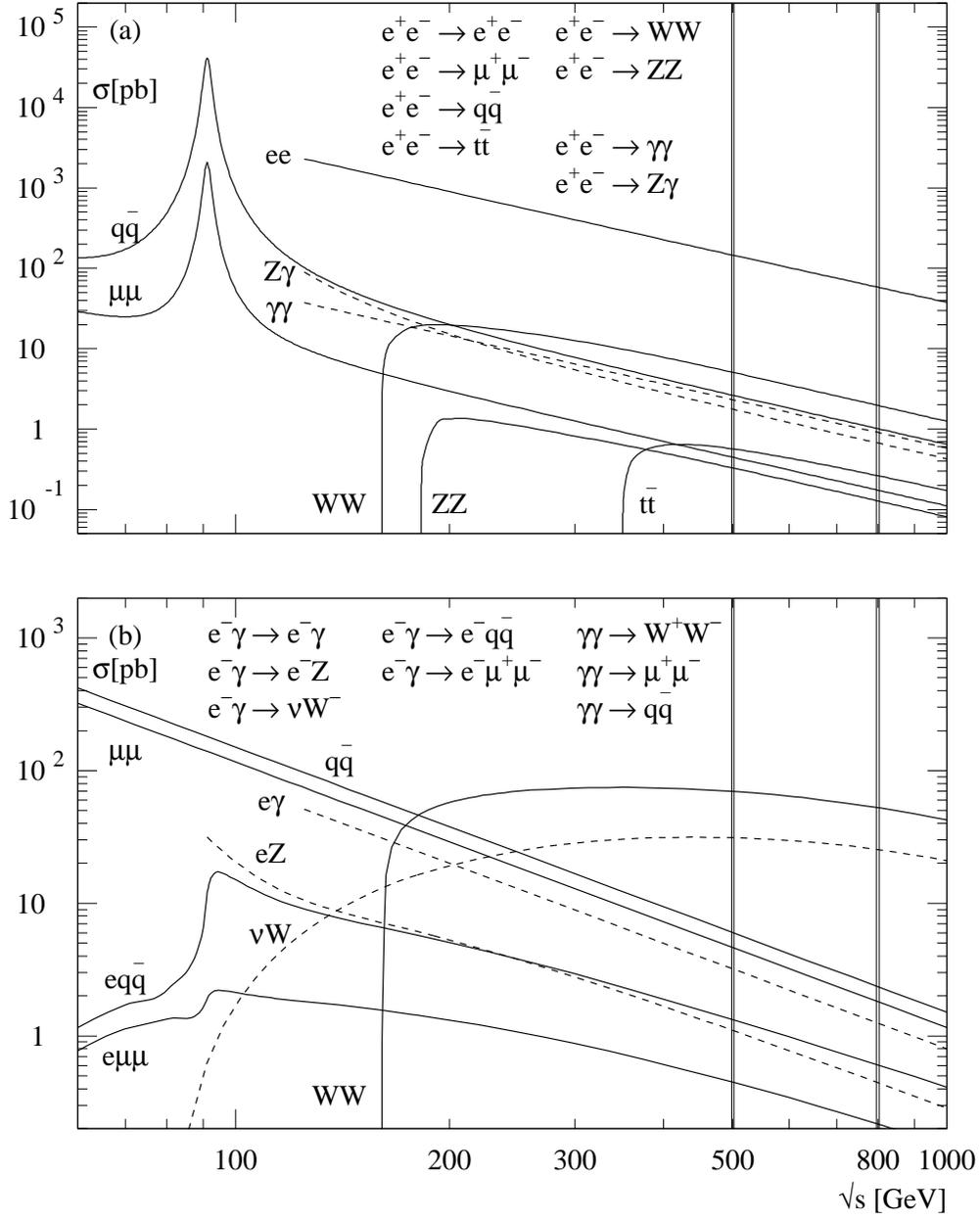, width= 14cm}
\end{center}
\vspace{-.5cm}
\caption[]{\it 
(a)
The basic processes of the Standard Model: 
\ee annihilation to pairs of fermions and gauge bosons. 
The cross sections are given for polar angles between 
$10^0 < \theta < 170^0$ in the final state.
(b)
Elastic/inelastic Compton scattering and $\gamma \gamma$ reactions.
$\sqrt{s}$ is the invariant $e \gamma$ and $\gamma \gamma$
energy. 
The polar angle of the final state particles is restricted as in (a); 
in addition, the invariant $\mu^+ \mu^-$ 
and $q \overline{q}$ masses in the
inelastic Compton processes are restricted to $M_{inv} > 50$ GeV.
\protect\label{f1}\label{2xSMcss}}
\end{figure}
\clearpage 

\vspace{11mm} \noindent
a) \underline{Rates of the Basic Standard Processes}

\STS The theoretical basis of the standard processes is familiar from
low-energy \ee collider experiments and will not be described in
detail here.

\STS The total cross sections are shown in Fig.\ref{f1}a for fermion
pair production in \ee annihilation: \ee $\rightarrow f \bar{f}$.
These processes are mediated by $s$--channel $\gamma$ and $Z$
exchanges, except for the Bhabha process which can also be generated
by $t$--channel $\gamma$ and $Z$ exchanges.  A cut in the polar angle
of the observed electrons and positrons in the final state, $10^0 <
\theta < 170^0$ corresponding to $| \cos \theta \mid < 0.985$, has
been introduced to remove the Rutherford pole; the size of the cut is
slightly larger than the masks for the detector around the beam pipe.
The magnitude of the cross sections, apart from the Bhabha process,
varies typically between 0.5 and 8 pb at an energy of $\sqrt{s}$ = 500
GeV, corresponding to 2,500 to 15,000 events for an integrated
luminosity of $\int{\LUM} =$~$50 \mbox{ fb}^{-1}$.  The cross section
for M{\o}ller scattering follows closely the Bhabha cross section.
[Program: CompHEP Ref.\cite{N9A}].

\STS The cross sections for \ee annihilation to pairs of gauge bosons,
\ee $\rightarrow \gamma \gamma, Z \gamma, ZZ$ and $W^+W^-$, are
presented in the same figure.  Since the angular distributions peak
strongly in the for\-ward/backward directions, the same cut in the
polar angle has been adopted as for Bhabha events.  The size of the
cross sections is similar to that for the fermionic annihilation cross
sections.

\STS The corresponding cross sections for initial state photons and
mixed electron-photon states are collected in Fig.\ref{f1}b: $e
\gamma \rightarrow e \gamma, eZ, e f \bar{f}$ and $\gamma \gamma
\rightarrow WW, f \bar{f}$.  [The cross sections of the $e \gamma$ and
$\gamma \gamma$ processes are shown for the invariant $e\gamma$ and
$\gamma \gamma$ energy $\sqrt{s}$; the results for the cross sections
after folding with the Weizs\"acker--Williams and Compton
back-scattering spectra are discussed later.]  The same polar-angle
cut has been applied as before.  Moreover, the difermion invariant
masses in the inelastic Compton processes have been restricted to $M[f
\bar{f}] > $ 50 GeV.  Though still of similar overall size, the cross
sections are in general slightly smaller than the annihilation cross
sections for the cuts applied in the present analysis.

\vspace*{11mm} \noindent
b) \underline{Polarization of Electron and Positron Beams}

\STS The polarization of the electron and positron beams gives a very
effective means to control the effect of the Standard Model processes
on the experimental analyses.  By choosing the polarizations
appropriately, different mechanisms which build up the Standard Model
processes, can be switched on and off so that the rates of the various
backgrounds can be studied and eventually much reduced.  This is
best-known for $W$ pair production in \ee annihilation, where the \cs
for right-handed electrons is much smaller than the \cs for
left-handed electrons.
\newpage

\begin{figure}[ht]
\begin{center}
\epsfig{file=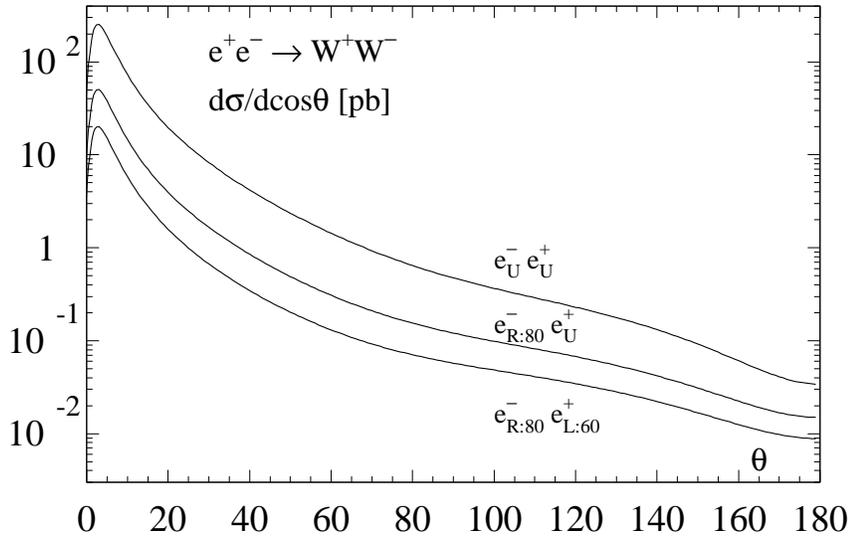 , width=13cm }
\end{center}
\vspace{-1cm}
\caption[]{\it 
  The effect of beam polarization on the cross section for the
  production of $W^+ W^-$ pairs.  $U$ denotes unpolarized electron and
  positron beams, R:80 denotes 80\% right-handedly polarized electron
  beams, and L:60 denotes 60\% left-handedly polarized positron
  beams.  \protect\label{f2}\label{eetoWW}}
\end{figure}

\STS Beam polarization is also an indispensable tool for the
identification and study of new \ps and their interactions.  In some
cases, the event rates can be increased considerably by choosing the
most suitable beam polarization for a specific reaction; for example,
the cross section for Higgs production in $WW$ fusion increases by a
factor 4 if the electron and positron beams are polarized.  In others,
the observation of polarization phenomena can add qualitatively new
information on the basic properties of particles and interactions; a
well-known example in this context is the analysis of mixed
gaugino/higgsino and L/R sfermion states in supersymmetric theories.

\STS In practice, the degree of polarization of electron beams is
expected to be approximately 80\%.  Polarized positron beams are more
difficult to generate, with a degree of polarization presumably in the
range of 60\% to 65\%.

\GS A few typical examples of Standard Model processes should
illustrate the impact of beam polarizations on the analysis.

\STS
\noindent
\underline{\it Fermion pair production \ee $\rightarrow f \bar{f}$}.
The dynamical impact of beam polarization on fermion-pair production
through the annihilation channel is very modest.  The polarization of
the electron determines the polarization of the positron to be
opposite in the annihilation process since gauge fields couple
chirally flipped particles and antiparticles.  Moreover, since the
photon couplings are left/right symmetric, as well as the $Z$
couplings for electrons/positrons in the axial limit $\sin^2 \theta_w
\rightarrow 1/4$, the polarization does not have a dynamical impact on
the total cross sections, but only the statistical weight affects the
cross sections.  If $\sigma_p$ is the annihilation cross section for
both beams polarized, the cross section for polarized
electrons/unpolarized positrons and for both beams unpolarized are
both given approximately by $\frac{1}{2} \sigma_p$.

\STS
\noindent
\underline{\it $W$ pair production \ee $\rightarrow W^+W^-$}.  This
process is mediated by $t$--channel $\nu_e$ exchange, and $s$--channel
$\gamma$ and $Z$ exchanges.  A large fraction of the events is
generated in the forward direction by the $t$--channel $\nu_e
$-exchange mechanism.  Choosing right-handedly polarized electrons,
this mechanism is switched off.  [Additional left-handed polarization
of the positrons is statistically helpful but dynamically not
required.]  Moreover, the $s$--channel exchange diagrams are switched
off at high energies for right-handedly polarized electrons; they do
not couple to the $W^3$ component of the gauge fields in the
intermediate state which is projected out by the charged $W$'s in the
final state.  The impact of the beam polarization on the differential
cross section is demonstrated in Fig.\ref{f2} where the cross sections
for (partially) polarized beams are compared with the unpolarized
cross section.

\STS
\noindent\underline{\it Single $W$ production}.
$W$ bosons are generated singly in the reactions \ee $\rightarrow e^+
\nu_e W^-$ and \ee $\rightarrow \overline{\nu}_e e^- W^+$.  These
reactions are almost exclusively generated by Weizs\"acker--Williams
photons $e \to e \gamma$ and the subsequent processes $\gamma e^-
\rightarrow W^- \nu_e$ and $e^+ \gamma \rightarrow \overline{\nu}_e
W^+$.  The electron and positron beams both must be polarized in the
right/left state to suppress this background reaction.  This is one of
the few cases where the suppression of a possible background requires
the polarization of both beams.

\vspace*{11mm}
\noindent
c) \underline{Photon Beams}

\STS Intense high-energy photon beams can be generated by
back-scattering of laser light off the incoming electrons and
positron \cite{N10}.  A large fraction of the energy can be
transferred from the leptons to the photons in this configuration.
The photon spectrum is rather broad however for unpolarized lepton and
laser beams.  The monochromaticity can be improved significantly if
the incoming leptons and laser photons have opposite helicities,
$P_eP_{\gamma} = -1$; the energy spectrum is given by:
\begin{equation}
P (y) = \frac{1}{1-y} 
      + 1 - y - 4 r (1-r) - 2 P_e P_\gamma x_0 r (2r - 1) (2-y)
\end{equation}
The fraction of energy transferred from the lepton to the
final-state photon is denoted by $y$ and $r = y/[(1-y)x_0]$; 
the maximum value of $y$ follows from $ y \le x_0 / (1 + x_0)$
with $x_0 = 4 E \omega_0  / m^2_e$. 
By tuning the frequency $\omega_0$ of the laser, the parameter
$x_0$ must be chosen less than
4.83 to suppress kinematically the copious \ee pair
production in the collision between the primary laser and
the secondary high-energy photons.

\begin{figure}[ht]
\begin{center}
\epsfig{file=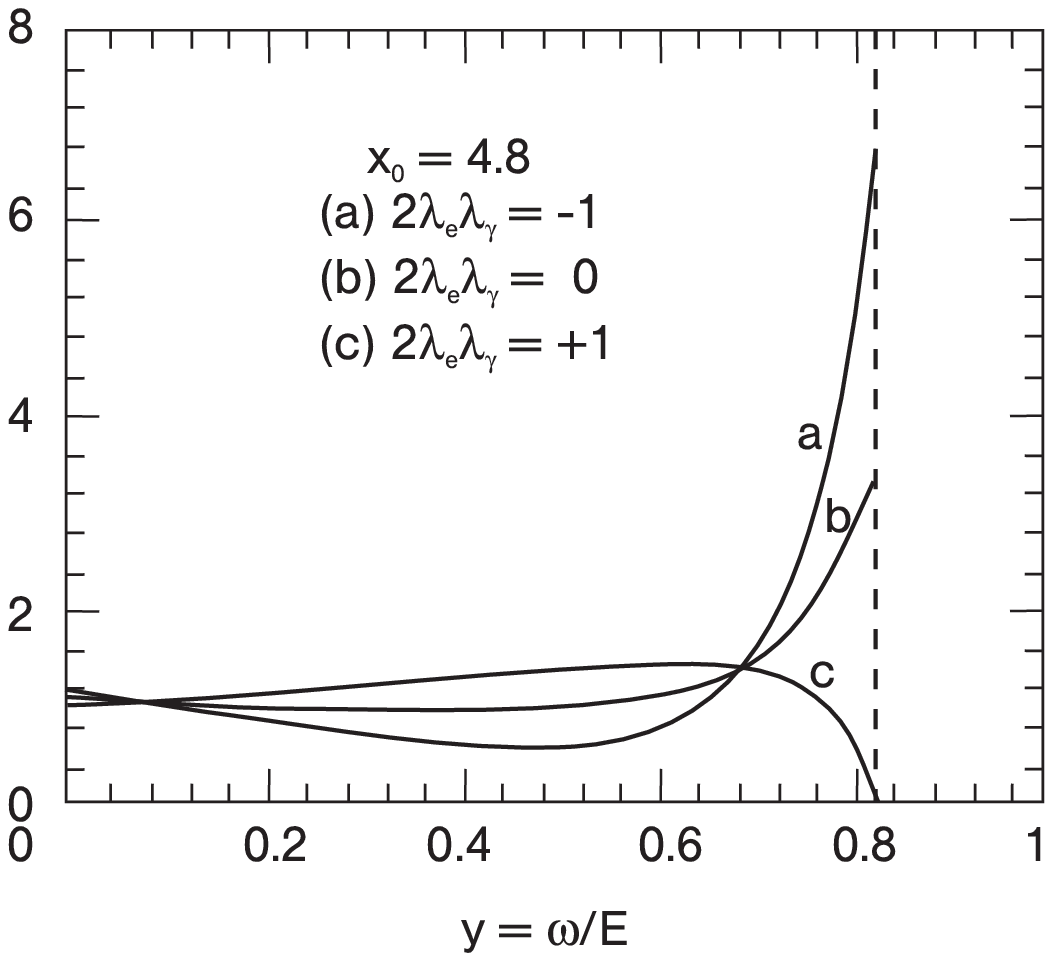,width=7.2cm,angle=0 }
\raisebox{0mm}{\epsfig{file=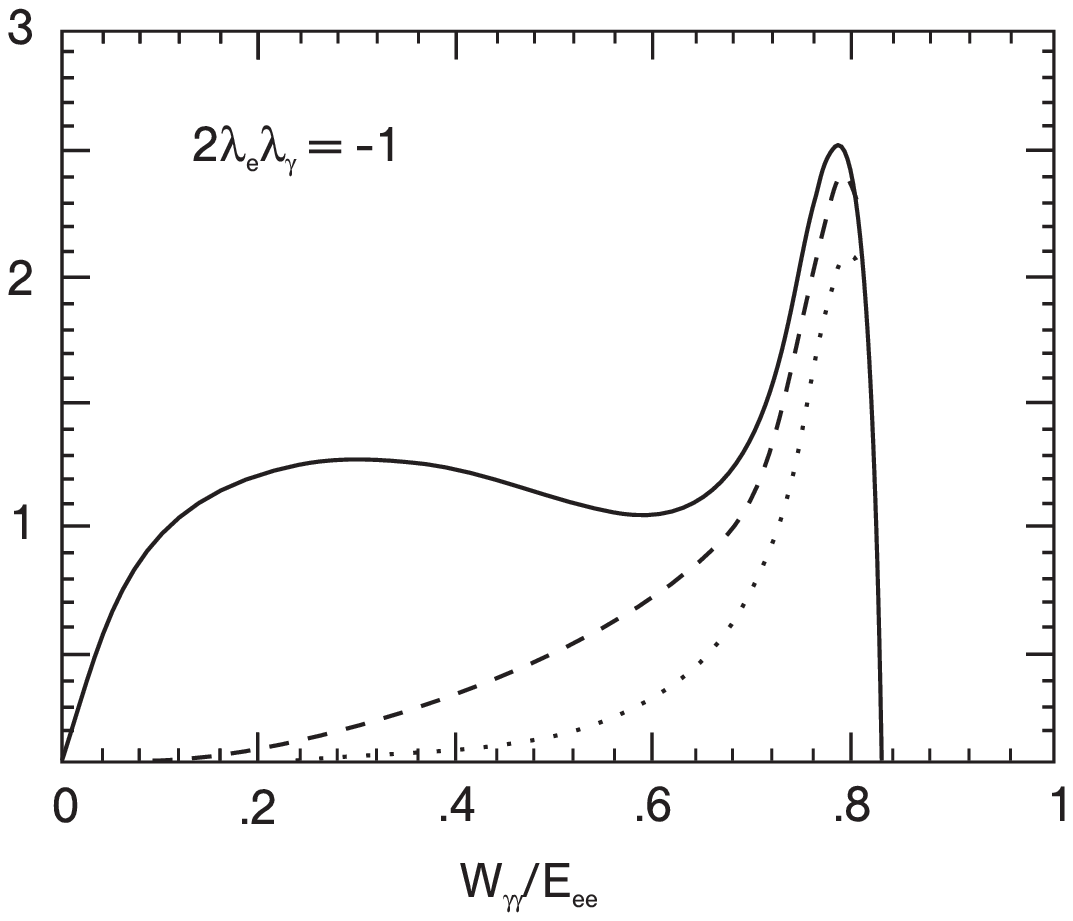,width=7.7cm}}
\end{center}
\vspace{-.5cm}
\caption[]{\it 
  Left: The $\gamma$ energy spectrum in Compton back-scattering of
  laser light for three values of initial laser and electron beam
  helicities \protect\cite{N11}.  Right: The distribution of the
  $\gamma \gamma$ invariant mass in Compton back-scattering of laser
  light with opposite laser/electron helicities.  The dashed curves
  demonstrate how the monochromaticity can be sharpened by separating
  the conversion from the collision point; c.f.
  Ref.\protect\cite{N10}.
\label{f3}}
\end{figure}

\STS The high-energy photon spectrum is shown for different
helicities in Fig.\ref{f3}(left).  The resulting $\gamma \gamma$
luminosity for the favorable case of opposite initial-state
helicities \cite{N11} is shown in Fig.\ref{f3}(right).  A clear,
nearly mono-energetic maximum of the $\gamma \gamma$ luminosity is
obtained, which is close to the maximum possible $\gamma \gamma$
invariant mass; the monochromaticity can be sharpened geometrically by
choosing non-zero conversion distances from the $\gamma \gamma$
collision points.

\GS High-energy $e \gamma$ and $\gamma \gamma$ collisions can be
applied to investigate problems in many areas of particle physics.
Outstanding examples are the production of Higgs bosons in $\gamma
\gamma$ collisions to measure the $\gamma \gamma$ widths, the
production of $W^+W^-$ pairs to determine the static magnetic and
electric multipole moments of the $W$ bosons, and the photon structure
functions and parton densities which provide deep insight into the
structure of QCD.  The cross sections for typical processes in the
Standard Model are exemplified in Table~\ref{t1} for two cases, with
the $\gamma$ beams generated by Weizs\"acker--Williams radiation and
with the Compton $\gamma$ spectrum generated in unpolarized electron
and laser beams.  

\begin{table}[ht!]
\begin{center}
\begin{tabular}{|l|c||ccc|cccc|} \hline
\rule[-3mm]{0mm}{8mm}
 & c.m. Energy  & \multicolumn{3}{|c|}{Cross Section $\sigma$[pb]} &  
\multicolumn{4}{|c|}{Cross Section $\sigma$[pb]}      \\
\hline
\hline
&\rule[-3mm]{0mm}{8mm} $\sqrt{s_{ee}}$ &  $ \gamma \gamma \rightarrow $ 
& &  & $  e \gamma \rightarrow $ & & & \\
 &\rule[-3mm]{0mm}{8mm} & $ \mu^+ \mu^- $ & $ u \bar{u}$  &
$ W^+ W^- $ & $ \nu_e W $  & $ eZ $ &
$ e \mu^+ \mu^- $ & $ e u \bar{u} $ \\
\hline
WWR &\rule[-3mm]{0mm}{8mm} 500 GeV 
& 2.4 & 1.4 & 0.2 & 2.9 & 0.3 & 0.1 & 0.2   \\
 &\rule[-3mm]{0mm}{8mm} 800 GeV 
& 3.1 & 1.9 & 0.5 & 4.9 & 0.3 & 0.1 & 0.1   \\
\hline
CBS &\rule[-3mm]{0mm}{8mm} 500 GeV 
& 33  & 20  & 40  & 28  & 1.8 & 0.6 & 0.7   \\
 &\rule[-3mm]{0mm}{8mm} 800 GeV 
& 17  & 10  & 49  & 32  & 0.9 & 0.3 & 0.3  \\
     \hline
\end{tabular}
\caption[]{{\it Cross sections of typical SM processes in
    $\gamma \gamma$ and $e\gamma$ collisions with the $\gamma$ beams
    generated by Weizs\"acker--Williams radiation (WWR) and Compton
    back-scattering of laser light with the frequency parameter
    $x_0=4.83$ (CBS). The cross sections are given for polar angles of
    the visible particles between $10^0 < \theta < 170^0$ in the final
    state; in addition, the invariant $\mu^+ \mu^-$ and $q
    \overline{q}$ masses are restricted to $M_{inv} > 50$ GeV.
\label{t1} }}
\end{center}
\end{table}

\section[Top Quark Physics]{Top Quark Physics}

Top quarks are the heaviest matter particles in the 3--family Standard
Model.  Introduced to incorporate ${\cal CP}$ violation \cite{N12},
indirect evidence for the top quark had been accumulated quite early.
After the isospin of the left-handed $b$ quarks was measured to be
$I_3 (b_L) = - 1/2$, derived from the $Z \rightarrow b \bar{b}$ width
and the forward-backward asymmetry of $b$ jets in \ee annihilation,
it was manifest that the symmetry pattern of the Standard Model
required the existence of the top quark \cite{N13}.  The top mass
enters quadratically through radiative corrections \cite{14A} into the
expression for the $\rho$ parameter, the relative strength between
weak neutral and charged current processes. The high-precision
measurements of the electroweak observables, in particular at the \ee
colliders LEP1 and SLC, could be exploited to determine the top mass
\cite{N14}: $m_t = 173 \, \pm \, 23$~GeV.  This prediction has
recently been confirmed by the direct observation of top quarks at the
Tevatron \cite{N15} with a mass of $m_t = 174 \, \pm \,6$~GeV, which
is in striking agreement with the earlier electroweak analysis.

\STS The large mass renders the top quark a very interesting object,
the properties of which should be studied with high precision.  Being
the leading particle in the fermion spectrum of the Standard Model, it
likely plays a key role in any theory of flavor dynamics.  Moreover,
due to the large mass, its properties are most strongly affected by
Higgs particles and nearby new physics scales.  High-precision
measurements of the properties of top quarks are therefore mandatory
at any future collider.

\GS Since the lifetime of the $t$ quark is much shorter than the time
scale $\Lambda^{-1}_{QCD}$ of the strong interactions, the impact of
non-perturbative effects on the production and decay of top quarks
can be neglected to a high level of accuracy \cite{N16}.  The short
lifetime provides a cut-off $k > \sqrt{2m_t \Gamma_t}$ for any soft
non-perturbative and infrared perturbative interactions.  The $t$
quark sector can therefore be analyzed within perturbative QCD.
Unlike light quarks, the properties of $t$ quarks are reflected
directly in the distributions of the decay jets and $W$ bosons, and
they are not affected by the obscuring confinement and fragmentation
effects.

\GS \ee colliders are the most suitable instruments to study the
properties of top quarks.  Operating the machine at the $t \bar{t}$
threshold, the mass of the top quark can be determined with an
accuracy that is an order of magnitude superior to measurements at
hadron colliders.  The static properties of top quarks, magnetic and
electric dipole moments, can be measured very accurately in continuum
top-pair production at high $e^+e^-$ colliders.  Likewise, the
chirality of the charged top-bottom current can be measured
accurately in the decay of the top quark.  In extensions of the
Standard Model, supersymmetric extensions for example, top decays into
novel particles, charged Higgs bosons and/or stop/sbottom particles,
may be observed.

\STS
\subsection[The Profile of the Top Quark: Decay]{The Profile of the 
Top Quark: Decay}

\GS
\noindent
a) \underline{The Dominant SM Decay}

\STS
\noindent
With the top mass established as larger than the $W$ mass, the channel
\[
t \rightarrow b + W^+
\]
is the dominant decay mode, not only in the Standard Model but also in
extended scenarios.  The top quark width grows rapidly to $\sim
1.4$~GeV in the mass range $m_t \sim 175$ GeV \cite{N16}:
\begin{equation}
\Gamma (t \rightarrow b + W^+) =
\frac{G_F m^3_t}{8 \sqrt{2} \pi}
\left[ 1 - \frac{m^2_W}{m^2_t} \right]^2
\left[ 1 + 2 \frac{m^2_W}{m^2_t} \right]
\end{equation}
approximately given by $\Gamma_t \simeq 175 \mbox{ MeV} \cdot [m_t /
M_W]^3$.  A large fraction, $p_L = m^2_t / (m^2_t + 2 m^2_W) \approx
0.7$, of the decay $W$ bosons are longitudinally polarized.  The rapid
variation of $\Gamma_t$, proportional to the third power of $m_t$, is
expected from the equivalence theorem of electroweak symmetry breaking
in which the longitudinal $W$ component, dominating for large $t$
masses, can be identified with the charged Goldstone boson, the
coupling of which grows with the $t$ mass.  The width of the top quark
is known to one-loop QCD and electroweak corrections \cite{N17}.  The
QCD corrections are about --10\% for large top masses; the electroweak
corrections turn out to be small, $\approx +2\%$ for a Higgs mass of
$\sim$ 100 GeV.

\GS The direct measurement of the top quark width is difficult.  The
most promising method appears to be provided by the analysis of the
forward-backward asymmetry of $t$ quarks near the \ee production
threshold.  This asymmetry is generated by the overlap of parity-even
$S$-- and parity-odd $P$--wave production channels; it is therefore
sensitive to the width $\Gamma_t$.  Including the other threshold
observables, cross section and momentum distributions, a precision of
10 to 20\% can be expected for the measurement of $\Gamma_t$ in total
\cite{N18}.
 
\GS \underline{\it Chirality of the $(tb)$ decay current}.  The
precise determination of the weak isospin quantum numbers does not
allow for large deviations of the $(tb)$ decay current from the
left-handed prescription in the Standard Model.  Nevertheless, since
$\VA$ admixtures may grow with the masses of the quarks involved [$\sim
\sqrt{m_t / M_\ast}$ through mixing with heavy mirror quarks of mass
$M_*$, for instance], it is necessary to check the chirality of the
decay current directly.  The $l^+$ energy distribution in the
semileptonic decay chain $t \rightarrow W^+ \rightarrow l^+$ depends
on the chirality of the current; for $V \!\! - \!\! A$ couplings it is
given by $dN/dx_l \sim x^2_l (1 - x_l)$.  Any deviation from the
standard $V \!\! - \!\! A$ current would stiffen the spectrum, and it
would lead to a non-zero value at the upper end-point of the energy
distribution, in particular.  A sensitivity of 5\% to possible $\VA$
admixtures can be reached experimentally (see Ref.\cite{18A}).  The
sensitivity can be improved by analysing the decays of polarized top
quarks which can be generated in collisions of longitudinally
polarized electrons with un/polarized positrons.  

\GS
\noindent
b) \underline{Non--Standard Top Decays} 

\begin{figure}[ht]
\begin{center}
\epsfig{file=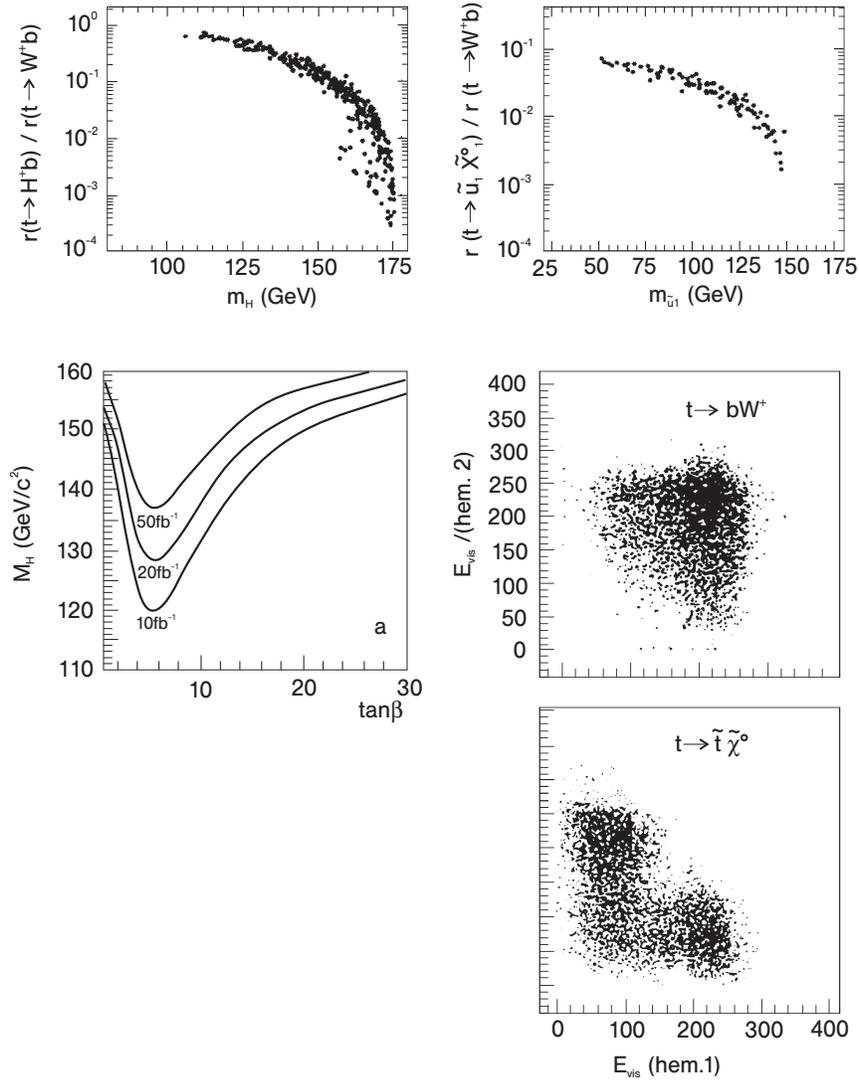,width=11.5cm}
\end{center}
\vspace{-.5cm}
\caption[]{\it 
  Left: Branching ratio of top decays to charged Higgs bosons, in
  supersymmetric theories.  Shown is also the range of charged Higgs
  masses as a function of the coupling $\tb$ that can be detected
  experimentally for a given luminosity of 10, 20, and 50~fb$^{-1}$.
  Refs.\protect\cite{N19,N20}.  Right: The decay of top quarks to stop
  particles and the lightest neutralino in supersymmetric theories.
  The lower plots present the energy distributions in the two event
  hemispheres for SM decays and SUSY decays which are characterized by
  missing energy due to escaping neutralinos.
  Refs.\protect\cite{N19,N20}.
  \protect\label{f4}\label{f5}\label{2xttobH}}
\end{figure}

\STS
\noindent
Such decays could occur, for example, in supersymmetric extensions of
the Standard Model: top decays into charged Higgs bosons and/or top
decays to stop particles and neutralinos or sbottom \ps and charginos:
\begin{eqnarray*}
t &\rightarrow & b + H^+ \hspace{+2mm} \\
t &\rightarrow & \tilde{t} + \tilde{\chi}^0_1 \quad \mbox{and} \quad 
              \tilde{b} + \tilde{\chi}^+_1
\end{eqnarray*}
If kinematically allowed, branching ratios for these decay modes could
be as large as 30\%, given the present constraints on supersymmetric
parameters, Fig.\ref{f4} \cite{N19}.  If LEP2 fails to discover
supersymmetric particles, stop decays would become very unlikely while
charged Higgs decays might still be frequent. The signatures for both
decay modes are very clear and they are easy to detect experimentally
\cite{N20}.  Charged Higgs decays manifest themselves through 
chargino+neutralino decays, and $\tau$
decays with rates 
which are different from the universal $W$ decay rates in
the Standard Model, 

\clearpage

\noindent
thus breaking $\tau$ {\it vs.} $e, \mu$
universality.  Final-state neutralinos, as the lightest
supersymmetric particles, escape undetected in stop decays so that a
large amount of missing energy would be observed in these decay modes.

\GS Besides breaking the $V \!\! - \!\! A$ law for the chirality of
the $ t \rightarrow bW$ decay current, mixing of the top quark with
other heavy quarks breaks the GIM mechanism if the new quark species
do not belong to the standard doublet/singlet assignments of isospin
multiplets.  As a result, FCNC ($tc$) couplings of order $\sqrt{m_t
  m_c / M^2_\ast}$ can be induced.  FCNC $t$ quark decays, for example
$t \rightarrow c \gamma$ or $cZ$, may therefore occur at the level of
a few permille; down to this level they can be detected experimentally
\cite{20A}. The large number of top quarks produced at the LHC 
allows however to search for rare FCNC decays with clean signatures, 
such as $t \rightarrow cZ$, down to a branching ratio of less than
$10^{-4}$.

\STS
\subsection[Continuum Production: Static $t$ Parameters]{Continuum 
Production: Static $t$ Parameters}

\begin{figure}[ht]
\vspace*{-2.cm}
\begin{center}
\epsfig{file= 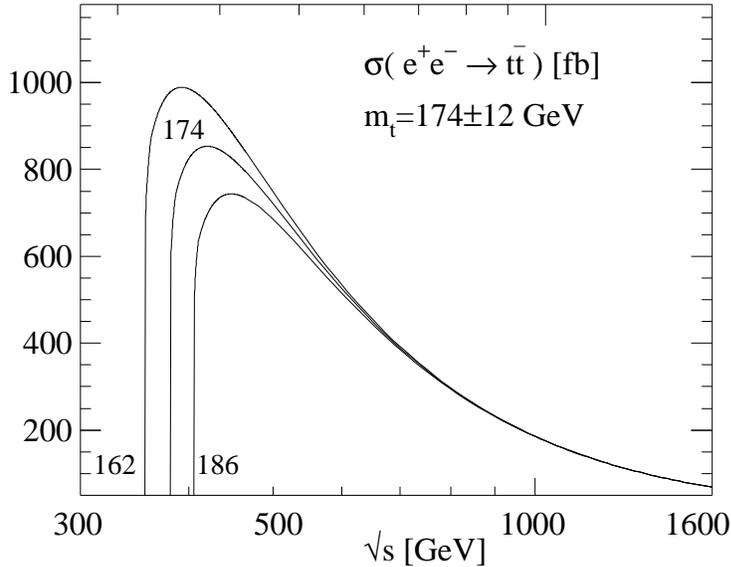, width=14cm}
\end{center}
\vspace{-2.cm}
\caption[]{\it 
  The cross section for the production of top-quark pairs in the
  continuum as a function of the total energy for three representative
  values of the top mass.  \protect\label{5t} }
\end{figure}

The main production mechanism for top quarks in \ee collisions is the
annihilation channel \cite{N21}
\[
e^+e^- \stackrel{\gamma, Z}{\longrightarrow} t \overline{t}
\]
As shown in Fig.\ref{5t}, the cross section
\begin{equation}
\sigma (e^+e^- \rightarrow t \bar{t}) = \beta
{\textstyle \frac{3 - \beta^2}{2}} \sigma^{VV} + \beta^3 \sigma^{AA}
\end{equation}

\vspace{-4mm}
\begin{eqnarray}
\sigma^{VV} & = & \frac{4 \pi \alpha^2 (s)}{s}
e^2_e e^2_t + \frac{G_F \alpha (s)}{\sqrt{2}}
e_e e_t v_e v_t \frac{m^2_Z}{s - m^2_Z}
+ \frac{G^2_F}{32 \pi} (v^2_e + a^2_e) v^2_t
\frac{m^4_Z s}{(s - m^2_Z)^2}   \nonumber \\
\sigma^{AA} & = & \frac{G^2_F}{32 \pi} (v^2_e + a^2_e)
a^2_t \frac{m^4_Z s}{(s - m^2_Z)^2}
\nonumber
\end{eqnarray}

\noindent
[$v_f, a_f$ being the $Z$ charges, and $\beta$ the velocity of the $t$
quarks] is of the order of 1 pb so that top quarks will be produced at
large rates in a clean environment at \ee linear colliders, about
50,000 pairs for an integrated luminosity of $ \int {\LUM} \sim$ $50
\makebox{ fb}^{-1}$.
  
\GS Since production and decay are not affected by the
non-perturbative effects of hadronization, the helicities of the top
quarks can be determined from the distribution of the jets and leptons
in the decay chain $t \rightarrow b + W^+ \rightarrow b+ f \bar{f}'$.
The form factors of the top quark \cite{N22} in the electromagnetic
and the weak neutral currents, the Pauli--Dirac form factors
$F_1^{\gamma, Z}$ and $F_2^{\gamma, Z}$, the axial form factor $F_A^Z$
and the ${\cal{CP}}$ violating form factors $D_A^{\gamma, Z}$, can
therefore be measured very accurately.  The form factors $F_1^{\gamma,
  Z}$ and $F_A^Z$ are normalized to unity ({\it modulo} radiative
corrections) and $F_2^{\gamma, Z}$ and $D_A^{\gamma, Z}$ vanish in the
Standard Model. Anomalous values, in particular of the static
magnetic- and electric-type dipole moments, could be a consequence of
electroweak symmetry breaking in non-standard scenarios or of
composite quark structures.  Deviations from the values of the static
parameters in the Standard Model have coefficients in the production
cross section which grow with the c.m.  energy.

\STS Among the static parameters of the top quark which can be
determined only at \ee linear colliders, the following examples are of
particular interest:

\STS
\noindent
\underline{\it $Z$ charges of the top quark}.  The form factors
$F_1^Z, F_A^Z$, or likewise the vectorial and axial $Z$ charges of the
top quark, $v_t = +1 - \frac{8}{3} \sin^2 \theta_w$ and $a_t = +1$,
can be determined from the $t\bar{t}$ production cross section
\cite{L27A}.  Moreover, the production of top quarks near the
threshold with longitudinally polarized beams leads to a sample of
highly polarized quarks.  The small admixture of transverse and normal
polarization induced by $S$--wave/$P$--wave interference, is extremely
sensitive to the axial $Z$ charge $a_t$ of the top quark \cite{27A}.

\STS
\noindent
\underline{\it Magnetic dipole moments of the top quark}.  If the
electrons in the annihilation process $e^+e^- \rightarrow
t\overline{t}$ are left-handedly polarized, the top quarks are
produced preferentially as left-handed particles in the forward
direction while only a small fraction is produced as right-handed
particles in the backward direction \cite{N23}.  As a result of this
prediction in the Standard Model, the backward direction is most
sensitive to small anomalous magnetic moments of the top quarks.  The
anomalous magnetic moments can be bounded to about 5 percent by
measuring the angular dependence of the $t$ quark cross section.

\STS
\noindent
\underline{\it Electric dipole moments of the top quark}.  Electric
dipole moments are generated by ${\cal{CP}}$ non-invariant
interactions.  Non-zero values of these moments can be detected
through non-vanishing expectation values of ${\cal{CP}}$--odd
momentum tensors such as $T_{ij} \sim (q_+ - q_-)_i (q_+ \times
q_-)_j$ or $A \sim p_+ \cdot (q_+ \times q_-)$, with $p_+,\, q_{\pm}$
being the unit momentum vectors of the initial $e^+$ and of the
$W$--decay leptons, respectively.  Sensitivity limits to $\gamma, Z$
electric dipole moments of $ d^{\gamma, Z}_t < 10^{-18}$ e$\,$cm can
be reached \cite{N24} for an integrated luminosity of $\int{\LUM}
=$~$20 \mbox{ fb}^{-1}$ at $\sqrt{s} = 500$ GeV if polarized beams are
available.

\STS
\subsection[Threshold Production: The Top Mass]{Threshold Production: 
The Top Mass}

Quark-antiquark production near the threshold in \ee collisions is,
quite generally, of exceptional interest.  For small quark masses, the
long time which the particles remain close to each other, allows the
strong interactions to build up rich structures of bound states and
resonances.  For the large top mass, the picture is different: The
decay time of the states is shorter than the revolution time of the
constituents so that toponium resonances can no longer form
\cite{N16}.  Traces of the $1S$ state give rise to a peak in the
excitation curve which gradually levels off for quark masses beyond
150 GeV.  Despite their transitory existence, the remnants of the
toponium resonances nevertheless induce a fast rise of the cross
section near the threshold.  The steep rise provides the best basis
for high-precision measurements of the top quark mass, superior to
the reconstruction of the top mass in the decay final states at hadron
colliders by more than an order of magnitude.

Since the rapid $t$ decay restricts the interaction region of the top
quark to small distances, the excitation curve can be predicted in
perturbative QCD \citer{N25,N27}.  The interquark potential is given
essentially by the short distance Coulombic part,
\begin{equation}
V(R) \simeq - \frac{4}{3} \, \,  \frac{\alpha_s (R)}{R}
\end{equation}
modified by the confinement potential $\sim \sigma R$ at intermediate
distances $R$ in the tail of the toponium resonances.

\begin{figure}[ht]
\begin{center}
\hspace*{-6cm}
\epsfig{file=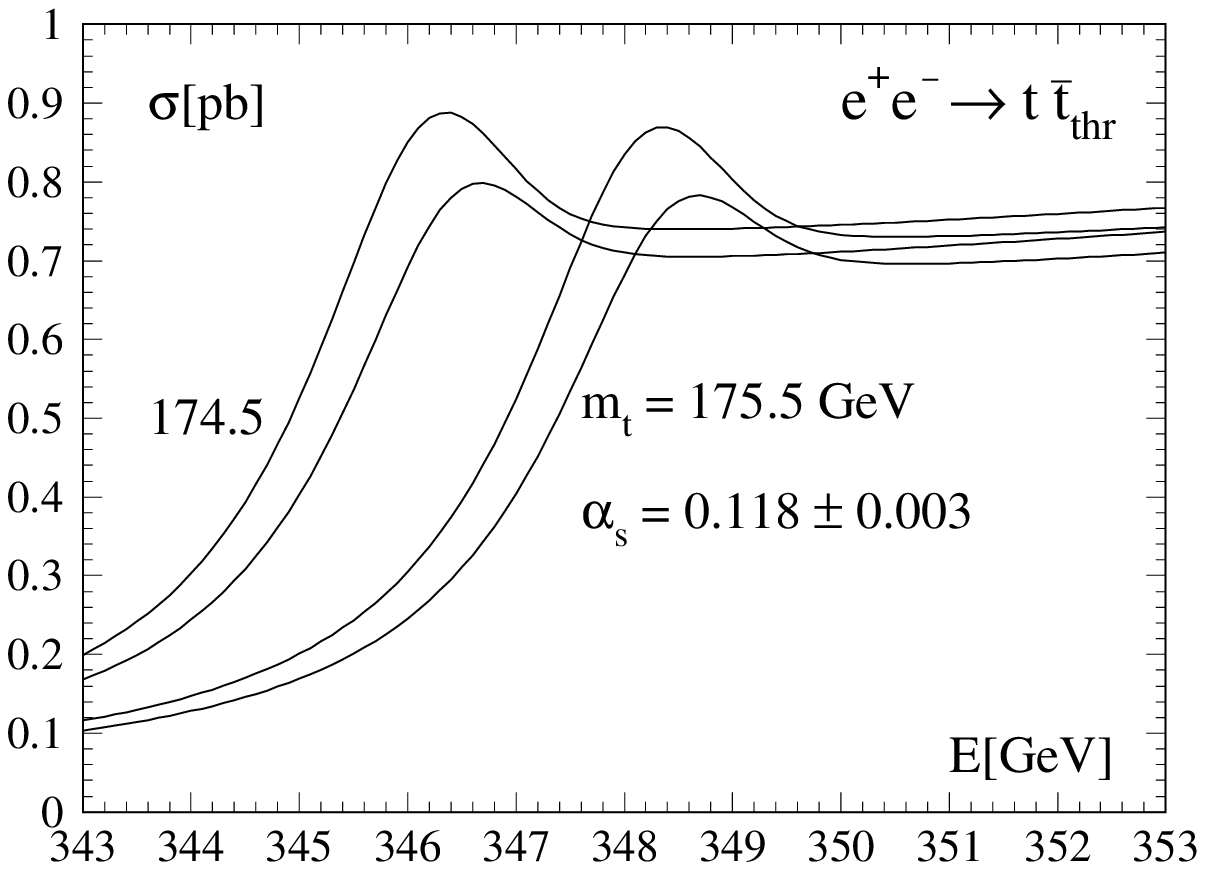,width= 13cm} \\
\vspace*{-6cm}
\hspace*{-6cm}
\epsfig{file=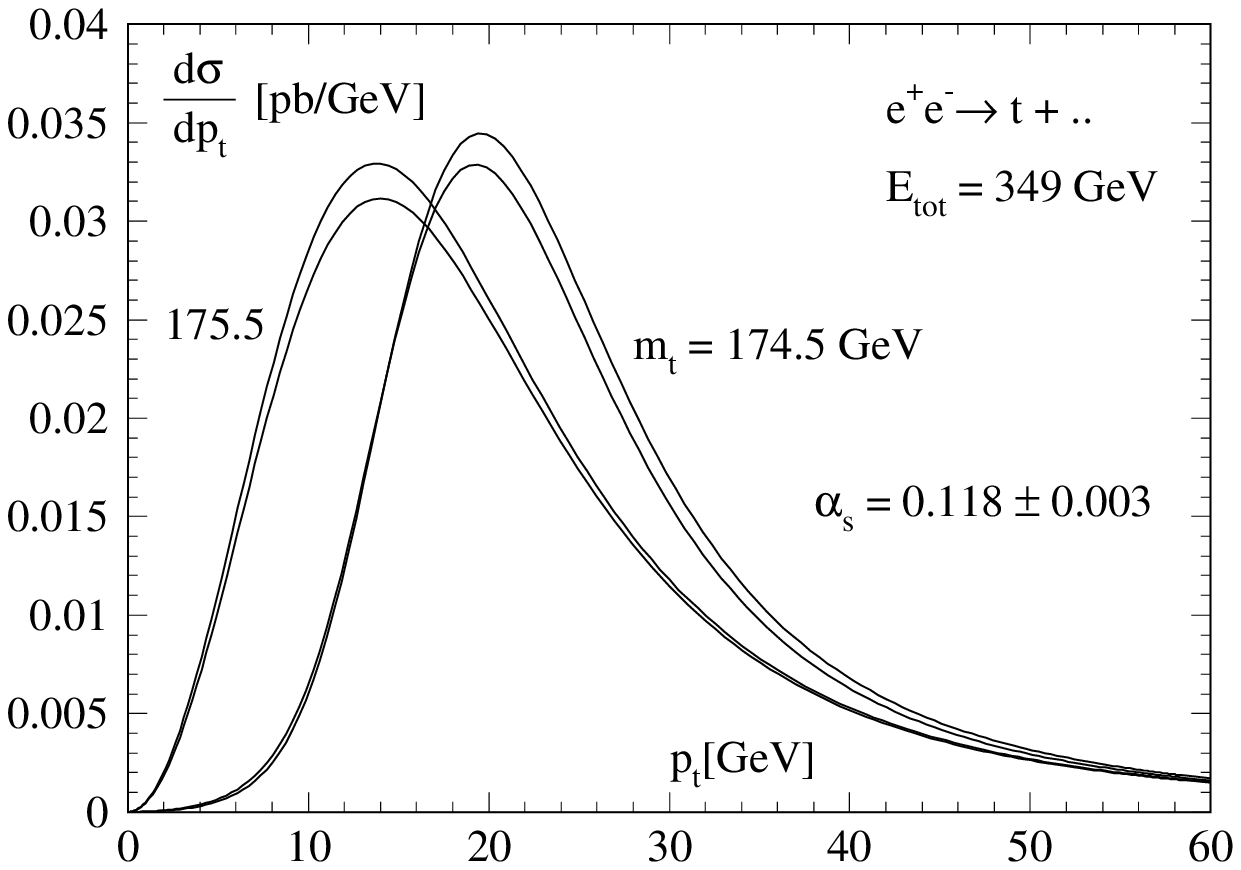,width= 13cm}
\end{center}
\vspace*{-9.cm}
\caption[]{\it 
  Upper part: The cross section for the production of top quarks near
  the threshold.  Demonstrated is the sensitivity of the cross section
  to the value of the top mass and the QCD coupling (normalized at the
  $Z$ mass). Lower part: The momentum spectrum of the top quarks near
  the threshold for a fixed total c.m. energy.  The momentum depends
  strongly on the top mass, yet less on the QCD coupling.
  Refs.\protect\cite{N26,N27}.  \protect\label{f6}\label{sigtot:exp}}
\end{figure}

\STS The excitation curve is built up primarily by the superposition
of the $nS$ states.  This sum can conveniently be performed by using
non-relativistic Green's function techniques:
\begin{equation}
\sigma (e^+e^- \rightarrow t \overline{t})_{thr} =
\frac{24 \pi^2 \alpha^2 e^2_t}{m^4_t}
{\cal I}m \, G (\vec{x}= 0; E + i \Gamma_t)
\end{equation}
The form and the height of the excitation curve are very sensitive to
the mass of the top quark, but less to the value of the QCD coupling,
Fig.\ref{f6}a.  Since any increase of the $t$ quark mass can be
compensated by a rise of the QCD coupling, which lowers the energy
levels, the measurement errors of the two parameters are positively
correlated.

\clearpage

\STS This correlation can partially be resolved by measuring the
momentum of the top quark \cite{N27} which is reflected in the
momentum distribution of the decay $W$ boson.  The $t$ momentum is
determined by the Fourier transform of the wave functions of the
overlapping resonances:
\begin{equation}
\frac{d \sigma}{d P_t} = \frac{24 \pi^2 \alpha^2 e^2_t}
{s} \frac{\Gamma _t}{m^2_t}
| \hat{G} (P_t, E + i \Gamma_t) |^2
\end{equation}
The top quarks, confined by the QCD potential, will have average
momenta of order \linebreak $\sim {\scriptstyle \frac{1}{2}} \alpha_s
m_t$; together with the uncertainty $\sim \sqrt{\Gamma_t m_t}$ due to
the finite lifetime, this leads to average momenta $<P_t>$ of about 15
GeV for $m_t \sim 175$~GeV.  The measurement of the top mass and the
QCD coupling by analysing the $t$ momentum spectrum is therefore
independent of the analysis of the excitation curve, Fig.\ref{f6}.

\STS The Higgs exchange between the top quarks generates a small
attractive Yukawa force which enhances the attractive QCD force
\cite{N28}.  Since the range of the Yukawa force is of order
$m^{-1}_H$, the effect on the excitation curve is small and restricted
to Higgs mass values of order 100 GeV.

\begin{figure}[ht]
\begin{center}
\epsfig{file=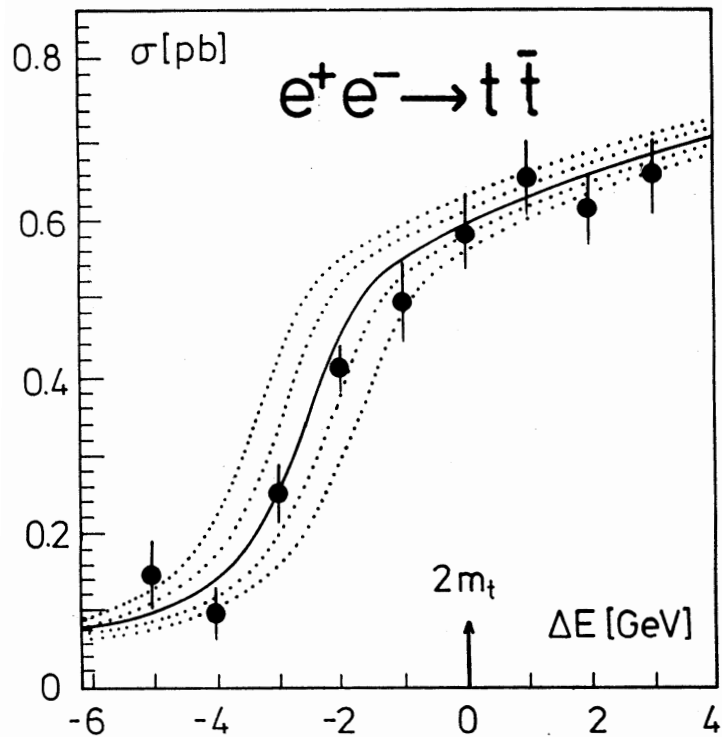,width=10cm,angle=89.5}
\end{center}
\vspace{-.5cm}
\caption[]{\it 
  Excitation curve of the top quarks including initial-state
  radiation and beamstrahlung.  The errors of the data points
  correspond to an integrated luminosity of $\int \LUM$ = 50 fb$^{-1}$
  {\it in toto}.  The dotted curves indicate shifts of the top mass by
  200 and 400 MeV.  Ref.\protect\cite{20A,N29}.  \protect\label{f7}}

\end{figure}

\GS Detailed experimental simulations at $m_t \sim 175$ GeV predict
the following sensitivity to the top mass and the QCD coupling,
Fig.\ref{f7}, when the measurements of the excitation curve and the
$t$ momentum spectrum are combined \cite{20A,N29}:
\begin{eqnarray}
\delta m_t & \approx & 120 \makebox{ MeV} \nonumber \\
\delta \alpha_s & \approx & 0.003
\nonumber
\end{eqnarray} 

\noindent
These errors have been derived for an integrated luminosity of $\int
{\LUM}$~$= 50 \makebox{ fb}^{-1}$.

\STS At proton colliders a sensitivity of about 2 GeV has been
predicted for the top mass, based on the reconstruction of top quarks
from jet and lepton final states.  Smearing effects due to soft stray
gluons which are radiated off the $t$ quark before the decay and off
the $b$ quark after the decay coherently, add to the complexity of the
analysis.  Thus, \ee colliders will improve our knowledge on the
top-quark mass by at least an order of magnitude.

\GS Why should it be desirable to measure the top mass with high
precision? Two immediate reasons can be given:

\STS {$(i)$} Top and Higgs particles affect the relations between
high-precision electroweak observables, $Z, W$--boson masses,
electroweak mixing angle and Fermi coupling, through quantum
fluctuations \cite{N30}.  The radiative corrections can therefore be
used to derive stringent constraints on the Higgs mass, $M_H = f(M_Z,
M_W, m_t )$, which must eventually be matched by the direct
measurement of the Higgs mass at the LHC and the linear collider.
Assuming a measurement of the $W$ mass with an accuracy of 15 MeV [see

\begin{figure}[ht]
\begin{center}
\epsfig{file=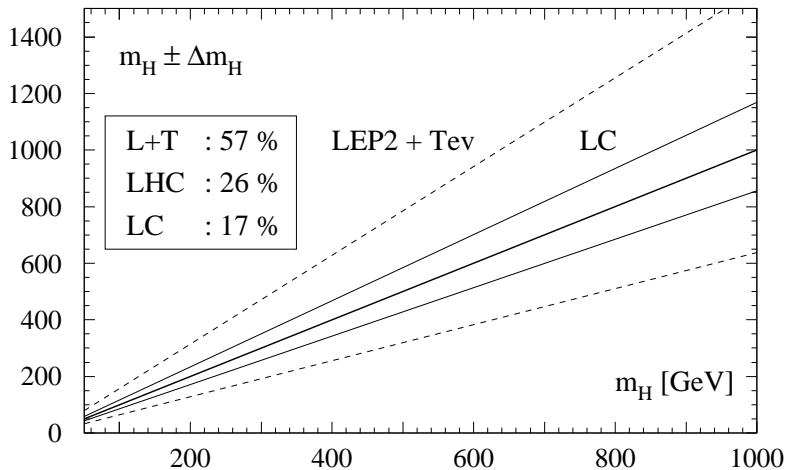,width=11cm}
\end{center}
\vspace{-.5cm}
\caption[]{\it 
  The error on the Higgs mass extracted from the radiative corrections
  to the high-precision electroweak observables: $M_H = \rho (M_Z,
  M_W, m_t)$.  The input values [$ \delta M_W, \delta m_t$] are
  assumed as follows: LEP2 + Tevatron = [30 MeV, 4 GeV]; LHC/Tev33 =
  [15 MeV, 2 GeV], and LC = [15 MeV, 200 MeV].  The LHC lines fall in
  between the dashed and thin solid lines.
  \protect\label{f9}\label{mHrad}}
\end{figure}

\noindent
later], tight constraints on the Higgs mass can be derived if the top
mass is measured with high accuracy.  This is demonstrated in
Fig.\ref{f9}, where the error on the predicted Higgs mass in the
Standard Model is compared for two different errors on the top mass,
$\delta m_t = 4 $~GeV and 200~MeV.  The error in $\alpha (M_Z^2)$ has
been assumed at the ultimate level of $3 \NT 10^{-4}$ \cite{33A}.
[Doubling this error to the present standard value does not have a
dramatic effect.]  It turns out that the Higgs mass can finally be
extracted from the high-precision electroweak observables to an
accuracy of about 17\%.  Thus, high precision measurements of the top
mass allow the most stringent tests of the mechanism breaking the
electroweak symmetries at the quantum level.

\STS {$(ii)$} Fermion masses and mixing angles are not linked to each
other within the general frame of the Standard Model.  This deficiency
will be removed when in a future theory of flavor dynamics, which may
be based for example on superstring theories, these fundamental
parameters are interrelated.  The top quark, endowed with the heaviest
mass in the fermion sector, will very likely play a key r\^ole in this
context.  In the same way as present measurements test the relations
between the masses of the electroweak $W, Z$ vector bosons in the
Standard Model, similar relations between lepton and quark masses will
have to be scrutinized in the future.  With a relative error of about
1 permille, the top mass will be the best-known mass value in the
quark sector, the only value matching the precision of the $\tau$ mass
in the lepton sector.

\section[QCD Physics]{QCD Physics}

\subsection[Annihilation Events]{Annihilation Events}

The annihilation of \ee into hadrons provides a high-energy source of
clean quark and gluon jets: $e^+e^- \to \gamma^*/Z^* \to
q\overline{q}, \ q \overline{q}g$ ...  This has offered unrivaled
opportunities for QCD tests at machines such as PETRA and LEP.  The
program will be continued at a linear collider, although separation
from new `backgrounds' such as top and $W/Z$ pair production will
require more delicate analyses of multijet events. Conversely, the
study of these other processes, as well as the new particle searches,
require a good understanding of the annihilation events.  Topics of
interest for QCD {\it per se} include the study of multijet
topologies, the energy increase of charged multiplicity, particle
momentum spectra and their scaling violations, angular ordering
effects, hadronization phenomenology (power corrections), and so on.

\begin{figure}[h]
\begin{center}
\mbox{\epsfig{file=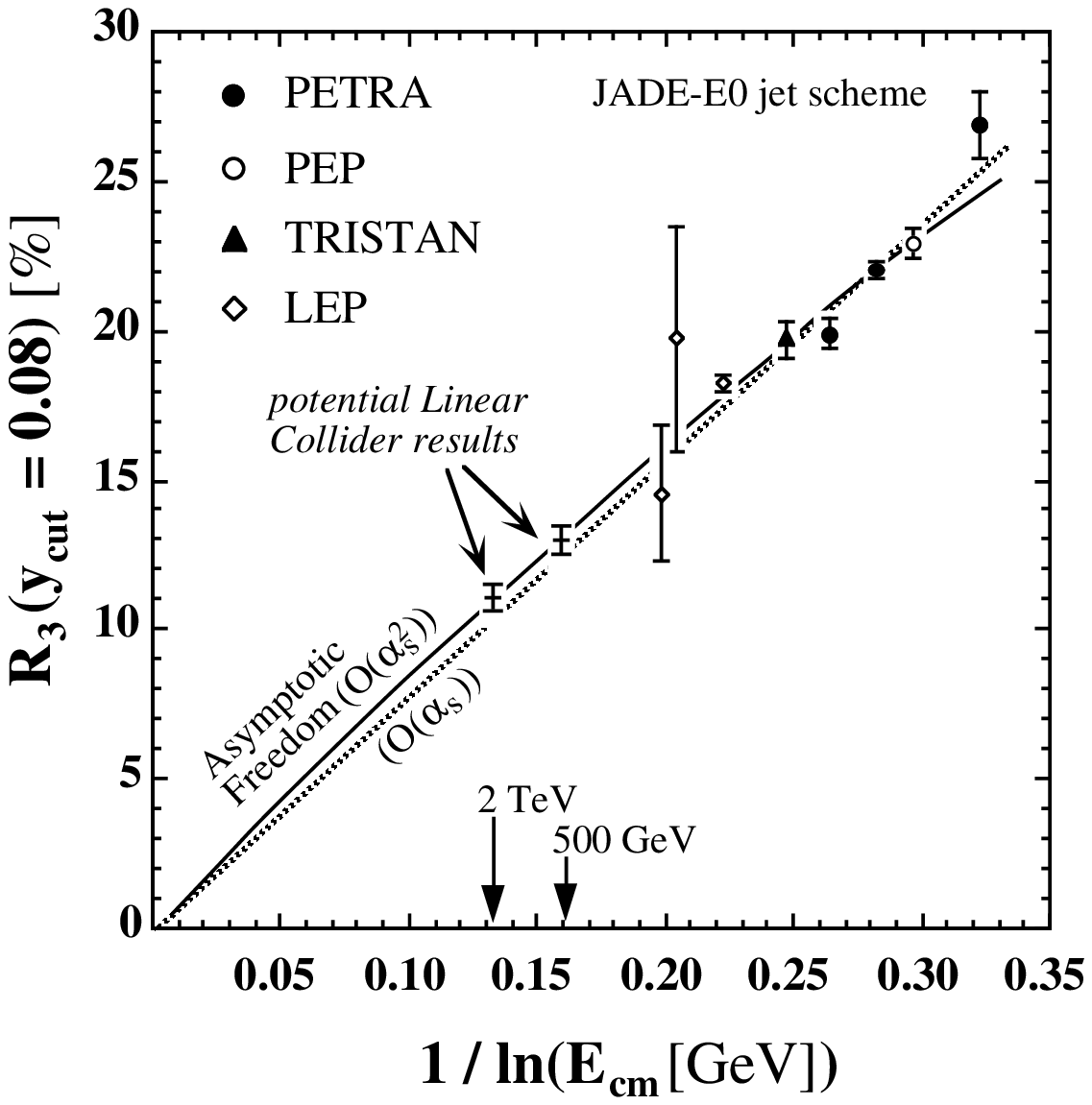,width=7.2cm}} 
\raisebox{6mm}{\epsfig{file=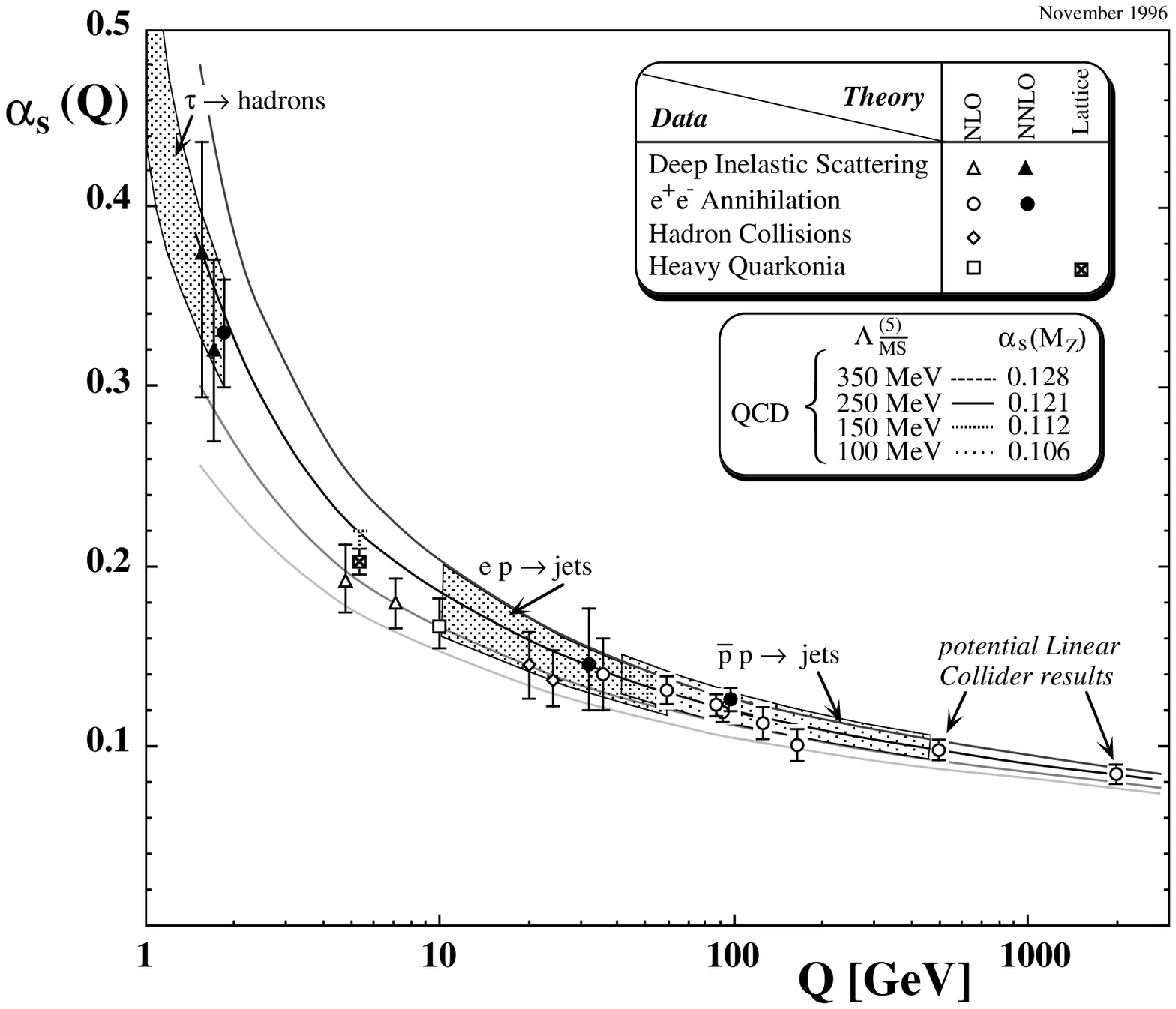,width=7.8cm}} 
\end{center}
\vspace{-5mm}
\caption[]{\it 
  Left: The energy dependence of the three-jet fraction in
  annihilation events.  Right: The energy dependence of $\alpha_s$.
  [Current data and simulated LC points; Ref.\protect\cite{N32}]
  \protect\label{QCD_R3_alphas}}
\end{figure}

\STS One of the key elements of quantum chromodynamics is asymptotic
freedom \cite{N31}, a consequence of the non-abelian nature of the
color gauge symmetry. This fundamental aspect has been tested in many
observables measured at $e^+e^-$ colliders and other accelerators
between a minimum $Q^2$ of order 4 GeV$^2$ up to $ 4 \NT 10^4
\makebox{ GeV}^2$, ranging from the $\tau$ lifetime to multi-jet
distributions in $Z$ decays.  The range of $Q^2$ can be extended at
$e^+e^-$ linear colliders by as much as two orders of magnitude to a
value $Q^2 \sim 4 \NT 10^6 \makebox{ GeV}^2$,
Fig.~\ref{QCD_R3_alphas}.  The most sensitive observable in this
energy range is the fraction of events with 2, 3, 4, \ldots\ jets in
the final state of $e^+e^- \rightarrow$ hadrons \cite{N32}.  The
results of the simulations can be nicely illustrated,
Fig.~\ref{QCD_R3_alphas}, by presenting the evolution of the three-jet
fraction in the variable $1/ \log Q^2$. Asymptotic freedom predicts
this dependence to be linear, modified only slightly by higher order
corrections.  Based on the present theoretical accuracy of the
perturbative jet calculations, the error with
which the QCD  coupling
at $\sqrt{s} = 500$~GeV can be measured, is expected to be $\delta
\alpha_s(M_Z^2) \simeq 0.005$ matching the error which can be expected
from the analysis of the top excitation curve at threshold.  If the
theoretical analysis of the jet rates can be improved, the error on
$\alpha_s$ can be reduced significantly.  

\STS
\subsection[$\gamma \gamma$ Events]{$\gamma\gamma$ Events}

$\gamma\gamma$ interactions provide a complementary way to study many
aspects of new physics. These applications are covered in the
respective physics sections. In addition, the objective of a
$\gamma\gamma$ physics program is to bring our understanding of the
photon to the same level as HERA is achieving for the proton. Since the
photon is the more complex of the two, as described below, this
will offer new insights in QCD \cite{N9,1A}.

\GS Linear $e^+e^-$ colliders offer three {\it sources of photons}:
$(i)$ bremsstrahlung \cite{1B}, $(ii)$ beamstrahlung \cite{1C} and
$(iii)$ potentially, from laser backscattering \cite{N10}. The
brems\-strahlung source provides a spectrum of different photon
energies and virtualities, but distributions are peaked at the lower
end so that the more interesting studies at higher $\gamma\gamma$
energies are limited by statistics.  Since beamstrahlung is a drawback
for the normal $e^+e^-$ physics program, current machine designs
attempt to reduce the beamstrahlung energy to a minimum, so that it
may not be interesting for $\gamma\gamma$ physics.  The laser
backscattering option, on the other hand, offers the prospect of
intense beams of real photons with an energy up to about 80\% of the
$e^{\pm}$ beam.  With one or both beams backscattered it would be
possible to study both deep inelastic scattering off a real photon and
the interactions of two real photons at very high energies. The
$\gamma\gamma$ studies are possible for both the $e^+e^-$ and the
$e^-e^-$ modes; the latter would have some advantages in terms of
lower backgrounds from other processes.

\GS (a) The {\it nature of the photon} is complex.  A photon can
fluctuate into a virtual $q\overline{q}$ pair. The low-end part of the
spectrum of virtualities is in a non-perturbative r\'egime, where the
Vector Meson Dominance (VMD) model can be used to approximate the
photon properties by those of mesons with the same quantum numbers as
the photon --- mainly the $\rho^0$. The high-end part, on the other
hand, is perturbatively calculable \cite{1D}. These `resolved' parts
of the photon with a spectrum of order $\alpha / \alpha_s$, can
undergo strong interactions of order $\alpha_s$. Therefore they can
dominate in cross section over the nonfluctuating `direct' photons,
whose interactions are of ${\cal O}(\alpha)$. The direct/resolved
subdivision of interactions is unambiguous only to leading order, but
also in higher orders it is possible to introduce a pragmatic
subdivision, as has been demonstrated for $\gamma p$ physics at HERA.
In the direct interactions the full photon energy is used to produce
(high-$p_{\perp}$) jets, whereas the resolved photon leaves behind a
beam remnant that does not participate in the primary interaction.

\GS (b) The {\it total cross section} of $\gamma\gamma$ interactions
is not understood from first principles. This situation is analogous
with that for $pp/p\overline{p}$ and $\gamma p$ cross sections, but
not identical.  Therefore the possibility of systematic comparisons
between $pp/p\overline{p}$, $\gamma p$ and $\gamma\gamma$ at a wide
range of energies could shed light on the mechanisms at play
\cite{1E}.\pagebreak

\begin{figure}[tbp]
\begin{center}
\mbox{\epsfig{file=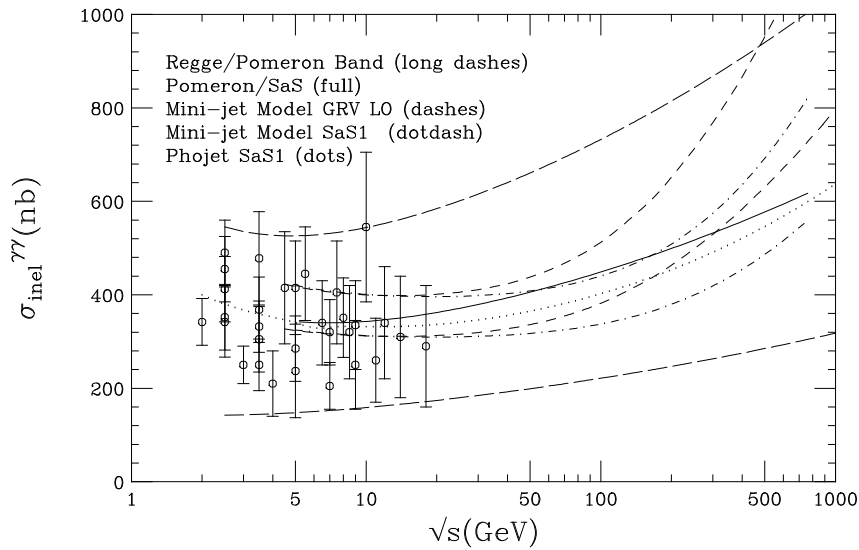,width=9cm}}
\end{center}
\vspace{-4cm}
\caption[]{\it 
  The inelastic $\gamma\gamma$ cross section, i.e. the major part of
  the total $\gamma\gamma$ cross section, as a function of the
  $\gamma\gamma$ c.m. energy $\sqrt{s}$; Ref.\protect\cite{2A}.
\label{gamgam_sigtot}}
\end{figure}

\STS The uncertainty in our current understanding is illustrated in
Fig.~\ref{gamgam_sigtot}, where three representative models are
compared with low-energy data \cite{2A}. The Pomeron/SaS model is
based on a simple ansatz with an $s^{0.08}$ asymptotic rise, in accord
with $pp/p\overline{p}$ and $\gamma p$ experience. The minijet model
is based on an eikonalization of the mini-jet cross section, with
parameters extrapolated from the $\gamma p$ case. In the `dual
topological unitarization' model of {\sc Phojet} also elastic and
diffractive topologies are included in an eikonalization approach.  It
is worth noticing that all three predictions, as well as current LEP
data, are consistent with a straightforward application of
factorization and Regge behavior.  The long-dashed region in the
figure is obtained from the ans\"atze
\begin{equation}
\sigma_{ab}^{tot}=X_{ab}s^{\epsilon}+Y_{ab}s^{-\eta}
\end{equation}
with $\epsilon=0.079$ and $\eta=0.46$ from the average for all high
energy cross-sections, X and Y extracted from pp and $\gamma p$ data,
according to $X_{\gamma \gamma}=X_{\gamma p}^2/X_{pp}$. The band
corresponds to the error induced by the uncertainty on $X_{\gamma p}$.

\STS The total cross section can be subdivided into several
components.  The elastic and diffractive ones correspond to events
like $\gamma\gamma \to \rho^0 \rho^0$, $\gamma\gamma \to \rho^0 X$ and
$\gamma\gamma \to X_1 X_2$. Studies of these would further probe the
nature of the photon and the Pomeron, \pagebreak
\noindent 
while $\gamma\gamma \to \pi^0 X$
and $\gamma\gamma \to \pi^0 a_2^0$ would probe the Odderon \cite{3A}.
A study of the $p_{\perp}$ dependence could highlight the transition
from the soft Pomeron to the perturbative one. Studies of rapidity gap
physics could provide further insights into an area that currently is
attracting intense interest at HERA. However, it will be difficult to
realize the full potential of many of these topics, since most of
these particles are produced at very small angles below 40~mrad, where
they will be undetectable.

\GS (c) The cross section for deep inelastic scattering off a real
photon, $e \gamma \to e' X$, is expressed in terms of the {\it
  structure functions} of the photon \cite{1D}. To leading order,
these are given by the quark content, e.g.
\begin{equation}
F_2^{\gamma}(x,Q^2) = \sum_q e_q^2 \left[ x q^{\gamma}(x,Q^2) +
x \overline{q}^{\gamma}(x,Q^2) \right] ~.
\end{equation}
The parton distributions obey $Q^2$ evolution equations which, in
addition to the homogeneous terms familiar for the proton, also
include inhomogeneous terms related to the $\gamma \to q\overline{q}$
branchings. Experimental input is needed to specify the initial
conditions at some reference scale $Q_0^2$.  Data at larger $Q^2$ and
smaller $x$ values than currently accessible would both provide
information on the quark/gluon structure of the photon and offer
consistency checks of QCD.

\begin{figure}[tbp]
\begin{center}
\mbox{\epsfig{file=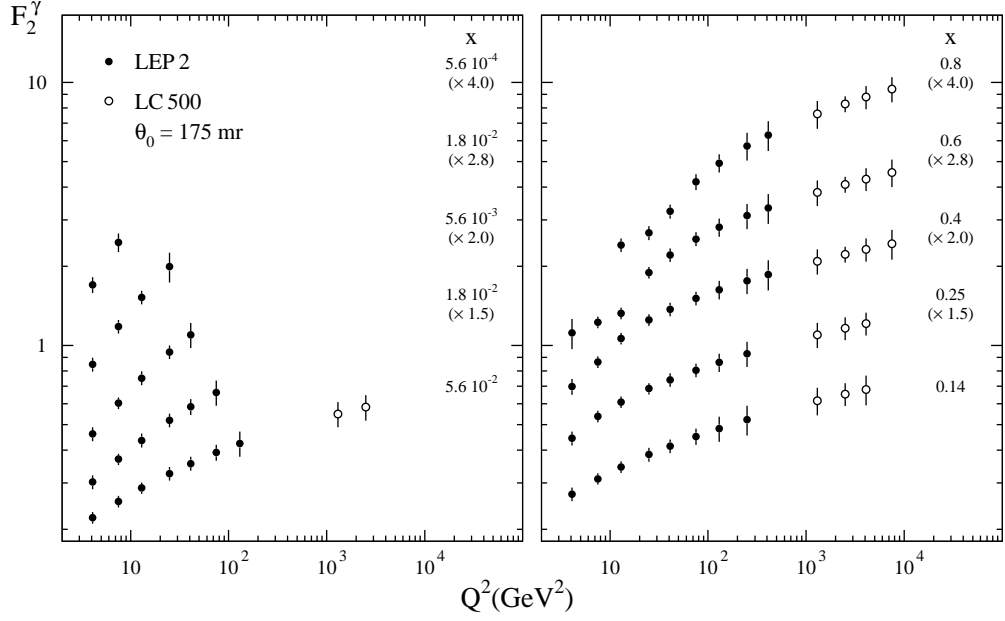,width=14cm}}\\[-2mm]
(a)\\[2mm]
\mbox{\epsfig{file=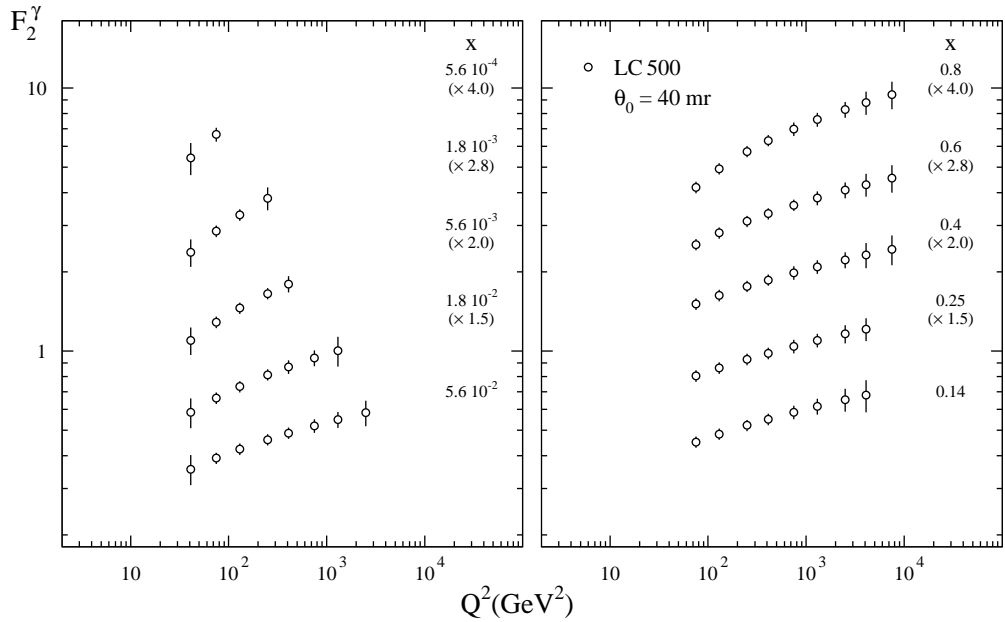,width=14cm}}\\[-2mm]
(b)
\end{center}
\vspace{-5mm}
\caption[]{\it 
  (a) Simulated data points for $e$ tagging outside 175~mrad for
  10~fb$^{-1}$ at 500~GeV, compared with LEP2 expectations;
  Ref.\protect\cite{4A}.  (b) Simulated data points if tagging is
  feasible outside 40~mrad.  \protect\label{gamgam_F2}}
\end{figure}

\STS Electron tagging outside of a cone of about 175 mrad will give
access to a previously unexplored high-$Q^2$ range,
Fig.~\ref{gamgam_F2}a \cite{4A}, but will give neither overlap with
LEP~2 results nor sensitivity to the small-$x$ region.  To achieve the
overlap with LEP~2, one needs an electron tagging device inside the
shielding, down to about 40 mrad, Fig.~\ref{gamgam_F2}b.  This is
however still not sufficient to unfold $x$ measurements in the region
$x \leq 0.1$, since small $x = Q^2/(Q^2 + W^2)$ correspond to large
$W^2$, where an unknown part of the hadronic system disappears
undetected in the forward direction.  In order to circumvent this
problem, and also for reducing the main systematic errors at high $x$,
the $e \gamma$ laser backscattering mode is ideal.

\begin{figure}[tbp]
\begin{center}
\raisebox{-8mm}{\epsfig{file=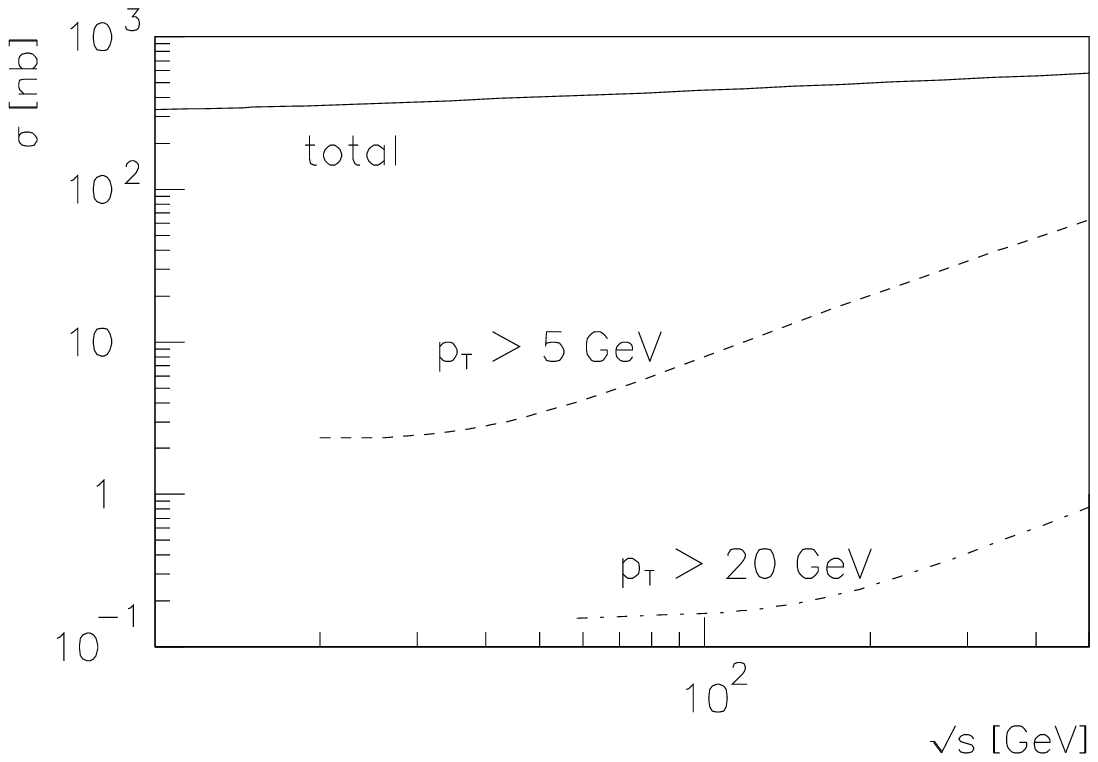,width=8.0cm}}
\mbox{\epsfig{file=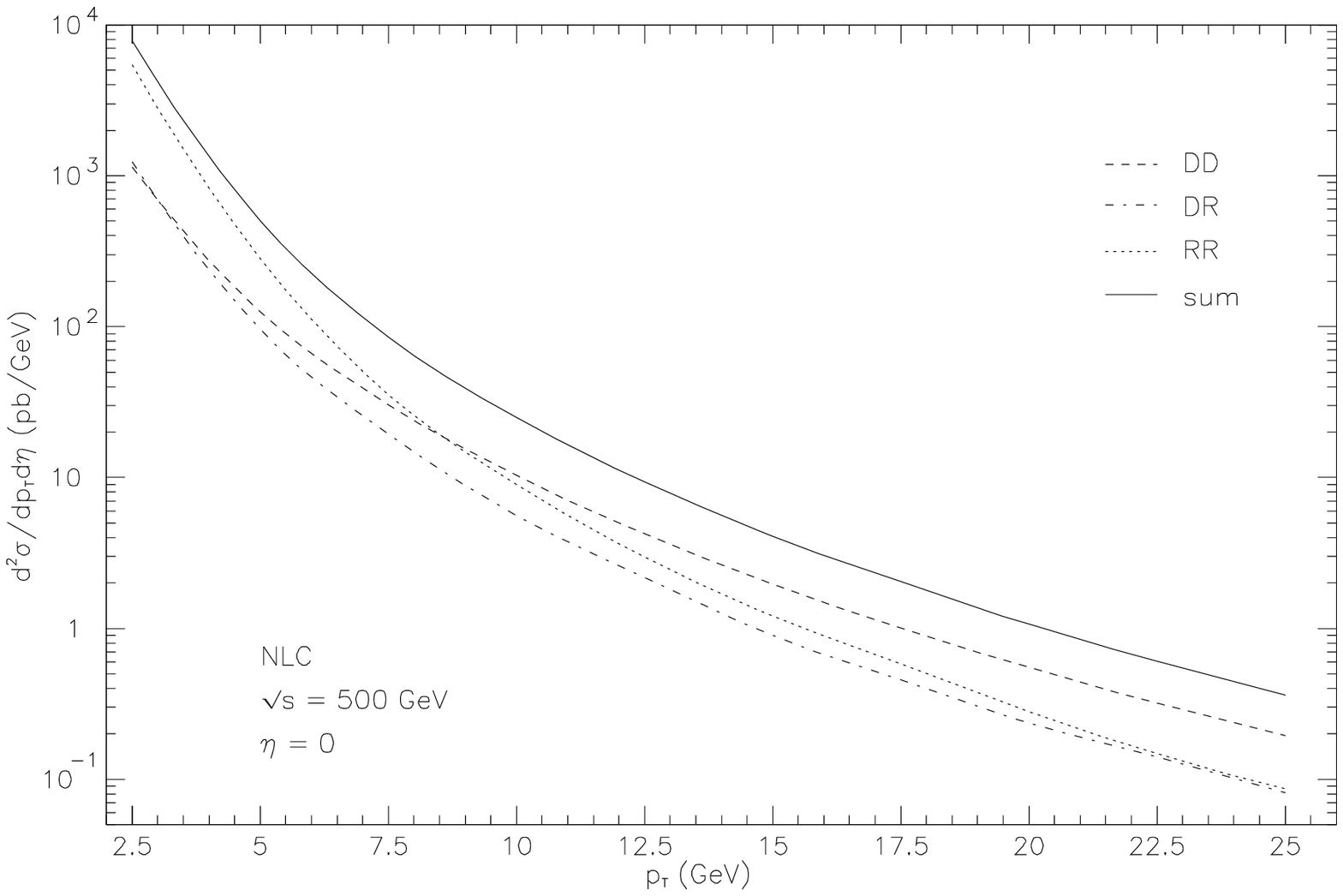,width=7.2cm}}
\end{center}
\vspace{-11mm}
\caption[]{\it 
  Left: Parton-level jet cross sections for interactions with
  transverse momenta above 5~GeV and 20~GeV, respectively, as a
  function of the $\gamma\gamma$ c.m. energy.  Each interaction gives
  two jets.  The total $\gamma\gamma$ cross section is shown for
  comparison.  Right: Next-to-leading order calculations of the jet
  $p_{\perp}$ spectrum \protect\cite{5A} at 500~GeV.  The
  $\gamma\gamma$ cross section has been convoluted with the photon
  flux from bremsstrahlung and beamstrahlung.  The spectrum is also
  shown subdivided into three components: direct (DD), once-resolved
  (DR) and twice-resolved (RR).  \protect\label{gamgam_jet}}
\end{figure}

\STS The longitudinal structure function $F_L$ of the photon, though
very interesting theoretically since it is scale-invariant in leading
order \cite{1F} in contrast to $F_2^{\gamma}$, appears to be very
difficult to measure, having a coefficient $y^2$ in the cross section,
the square of the scaled energy transfer which is generally small.

\GS (d) A non-negligible fraction of the total $\gamma\gamma$ cross
section involves the {\it production of jets},
Fig.~\ref{gamgam_jet}(left).  The jet events \cite{1G} may be
classified according to whether the two photons are direct or one or
both are resolved. The full photon energy is available for jet
production in direct processes, so this event class dominates at large
$p_{\perp}$ values, Fig.~\ref{gamgam_jet}(right).  Here our
understanding of the photon can be tested essentially parameter-free.
At lower $p_{\perp}$ values the resolved processes take over, since
the evolution equations build up large gluon densities at small $x$
and since the gluon-exchange graphs that dominate here are more
singular in the $p_{\perp} \to 0$ limit. The low-$p_{\perp}$ region
therefore is interesting for constraining the parton densities of the
photon. It complements the quark-dominated information obtained from
$F_2^{\gamma}(x,Q^2)$. The standard analysis strategy is based on jet
reconstruction, but alternatively the inclusive hadron production as a
function of $p_{\perp}$ could be used \cite{6A}.  These processes are
powerful instruments to constrain the gluon density of the resolved
photon.

\begin{figure}[tbp]
\begin{center}
\raisebox{63mm}{\epsfig{file=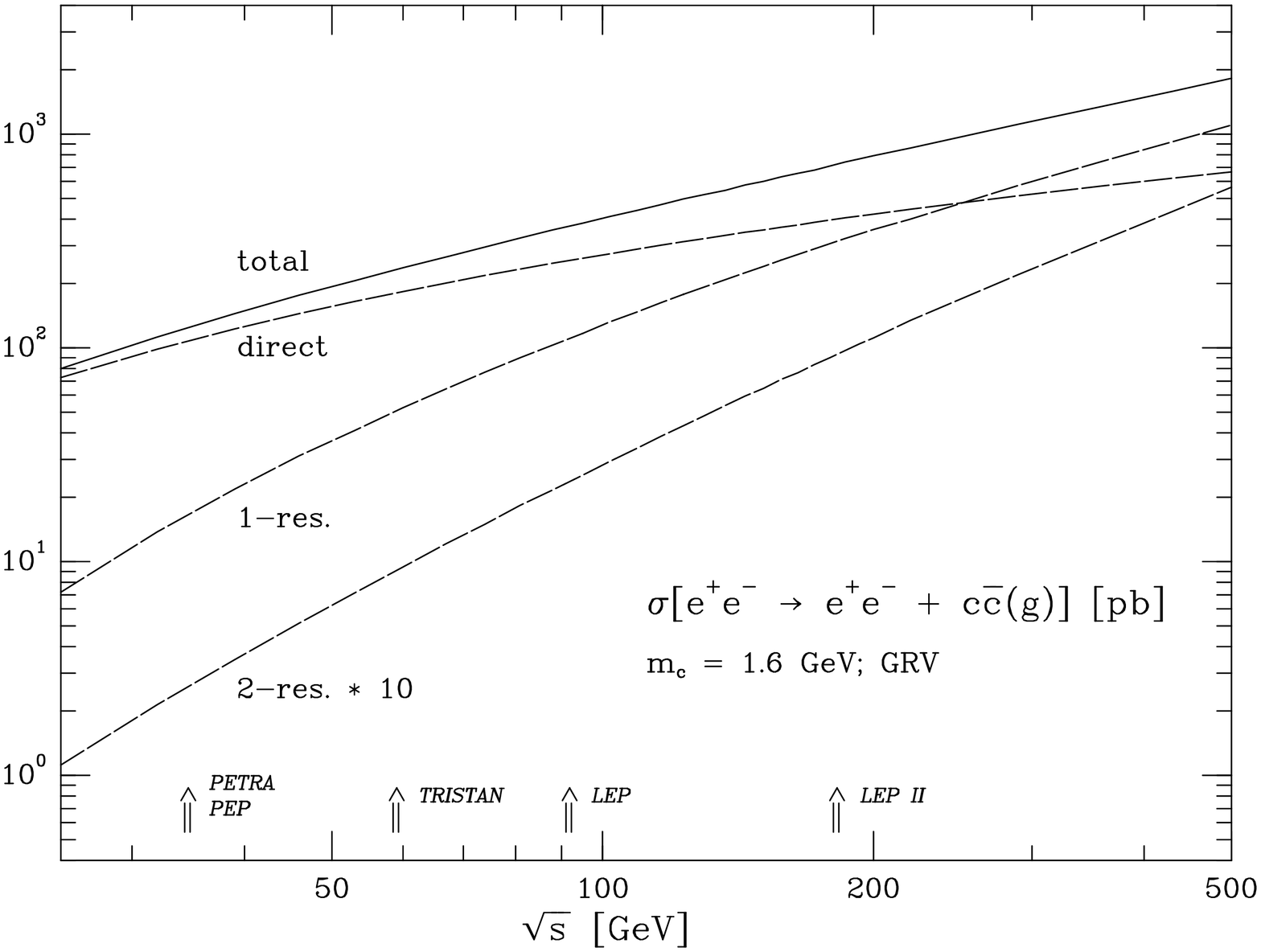,width=6.5cm,angle=-90}}
\mbox{\epsfig{file=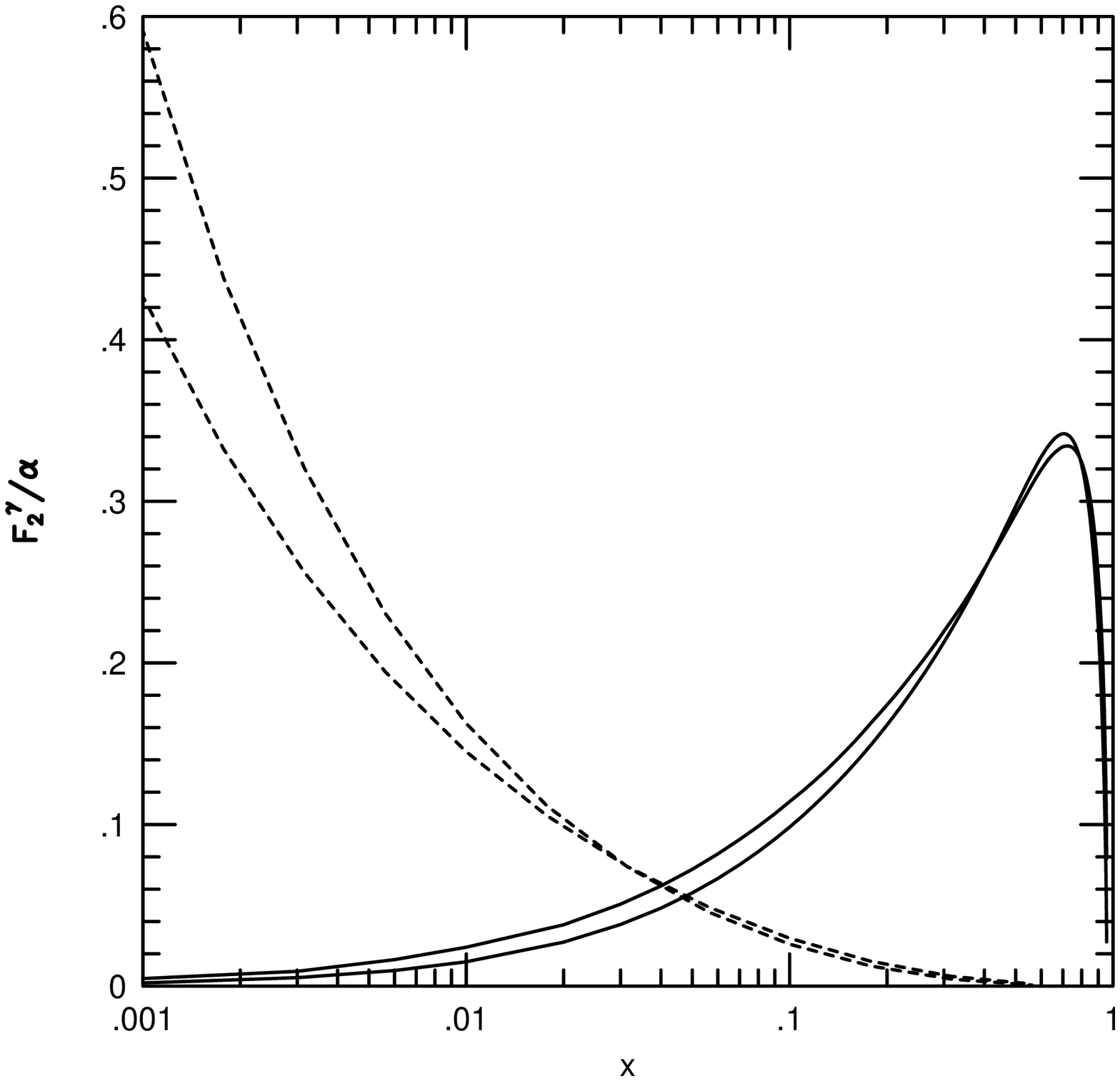,width=6.5cm}}
\end{center}
\vspace{-7mm}
\caption[]{\it 
  Left: The cross section for charm production as a function of
  $e^+e^-$ energy, total and subdivided; Ref.\protect\cite{7A}.
  Right: The hadronic (dashed line) and the pointlike component (solid
  lines) of $F_2^{\gamma,c\overline{c}}/\alpha$ at $Q^2 =
  200$~GeV$^2$.  The two sets of curves correspond to leading and
  next-to-leading order calculations; in the small--$x$ range: dashed,
  upper curve: LO, lower curve: NLO; solid, upper curve: NLO, lower
  curve: LO.  Ref.\protect\cite{8A}.  \protect\label{gamgam_charm}}
\end{figure}

\GS (e) The photon couplings favor {\it charm production}; in the
direct channel $\gamma\gamma \to q\overline{q}$ the charge factor is
$e_q^4$. Therefore significant charm rates can be expected,
Fig.~\ref{gamgam_charm}a. At 500~GeV the once-resolved processes
dominate, and thereby the parton content of the photon is probed.
$J/\Psi$ production is dominated by the process $\gamma + g
\rightarrow J/\Psi + g$, and thus probes the gluon content of the
photon specifically \cite{f51A}.  A related test is offered by the
charm component of $F_2^{\gamma}(x,Q^2)$, Fig.~\ref{gamgam_charm}b,
\cite{8A}. The pointlike part $\gamma^*\gamma \to c\overline{c}$ is
perturbatively calculable, while the hadronic one is dominated by
$\gamma^* g \to c\overline{c}$ and thus probes the gluon content of
the photon.

\GS (f) {\it Double-tagged $\gamma^*\gamma^*$ events} occur at low
rates.  Compared with the Born-term cross section for
$\gamma^*\gamma^* \to q\overline{q}$, the evolution of a BFKL-style
small-$x$ parton distribution inside the photon would boost event
rates by more than a factor 10.  In fact, this process may be
considered the ultimate test of BFKL dynamics \cite{1K}.  With tagging
down to 30--40 mrad it will be possible to detect such a phenomenon,
if present, but more detailed studies would be limited by the low
statistics \cite{9A}.

\STS
\section[Elektroweak Gauge Bosons]{Electroweak Gauge Bosons}

\subsection[$W, Z$ Bosons in the Standard Model]{$W, Z$ Bosons 
in the Standard Model}

The fundamental electroweak and strong forces appear to be of gauge
theoretical origin.  This is one of the outstanding theoretical and
experimental results in the past three decades.  While the
non-abelian symmetry of QCD, manifest in the self-coupling of the
gluons, has been successfully demonstrated in the distribution of
hadronic jets in $Z$ decays, only indirect evidence has been
accumulated so far for the electroweak $W^\pm, Z, \gamma$ sector,
based on loop corrections to electroweak low-energy parameters and
$Z$ observables.  The direct evidence from recent Tevatron and LEP2
analyses is still feeble.  Deviations from the prescriptions of gauge
symmetry manifest themselves in the \css with coefficients $(\beta
\gamma)^2$, destroying fine-tuned unitarity cancellations \cite{N33}
at high energies.  Since the deviations of the static parameters from
the SM values are expected to be of order $[M_W / \Lambda]^j$,
$\Lambda$ denoting the energy scale at which the Standard Model breaks
down, only the very high energies at the LHC and \ee linear colliders
will allow stringent direct tests of the self-couplings of the
electroweak gauge bosons.

\STS The gauge symmetries of the Standard Model determine the form and
the strength of the self-interactions of the electroweak bosons: the
triple couplings $WW \gamma, WWZ$ and the quartic couplings.
Deviations from the form and the strength of these vertices predicted
by the gauge symmetry, as well as novel couplings like $ZZZ$ and
$ZZZZ$ in addition to the canonical SM couplings, could however be
expected in more general scenarios, in models with composite $W, Z$
bosons, for instance.  Other examples are provided by models in which
the $W, Z$ bosons are generated dynamically or interact strongly with
each other.

\begin{figure}[ht]
\begin{center}
\epsfig{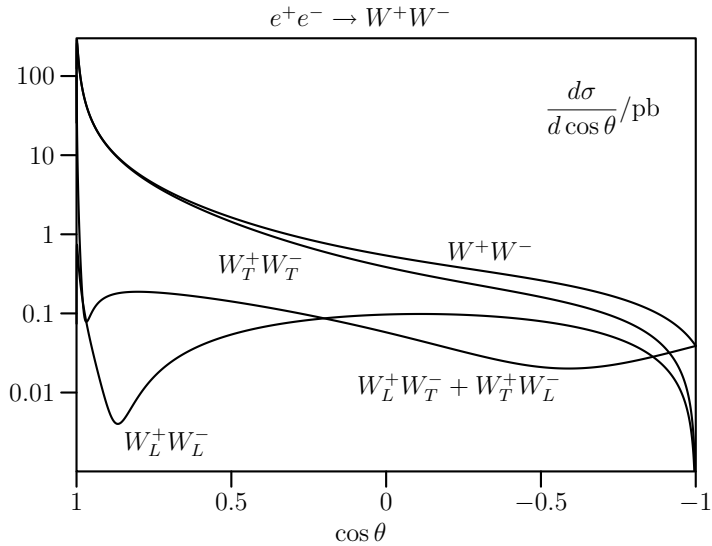}
\end{center}
\vspace{-.5cm}
\caption[]{\it 
  The cross section for $W^+ W^-$ pair production, broken down to the
  transverse and longitudinal helicity components of the $W$ bosons.
  \protect\label{f16}\label{eetoWWOhl}}
\end{figure}

\GS
Pair production of $W$ bosons in \ee collisions,
\[
e^+e^- \longrightarrow W^+ W^-
\]
is the best-suited process to study the electroweak gauge symmetries.
The high efficiency for reconstructing $W, Z$ bosons from hadronic and
leptonic decays in the clean environment of \ee collisions makes a 500
GeV collider superior to the LHC.  Deviations from the predictions of
the Standard Model for the total cross section \cite{N34}, c.f.
Fig.\ref{f16},

\footnotesize

\vspace{-3mm}
\begin{eqnarray}
\sigma (e^+e^- \rightarrow W^+W^-)  & = & 
 \frac{\pi \alpha^2(s)}{2s_w^4} \frac{\beta }{s}
 \left\{  \left[ 1+ \frac{2M_W^2}{s} + \frac{2M_W^4}{s^2} \right]
 \frac{1}{\beta } \log \frac{1+\beta }{1-\beta }
 - \frac{5}{4} \right. \nonumber \\
& & + \frac{M_Z^2(1-2s_w^2)}{s-M_Z^2}
 \left[ 2  \left( \frac{M_W^4}{s^2} + \frac{2M_W^2}{s} \right) 
 \frac{1}{\beta } \log \frac{1+\beta }{1-\beta }
 - \frac{s}{12 M_W^2} - \frac{5}{3} - \frac{M_W^2}{s}  \right] 
\nonumber \\
& & \left. + \frac{M_Z^4(8s_w^4-4s_w^2+1)\beta^2}{48(s-M_Z^2)^2}
 \left[ \frac{s^2}{M_W^4} + \frac{20s}{M_W^2} + 12 \right]  \right\}  
\end{eqnarray}

\pagebreak

\normalsize [$s_w^2 = \sin^2{\theta_w}$ and $\beta$ denoting the $W$
velocity] would signal non-standard self-couplings of the
electroweak gauge bosons.  The most stringent limits can be derived
from the angular distributions of the $W$ pairs and their helicities
\cite{N35} (derived from the decay angular distribu\-tions).  These
analyses can be carried out at collider energies of 500 GeV, promising
sensitivities to non-standard couplings of order 1 percent and
better.

\GS If light Higgs bosons do not exist, the electroweak $W$ bosons
become strongly interacting particles at high energies to comply with
the requirements of quantum-mechanical unitarity.  Strong
interactions between $W$ bosons can be studied in (quasi) elastic $WW
\rightarrow WW$ scattering \cite{N36,N37} and in $WW$ pair production
\cite{N38,58A} at energies
in the TeV range.

\GS
\noindent
a) \underline{High Precision Measurements of $W$--Mass 
and $\sin^2\theta_w$}\hfill \\

\STS
\noindent
The mass of the $Z$ boson has been measured in \ee collisions at LEP1
to an accuracy of 2~MeV.  To obtain a similar precision on the $W$
{\it mass} is therefore a natural goal of future experiments.  Several
methods can be used to measure the \W mass at \ee linear colliders.
Since these machines can be operated near the \WW threshold with high
luminosity, one of the promising methods~\cite{N39} is the scan of the
threshold region near $\sqrt{s}=161$~GeV where the sensitivity of the
cross section to the \W mass is maximal.  [This scan will eventually
allow to measure also the \W width].  Since the uncertainty on the
beam energy is expected to be reduced, using high-precision analyses
of $Z\gamma$ and $ZZ$ events, to well below 10~MeV and the uncertainty
in the measurement of the cross section to well below one percent, an
accuracy
\[
\delta M_W \approx 15 {\rm MeV} 
\]
should finally be reached for the \W mass.

\STS The same accuracy will also be achieved by reconstructing the $W$
bosons in mixed lepton/jet $WW$ final states.  With an experimental
resolution of 3 to 4~GeV on an event-by-event basis, the final error
on the $W$ mass can be expected below $\delta M_W \sim 15 $~MeV for an
integrated luminosity of 50 to 100 fb$^{-1}$ at energies $\sqrt{s}$ of
350 and 500~GeV \cite{39A}.  This measurement of the $W$ mass can be
performed in parallel to other experimental analyses so that the
luminosity requirement for this standard channel remains within the
anticipated frame.

\GS A corresponding high-precision measurement of the {\it electroweak
  mixing angle} $\sin^2\theta_w$ can be performed by operating the
collider at the $Z$ mass where about $10^7$ $Z$ bosons can be expected
in two months of running.  The most sensitive observable for measuring
$\sin^2\theta_w$ is the left/right asymmetry for polarized
electron/positron beams,
\begin{equation}
A_{LR} = \frac{2 v_e a_e}{v_e^2 +a_e^2}
\end{equation}
where $v_e = -1 +4\sin^2\theta_w$ and $a_e = -1$ are the vectorial and
axial $Z$ charges of the electron, respectively.  For $\sin^2\theta_w$
close to 1/4, the sensitivity is enhanced by nearly a full order of
magnitude, $\delta \sin^2\theta_w \approx \frac{1}{8} \delta A_{LR}$.
However, the experimentally measured asymmetry is affected by the
average polarization $P=(P_+ + P_-)/(1+P_+P_-)$ that rises with the
degree of polarization $P_{\pm}$ of the $e^+$ and $e^-$ beams:
$A_{LR}^{{\rm exp}} = PA_{LR}$.  Adopting a sequence of cross section
measurements $e_L^-e_L^+ / e_R^-e_L^+ / e_L^-e_R^+ / e_R^-e_R^+ $
similar to Ref.\cite{39B}, both the degrees of polarization $P_{\pm}$
and the asymmetry $A_{LR}$ can be determined at the same time:
\begin{equation}
\sigma[P_+,P_-]=\sigma_u \left[ 1+P_+P_- +(P_+ +P_-)A_{LR}\right]
\nonumber
\end{equation}
The degree of polarization $P_{\pm}$ however may also be measured by
conventional laser Compton scattering: the error on $P$ is expected of
order $4\NT 10^{-3}$ for $P_- \sim 80$\% and $P_+ \sim 50$\%.  The
systematic error on $A_{LR}^{exp}$ should therefore be close to $7\NT
10^{-4}$.  With a luminosity of $\LUM =$~$10^{32}{\rm cm}^{-2} {\rm
  s}^{-1}$ at $E_{tot}=M_Z$, a sample of $10^7 \ Z$ events can be
collected within two months, giving a statistical error of $3\NT
10^{-4}$ on $A_{LR}$.  From the overall error of $8\NT 10^{-4}$ on
$A_{LR}^{exp}$ \cite{39C}, the absolute error on the electroweak
mixing angle can be reduced to
\[
\delta \sin^2\theta_w \lessim 0.0001 \ 
\]
These are analyses similar to those at LEP1/2 for $\sin^2\theta_w$ and
$M_W$, the increased accuracy of $\sin^2\theta_w$ being matched by the
increased accuracy on $M_W$ at \ee linear colliders.

\vspace{11mm}
\noindent
b) \underline{The Triple Gauge Boson Couplings}\hfill \\

\STS
\noindent
In the most general case the couplings $W^+W^-\gamma$ and $W^+W^-Z$
are each described by seven parameters. Assuming ${\cal C},{\cal P}$
and ${\cal T}$ invariance in the electroweak boson sector, the number
of parameters can be reduced to three \cite{N40},
\begin{eqnarray}
{\cal L}_\gamma /g_\gamma &=& ig_\gamma^1 W^*_
{\mu \nu} W_\mu A_\nu + {\rm
h.c.} \ + \ i \kappa_\gamma W^*_\mu W_\nu  F_{\mu \nu} \ + \ i
\frac{\lambda_\gamma}
{M_W^2} W^*_{\rho \mu} W_{\mu \nu} F_{\nu \rho} \nonumber \\
{\cal L}_Z /g_Z &=& [\gamma \ra Z ]
\end{eqnarray}
The usual couplings $g_\gamma = e$ and $g_Z = e\cot\theta_W$ in the
\SM have been factored out.  In the static limit the $\kappa, \lambda$
parameters $(\Delta \kappa = \kappa -1)$ can be identified with the
$\gamma, Z$ charges of the $W$ bosons and the related magnetic dipole
moments and electric quadrupole moments,
\begin{eqnarray}
\mu_\gamma &=& \frac{e}{2M_W}
\Big[ 2+\Delta \kappa_\gamma +\lambda_\gamma  \Big]  \hspace{7.5mm}
{\rm and}
\ \
\gamma \ra Z     \nonumber \\
Q_\gamma &=&-\frac{e}{M_W^2}
\Big[ 1+\Delta \kappa_\gamma - \lambda_\gamma  \Big]  \hspace{6.5mm}
{\rm and} \ \   \gamma \ra Z
\nonumber
\end{eqnarray}
The gauge symmetries of the \SM demand $\kappa = 1$ and $\lambda = 0$,
i.e. $\mu_\gamma = e/M_W$ and $Q_\gamma = -e/M^2_W$ etc.

\GS The magnetic dipole and the electric quadrupole moments can be
measured {\it directly} through the production of $W \gamma$ and $WZ$
pairs at $p\overline{p} / pp$ colliders and $WW$ pairs at
\ee~colliders.  Detailed experimental simulations have been carried
out for the mixed lepton/jet reaction
\[
e^+e^- \to W^+W^- \to (l \nu_l)(q\bar{q}')
\]
Beam polarization is very useful for disentangling the parameters.
The most stringent bounds on anomalous couplings can be derived from
the measurement of the \W decay angular distributions which reflect
the helicities of the \W bosons.  Bounds of order $10^{-3}$ to
$10^{-4}$ can be reached if the \ee energy is raised at energies of
500~GeV and beyond \cite{40A,f68A}.  The scale $\Lambda$ which can be
probed, extends beyond the energy scale which is accessible directly.

\STS These bounds can be supplemented, separately for $\Delta
\kappa_{\gamma}, \lambda_{\gamma}$ and $\Delta \kappa_Z, \lambda_Z$,
by studying \W pair production in $\gamma\gamma$ Compton colliders
\cite{N41} and single $\gamma/Z$ production in the process $e^+e^- \to
\nu \bar{\nu} \gamma/Z$ \cite{N42}.

\GS
\noindent
\underline{\it Models with Higgs bosons}.  A theoretically plausible
concept for the experimental analysis is based on the assumption that
any deviations from the Standard Model due to new physics manifest
themselves in ${\rm SU}(3) \times {\rm SU}(2) \times {\rm U}(1)$ gauge
invariant SM singlet operators \cite{N43}.  To the extent that the
operators affect the gauge boson propagators, they are stringently
constrained by the high-precision data from $Z$ boson physics etc.
These operators affect the triple boson couplings at a level of less
than $10^{-3}$.  However, there are sets of operators which are only
weakly constrained by propagator effects so that deviations from the
Standard Model of order $10^{-2}$ cannot be excluded \cite{N44} {\it
  a~priori}.  Classifying these operators as
\begin{equation}
\delta {\cal L} = \frac{f_{WWW}}{\Lambda^2} {\cal O}_{WWW}
                 + \frac{f_{W\Phi}}{\Lambda^2} {\cal O}_{W\Phi} +
                  \frac{f_{B\Phi}}{\Lambda^2 } {\cal O}_{B\Phi}
\end{equation}
with ${\cal O}_{WWW} = \mbox{tr } [W^3]$ and ${\cal O}_{W/B\Phi} = (D
\Phi)^*(W/B)(D \Phi)$, the five triple boson couplings can be
expressed by three parameters (see also Ref.\cite{L75A}),

\begin{figure}[ht]
\begin{center}
\epsfig{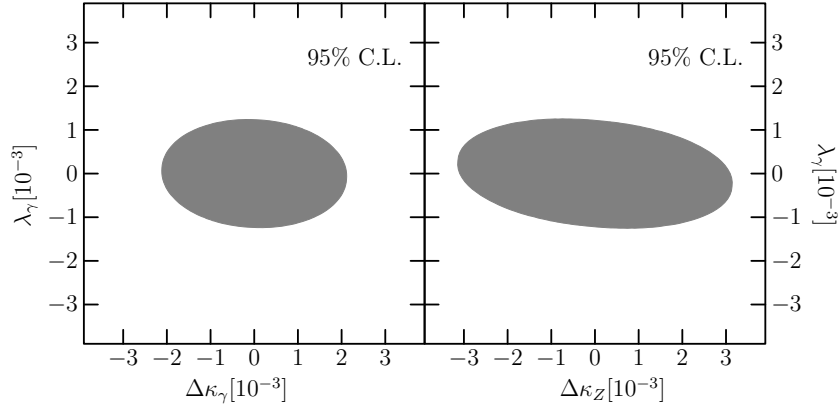}
\end{center}
\vspace{-0.5cm}
\caption[]{\it 
  Measurement of the anomalous couplings $\Delta \kappa_{\gamma}$,
  $\Delta \kappa_Z$, and $\lambda$ of the electroweak gauge bosons.
  The deviations from the Standard Model are taken as singlets under
  the SM symmetry group; $\int \LUM$ = 50 fb$^{-1}$ at $\sqrt{s}$ =
  500 GeV. Ref.\protect\cite{f68A}.  \protect\label{17}\label{}}
\end{figure}

\begin{figure}[ht]
\begin{center}
\epsfig{file=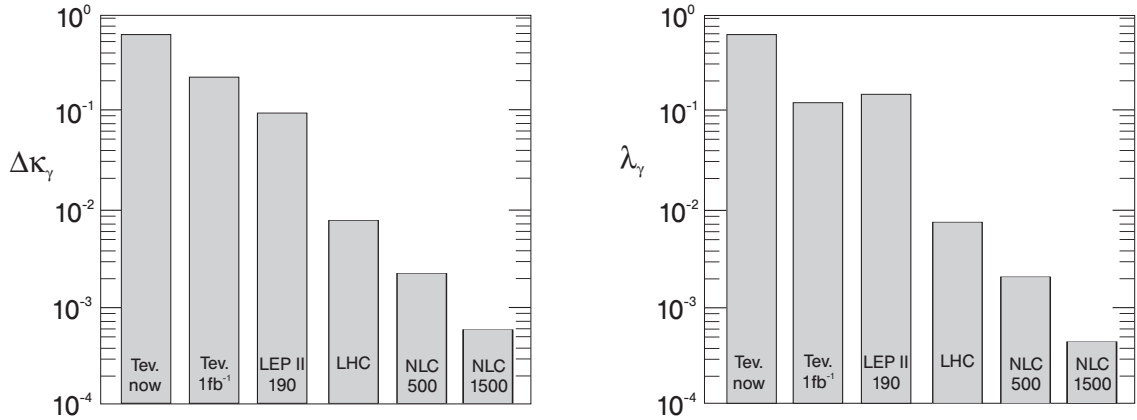,width=15cm}
\end{center}
\vspace{-.5cm}
\caption[]{\it 
  Comparison of the sensitivity on anomalous trilinear \cps of the
  electroweak gauge bosons at various colliders;
  Ref.\protect\cite{803}.  \protect\label{f801}\label{delkap}}
\end{figure}
\clearpage
\begin{eqnarray}
\Delta g^Z_1  & = & \displaystyle \frac{m^2_Z}{2 \Lambda^2} f_{W\Phi}
\makebox[3.0cm]{}
\Delta \kappa_Z  =  \frac{m^2_Z}{2 \Lambda^2}
\left[f_{W\Phi} - s^2_W (f_{B\Phi} + f_{W\Phi})\right] 
\nonumber \\  
& & \nonumber \\
\lambda_Z  & = & \lambda_\gamma = \displaystyle \frac{3m^2_W g^2}
{2 \Lambda^2} f_{WWW}  \makebox[0.8cm]{}
\Delta \kappa_{\gamma}  =  \frac{m^2_Z}{2 \Lambda^2}
c^2_W (f_{B\Phi} + f_{W\Phi})
\nonumber
\end{eqnarray}
The result of a fit for the pairs ($\Delta \kappa, \lambda$) is shown
in Fig.\ref{17}.  The fits give very stringent bounds on the boson
couplings for $\sqrt{s}=500 $~GeV and $\int \LUM =$~50~fb$^{-1}$:
\begin{equation}
\Delta \kappa_\gamma \leq 2\cdot 10^{-3}, \quad
\Delta \kappa_Z \leq 3\cdot 10^{-3} \quad \mbox{and} \quad
\lambda_\gamma \leq 1\cdot 10^{-3} 
\nonumber
\end{equation}
Exploiting the large number of $WW$ events at the LC experiments, the
systematic analysis of the full correlation matrix becomes possible
\cite{f68A,47A}.

\GS The \ee 
colliders are significantly better suited for high-precision analyses
of the self-couplings of the electroweak gauge bosons.  This is a
consequence of the highly efficient reconstruction of $W$ bosons from
the hadron decays in the clean environment of \ee collisions.  A
comparison between the machines, based on two-parameter variations of
the self-couplings, has been performed in Ref.\cite{803}; the result
is reproduced in Fig.\ref{f801}.  At $\sqrt{s} = 1.5 $~TeV the
precision achieved at the LC is one order of magnitude better than at
the LHC.

\begin{figure}[ht]
\begin{center}
\epsfig{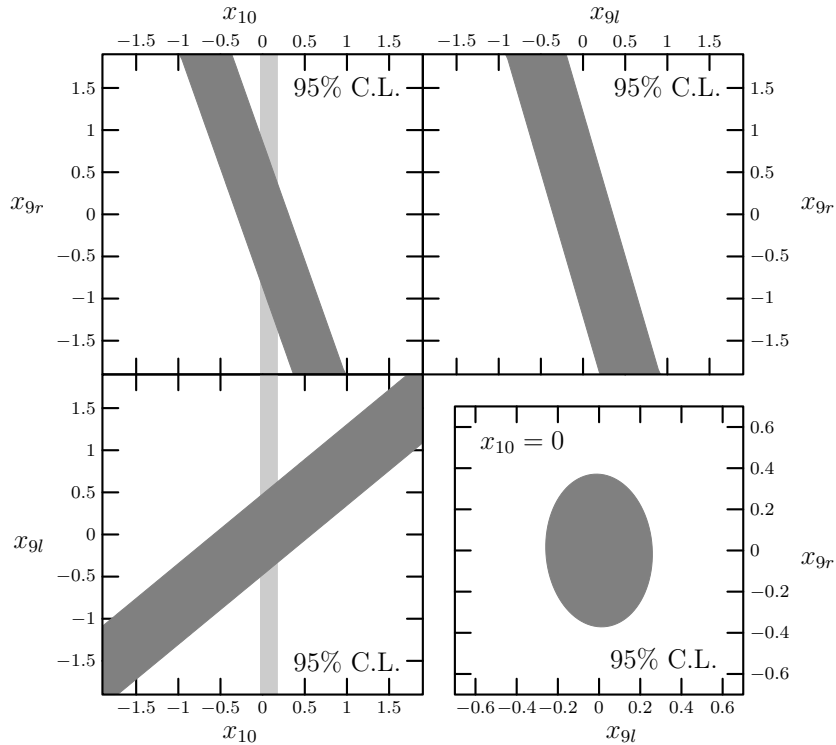}
\end{center}
\vspace{-0.5cm}
\caption[]{\it 
  The coefficients $x_{9R}$, $x_{9L}$ and $x_{10}$ in non-linear
  realizations of the SM gauge symmetries for $\int \LUM =$~$200 \ 
  {\rm fb}^{-1}$ at $\sqrt{s}$ = 800 GeV; Ref.  \protect\cite{49A}.
  The vertical bars indicate the constraint on $x_{10}$ derived from
  LEP analyses.  \protect\label{17t}\label{OhlNippe}}
\end{figure}

\GS
\noindent
\underline{\it Models without Higgs bosons}.  In theories without
light Higgs particles, the electroweak gauge bosons interact strongly
with each other at energies above $\sim$~1~TeV.  Such a scenario can
be described by a non-linear realization of the symmetry in a chiral
Lagrangian formalism \cite{N47},
\begin{eqnarray}
\delta {\cal L} & = & -i \frac{x_{9L}}{16 \pi^2} \mbox{tr }
[g W_{\mu \nu} D_\mu U^+ D_\nu U]
- i \frac{x_{9R}}{16 \pi^2} \mbox{tr } [g' B_{\mu \nu}
D_\mu U^+ D_\nu U] \nonumber \\
 & & \mbox{} + \frac{x_{10}}{16 \pi^2} \mbox{tr }
 [U^+ g' B_{\mu \nu} U g W_{\mu \nu}] \phantom{BBB}
\end{eqnarray}
where $U$ corresponds to the (exponentiated) longitudinal \W field.
Dimensional analysis suggests that the natural size of the
coefficients is $x_i \sim {\cal O} (1)$ for any strongly interacting
field theory so that the corresponding anomalous moments are of order
$10^{-2}$.  Experimental simulations have shown that for $\sqrt{s} =
800$ GeV the parameters $x_{9R}, x_{9L}$ and $x_{10}$ can be
constrained to values of order unity and less, c.f. Fig.\ref{17t},
Ref.\cite{49A}.

\GS The measurement of the quartic couplings requires the production
of three gauge bosons in \ee annihilation \cite{74A} which is
suppressed however by the electroweak couplings and phase space.
Alternatively, part of these \cps can be studied in $\gamma \gamma$
collisions to pairs of gauge bosons \cite{74B}.

\vspace{11mm}
\noindent
c) \underline{Strongly Interacting \W Bosons: \WW Scattering}\hfill \\

\STS
\noindent
If the scenario in which $W/Z$ bosons and light Higgs bosons are
weakly interacting up to the GUT scale is not realized in Nature, the
next attractive physical scenario is a strongly interacting $W/Z$
sector. Without a light Higgs particle with a mass of less than 1 TeV,
the electroweak bosons must become strongly interacting at energies of
about 1.2 TeV to fulfill the requirements of quantum-mechanical
unitarity for $WW$~scattering amplitudes \cite{N49}.  A novel type of
strong interactions may be the physical {\it raison d'\^etre} of these
phenomena.  In such scenarios, new resonances could be realized
already in the ${\cal O}$(1 TeV) energy range.

\STS In scenarios of strongly interacting vector bosons, $W_LW_L$
scattering must be studied at energies of order 1 TeV which requires
the highest energies possible in the 1 to 2~TeV range at \ee and \emem
colliders.  (Quasi)elastic $WW$ scattering can be analyzed by using
$W$ bosons radiated off the electron/positron beams \cite{N36,N37}, or
by exploiting final state interactions in the \ee annihilation to $W$
pairs \cite{N38,58A}.  All possible combinations of weak isospin and
angular momentum [$I,J$] in the $WW$ scattering amplitudes $a_{IJ}$
can be realized in the first process.  The cross sections however are
small until resonances are formed.  Adopting the complementary
rescattering method, the phase shift of the $[I,J]=[1,1]$~\WW~channel
enters the \cs for $e^+e^-$~annihilation to $W^+W^-$~pairs through the
Mushkelishvili-Omn\`es factor
\begin{equation}
a_{11}=a_{11}^0 
\exp\left[\frac{s}{\pi}\int\frac{ds'\delta_{11}(s')}{s'(s'-s)}
 \right] 
\end{equation}
This classical method provides a powerful probe of the
\WW~interactions since the \W bosons are (re)scattered at the maximum
possible energy and the restrictive final-state kinematics allows for
an experimentally clean analysis.

\GS  Non-perturbative interactions of $W, Z$ bosons at a scale of
$\sim$ 1 TeV can be studied better at the LC in the high energy phase
than at the LHC. The LHC is superior for the search of multi-TeV
resonances.  In \ee collisions the $W/Z$ bosons can be reconstructed
from jet decays, while the jetty background at LHC allows only to
trace back $WW$ pairs from mixed hadronic/leptonic decays, involving
undetectable neutrinos. 

\begin{figure}[ht]
\begin{center}
\hspace*{-0.5cm}
\epsfig{file=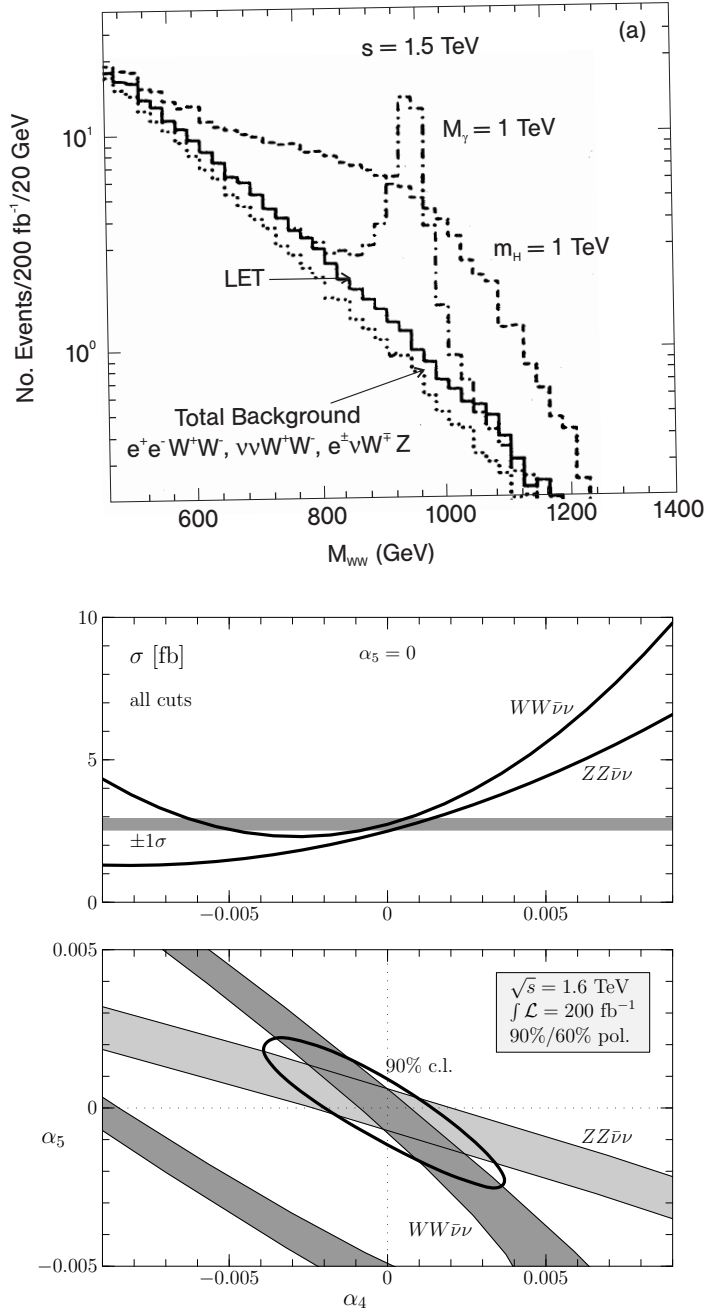,width=9.4cm,angle=1} \\
\vspace{0.5cm}
\hspace{0.6cm}
\epsfig{file=ww_strong_thr.eps,width=7.6cm }
\end{center}
\caption[]{\it 
  Upper part: The distribution of the $WW$ invariant energy in $e^+e^-
  \rightarrow \overline{\nu} \nu WW$ for scalar and vector resonance
  models [$M_H, M_V$ = 1 TeV], as well as for non-resonant $WW$
  scattering in chiral models near the threshold;
  Ref.\protect\cite{N36}.  Lower part: Sensitivity to the expansion
  parameters in chiral electroweak models of $WW \to WW$ and $WW \to
  ZZ$ scattering at the strong-interaction threshold;
  Ref.\protect\cite{N37}.  \protect\label{17tt}\label{PKB}}
\end{figure}

\STS Generating the longitudinal degrees of freedom of the massive
electroweak bosons by absorbing the Goldstone bosons associated with
the spontaneous symmetry breaking of the underlying strong-interaction
theory, the first term in the energy expansion of the \WW scattering
amplitudes $a_{IJ}$ is determined independent of dynamical details:
$a_{00}=+6, a_{11}=+1, a_{20}=-2$ in units of $1/96\pi v^2$.  While in
the isospin $I=2$ channel the $WW$ interaction is repulsive, the
attractive $I=0$ and $I=1$ channels may form Higgs and $\rho$--type
resonances at high energies.  The $H$-- and $\rho$--type resonances
would modify the scattering amplitudes dramatically compared with the
predictions of the light-Higgs scenario, yet the threshold terms
affect the cross sections significantly, too.  For $\sqrt{s}=$~1.5~TeV
the predictions for the $WW$~scattering cross sections in the weak
scenario with a light Higgs mass are confronted with possible strong
scenarios in the upper part of Fig.\ref{17tt} \cite{N36} for vector
and scalar resonances.  The signal of S--wave resonances is enhanced
by the additional $ZZ$ channel.  The sensitivity to the
next-to-leading terms which preserve the custodial SU(2)$_c$ symmetry
in the effective $W^4$ Lagrangian below the resonance region, is
demonstrated in the lower part of Fig.\ref{17tt} \cite{N37} for
$\sqrt{s}=$~1.6~TeV; it follows that the leading chiral contributions
to the \WW scattering amplitudes, which are free of any adjustable
parameters, can be measured to an accuracy of about 10 percent.
[Scalar Higgs-type resonances may also have a large impact on the
production of top-quark pairs in $WW$ collisions, c.f.
Ref.\cite{75A}.]

\STS Similar effects would also be observed in $WW$ pair production,
$\epem \rightarrow W^+W^-$.  This process is very sensitive to the
formation of $[I,J] = [1,1]$ resonances, even if for $M_V > \sqrt{s}$
the new intermediate vector bosons $V$ remain virtual.  A quantitative
analysis has been performed within the BESS model \cite{N50}.  \hfill
The model describes the
\clearpage 
\noindent
interactions of the Goldstone bosons [which are
associated with the spontaneous chiral symmetry breaking and
transformed to the $W_L$ components] with the heavy vector bosons of
the underlying new strong interactions in the most general way.
Disregarding fermion interactions which can readily be incorporated,
the interactions of the new massive vector bosons $V$ among each other
and with the $W$ bosons are characterized by one (gauge) coupling.
The system can therefore be described by two parameters: the mass
$M_V$ and the $V \rightarrow WW$ decay width $\Gamma_V$.  In analogy
to the measurements of the $\rho$--meson parameters in the process
$\epem \rightarrow \rho \rightarrow \pi^+ \pi^-$, the properties of
the vector bosons $V$ can be studied in the reaction $\epem
\rightarrow V \rightarrow W^+W^-$.
  
\begin{figure}[hb]
\begin{center}
\epsfig{file=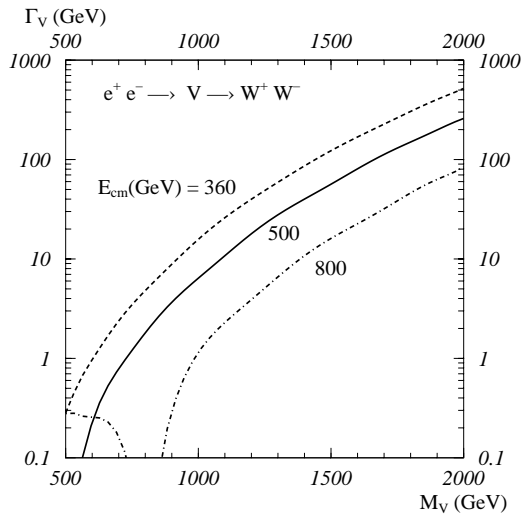,width=7cm}
\end{center}
\vspace{-.5cm}
\caption[]{\it 
  Sensitivity to the mass and the width of new heavy vector resonances
  in rescattering corrections to $e^+e^- \rightarrow W^+ W^-$ pair
  production; the analysis is based on the BESS model.  Shown are the
  90\%~CL contour lines derived from differential \css and left-right
  asymmetries; the $W$ polarizations are reconstructed from the decay
  leptons and quark jets.  No direct \cp of the new vector bosons to
  fermions has been included.  Energies and luminosities have been
  chosen as follows: [$\sqrt{s}, \int{\LUM}$]~=~[360~GeV,
  10~fb$^{-1}$] dashed; [500, 20] solid; [800, 50] dash-dotted.
  \protect\label{f0.18}}
\end{figure}

\STS The regions of the parameter space
$[M_V,\Gamma_V]$ which can be probed at $\sqrt{s} = 360, 500$ and
800~GeV are shown in Fig.\ref{f0.18}.  The sensitivity of $WW$ pair
production at high 
energies exceeds  the sensitivity which could be
reached at LEP1, for $M_V \lessim 0.8$~TeV at $\sqrt{s}=500$~GeV; at
$\sqrt{s}=800$~GeV the sensitivity exceeds the LEP1 range for all mass
values of the vector boson $V$.  The area in parameter space which
will be covered at linear colliders, is also larger than the region
accessible at LHC if the mass $M_V$ is larger than 1~TeV.

\vfill  
\clearpage


\newpage
\vspace{11mm}
\subsection[Extended Gauge Theories]{Extended Gauge Theories}

Despite its tremendous success in describing the experimental data
within the range of energies available today, the Standard Model,
based on the gauge symmetry ${\rm SU}(3) \times$ ${\rm SU}(2) \times
{\rm U}(1)$, cannot be the ultimate theory.  It is expected that in a
more fundamental theory the three forces are described by a single
gauge group at high energy scales.  This grand unified theory would be
based on a gauge group containing ${\rm SU}(3) \times {\rm SU}(2)
\times {\rm U}(1)$ as a subgroup, and it would be reduced to this
symmetry at low energies.

\GS Two predictions of grand unified theories may have interesting
phenomenological consequen\-ces in the energy range of a few hundred
GeV \cite{N51}:

\GS {$(i)$} The unified symmetry group must be broken at the
unification scale $\Lambda_{\rm GUT} \gtrsim 10^{16}$~GeV in order to
be compatible with the experimental bounds on the proton lifetime.
However, the breaking to the SM group may occur in several steps and
some subgroups may remain unbroken down to a scale of order 1~TeV.  In
this case the surviving group factors allow for {\it new gauge bosons}
with masses not far above the scale of electroweak symmetry breaking.
Besides ${\rm SU}(5)$, two other unification groups have received much
attention: In ${\rm SO}(10)$ three new gauge bosons $W^\pm_R, Z_R$ may
exist, in E$_6$ a light neutral $Z'$ in the TeV range.

\begin{figure}[ht]
\begin{center}
\epsfig{file= 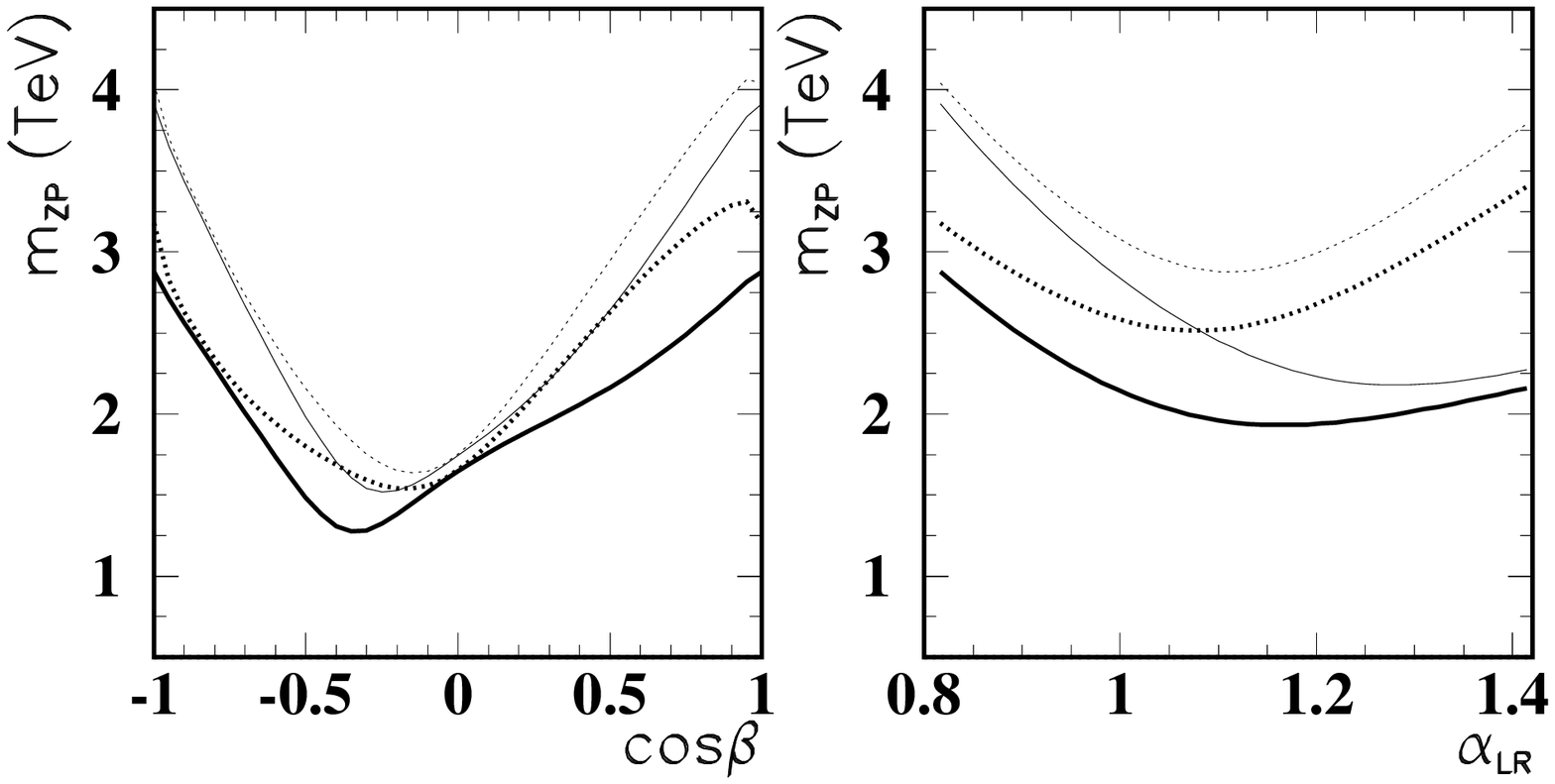,width=13.5cm}
\epsfig{file= 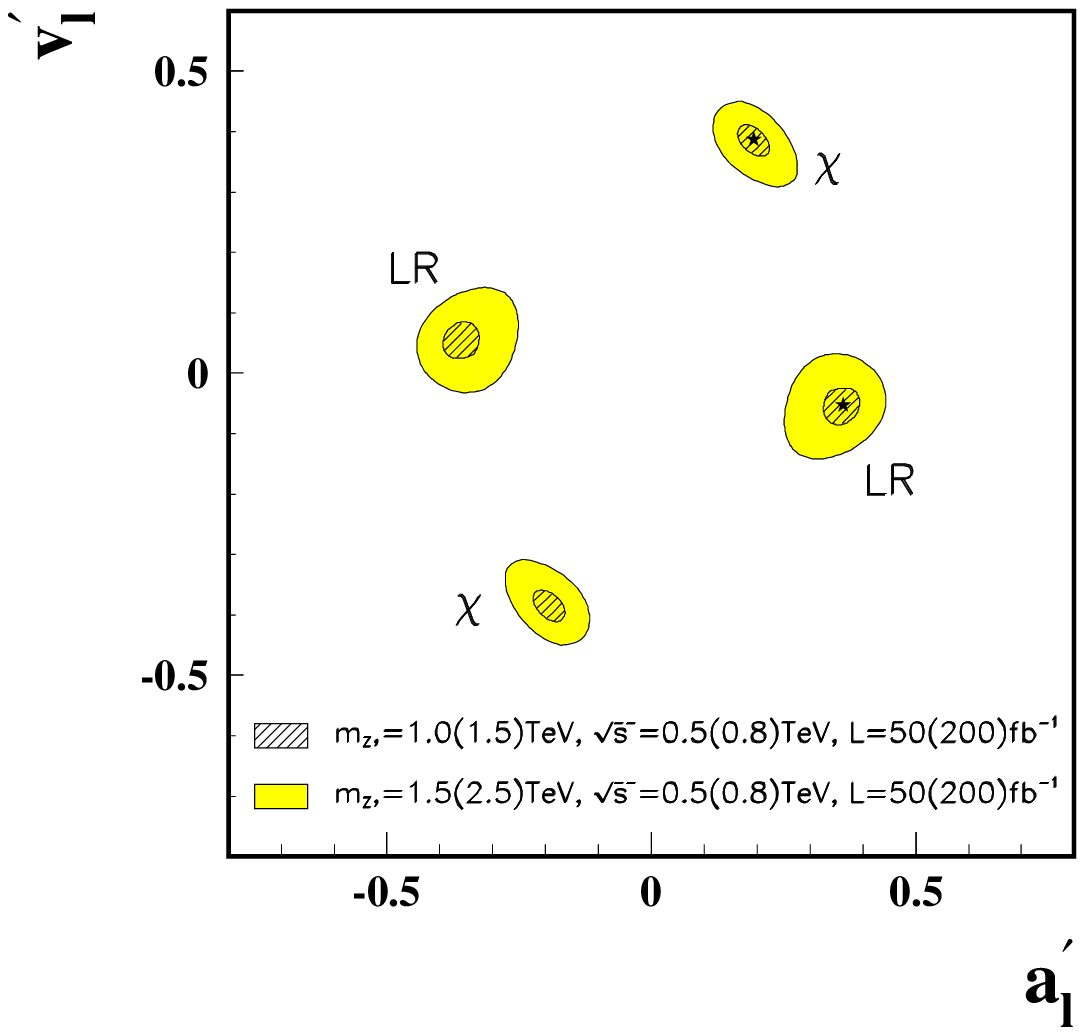,width=7.5cm}
\end{center}
\vspace{-1.cm}
\caption[]{\it 
  Upper part: $Z'$ mass limits in $E_6$ and left-right models as a
  function of the mixing parameters $\cos \beta$ and $\alpha_{LR}$,
  respectively.  Shown are 95\% confidence limits based on an
  integrated luminosity of 20 ${\rm fb}^{-1}$ at a c.m. energy of 500
  GeV.  The thick solid curve is the result of combining the
  measurements of $\sigma^{\rm lept}, R = \sigma^{\rm had} /
  \sigma^{\rm lept}$ and $A^{\rm lept}_{FB}$.  The thick dotted curve
  assumes longitudinal polarization and includes the measurement of
  $A^{\rm had}_{\rm LR}$ and $A^{\rm lept}_{\rm LR}$.  The
  corresponding thin curves include only the effects of statistics.
  Ref.\protect\cite{N52}.  Lower part: Resolution power for $M_{Z'} =
  1.5$ TeV at $\sqrt{s} = 500$ GeV, 95\% CL, for the $E(6)\chi$ and
  $LR$ models, based on the measurements of leptonic vector and axial
  charges; Ref.\protect\cite{N53}.  \protect\label{12}\label{SL}}
\end{figure}

\STS The virtual effects of a new $Z'$ or $Z_R$ vector boson
associated with the most general effective theories which arise from
breaking E(6) $\ra {\rm SU(3) \times SU(2) \times U(1) \times
  U(1)_{Y'}}$ and ${\rm SO(10) \ra}$ ${\rm SU(3) \times SU(2)_L \times
  SU(2)_R \times U(1)}$, have been investigated in
Refs.~\cite{N52,N53}.  Assuming that the $Z' (Z_R)$ are heavier than
the available c.m. energy, the propagator effects on various
observables of the process
\[
\epem \stackrel{\gamma, Z, Z'}{\longrightarrow} f\bar{f}
\]
have been studied.  The effects of the new vector bosons with mass
$M_{Z'}$ between 1.5 and 3.5~TeV can be probed at a 500 GeV collider,
Fig.\ref{12} (upper part) and Table~\ref{t2}.  They can be produced
directly up to $M_{Z'} \sim$~5~TeV at hadron colliders. However, \ee
colliders can help identify the physical nature of the new boson by
measuring the \cps to leptons and quarks, Fig.\ref{12} (lower part).
At 1.5 TeV \ee colliders, the mass window can be extended to 6 ....
11~TeV, depending on the nature of the vector boson, i.e., far beyond
the reach of proton colliders.

\begin{table}[ht]
\begin{center}
\begin{tabular}{|rr||rrr|r|}
\hline
$\sqrt{s}$\rule[-4mm]{0mm}{10mm}&$\int{\cal L}$
&$\chi$&$\psi$&$\eta$&   LR \\
\hline
\hline
  500~GeV \rule[0mm]{0mm}{7mm}&  50~fb$^{-1}$ 
&  3400 & 1850 & 2020 & 2720 \\
  800~GeV & 200~fb$^{-1}$ &  5700 & 3130 & 3350 & 4550 \\
 1600~GeV \rule[-3mm]{0mm}{4mm}& 800~fb$^{-1}$ 
& 11100 & 6260 & 6610 & 9040 \\
\hline
\end{tabular}
\parbox{14cm}{
\caption[]{\it 
  Lower bounds (95\% CL) on the $Z', Z_R$ masses in E(6) [$\chi , \psi
  ,\eta $ realization] and left/right symmetric models; $M_{Z', Z_R}$
  are given in GeV.  Ref.\protect\cite{N53}.  \protect\label{t2} }}
\end{center}
\end{table}
\vspace{5mm}

\GS {$(ii)$} The grand unification groups incorporate extended fermion
representations in which a complete generation of SM quarks and
leptons can be naturally embedded.  These re\-pre\-sentations
accommodate a variety of additional {\it new fermions}.  It is
conceivable that the new fermions [if they are protected by
symmetries, for instance] acquire masses not much larger than the
Fermi scale.  This is necessary, if the predicted new gauge bosons are
relatively light.  SO(10) is the simplest group in which the 15 chiral
states of each SM generation of fermions can be embedded into a single
multiplet.  This representation has dimension {\bf 16} and contains a
right-handed neutrino.  The group E(6) contains $\rm SU(5)$ and $\rm
SO(10)$ as subgroups, and each quark-lepton generation belongs to a
representation of dimension {\bf 27}.  To complete this
representation, twelve new fields are needed in addition to the SM
fermion fields.  In each family the spectrum includes two additional
isodoublets of leptons, two isosinglet neutrinos and an isosinglet
quark with charge $-1/3$.

\STS If the new particles $F$ have non-zero electromagnetic and weak
charges, they can be pair-produced if their masses are smaller than
the beam energy of the $\epem$ collider.  In general, these processes
are built up by a superposition of $s$--channel $\gamma$ and $Z$
exchange, but additional contributions could come from the extra
neutral bosons if their masses are not much larger than the
c.m.~energy \cite{N54}:
\[
\epem \stackrel{\gamma, Z, Z'}{\longrightarrow} F\bar{F}
\]
At 500~GeV colliders, the cross sections are fairly large, apart from
phase space suppression factors, of the order of the point-like QED
cross section $\sigma(e^+ e^- \rightarrow F \overline{F}) \sim
\sigma_0 \simeq 400$ fb.  This leads to samples of several thousands
of events, with clear signatures from decays like $F \rightarrow f'+W$
etc.  The large number of events allows to probe masses up to the
kinematical limit of 250 GeV for $\sqrt{s} = 500$ GeV.

\STS Fermion mixing, if large enough, gives rise to an additional
production mechanism for the new fermions, single production in
association with their light partners:
\[
\epem \stackrel{Z, Z'}{\longrightarrow} F\bar{f}
\]
In this case, masses very close to the total energy of the $\epem$
collider can be reached if the mixing is large enough.  For the second
and third generation of leptons [if inter-generational mixing is
neglected] and for quarks, the process proceeds only through \linebreak 

\clearpage

\noindent
$s$--channel $Z$ (or $Z'$) exchange, so that the cross sections are
relatively small.  But for the first generation leptons, additional
$t$--channel exchanges [$W$ exchange for neutral leptons and $Z$
exchange for charged leptons] are present, increasing the cross
sections eventually by several orders of magnitude to a level of
$10^2$ to $10^3$~fb.

\GS Extended gauge theories can lead to additional exciting phenomena
which are quite foreign to the observations in the \SM.  This may be
illustrated by two examples.  In left-right symmetric theories based
on $\rm SO(10)$, heavy Majorana neutrinos may exist.  The $t$--channel
exchange of these particles can induce the lepton-number violating
process
\[
e^-e^- \ra W^-W^- 
\]
in electron-electron collisions \cite{N55}, probing Majorana masses
up to 20~TeV for neutrino mixings of order $10^{-3}$.  The second
example in such a scenario is the production of doubly-charged Higgs
bosons $\Delta^{--}$ in $e^-e^-$~collisions \cite{N56},
\[
e^-e^- \to \Delta^{--} 
\]
Additional production channels of this particle, based on the
conversion $\gamma \rightarrow e^-$ in $e^-\gamma$ collisions, are
discussed in Ref.\cite{82A}.

\vspace*{-1mm}
\section[The Higgs Mechanism]{The Higgs Mechanism}

\vspace*{-1mm}
The Higgs mechanism is the third building block in the electroweak
sector of the Standard Model.  The fundamental particles, leptons,
quarks and weak gauge bosons, acquire masses through the interaction
with a scalar field of non-zero field strength in the ground state
\cite{N2}.

\STS To accommodate the well-established electromagnetic and weak
phenomena, the Higgs mechanism requires the existence of at least one
weak isodoublet scalar field.  After absorbing three Goldstone modes
to build up the longitudinal polarization states of the $W^\pm/Z$
bosons, one degree of freedom is left over, corresponding to a real
scalar particle.  The discovery of this Higgs boson and the
verification of its characteristic properties is crucial for the
theory of the electroweak interactions.  The physical implications
reach far beyond the canonical formulation of the Standard Model.

\begin{figure}[ht]
\begin{center}
\hspace*{-5mm}
\epsfig{file= 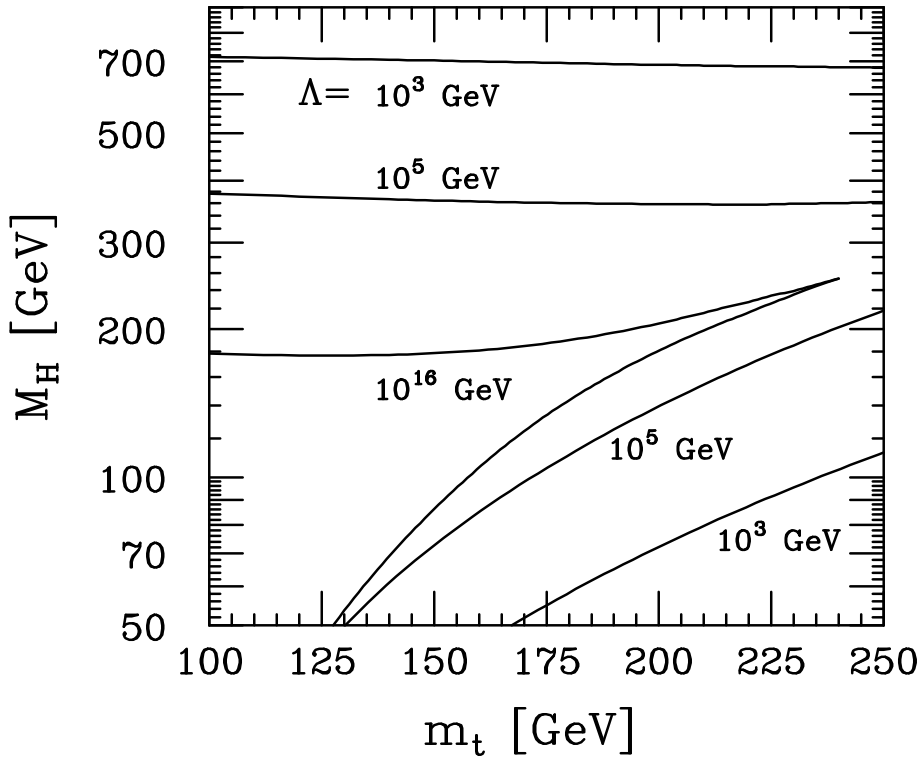,width=10cm,angle=90}
\end{center}
\caption[]{\it 
  Bounds on the mass of the Higgs boson in the Standard Model.
  $\Lambda$ denotes the energy scale at which the Higgs-gauge boson
  system of the \SM would become strongly interacting (upper bound);
  the lower bound follows from the requirement of vacuum stability.
  Refs.\protect\cite{502,503}.  \protect\label{13}\label{Lindner}}
\end{figure}

\STS The only unknown parameter in the Higgs sector of the \SM is the
mass of the Higgs particle.  Stringent constraints however can be
derived from the scale $\Lambda$ up to which the \SM is assumed to be
valid before the gauge and Higgs particles become strongly interacting
and new physics phenomena may emerge \cite{502}, c.f. Fig.\ref{13}.
The strength of the Higgs self-interaction is determined by the Higgs
mass itself at the scale $v= $~246~GeV, the value of the Higgs field
in the ground state which characterizes the spontaneous breaking of
the electroweak gauge symmetries.  Increasing the energy scale, the
quartic self-coupling of the Higgs field increases logarithmically,
in a similar way to the electromagnetic coupling in QED.  
If the Higgs mass
is small, the energy cut-off $\Lambda$ at which the coupling grows
beyond any given bound, is large; conversely, if the Higgs mass is
large, the cut-off $\Lambda$ is small.  The condition $M_H < \Lambda$
sets an upper limit on the Higgs mass in the Standard Model.  It has
been shown in lattice analyses, which account properly for the onset
of the strong interactions in the Higgs sector, that this condition
leads to an estimate of about 700~GeV for the upper limit on $M_H$
\cite{503A}.  [These analyses are based on the orthodox $\Phi ^4$
formulation of the Standard Model.  Therefore, they do not exclude
higher values for Higgs masses in any extension of the Standard
Model.]

\STS However, if the Higgs mass is less than 180 to 200~GeV, the \SM
can be extended up to the grand unification scale $\Lambda_{\rm GUT}
\sim 10^{16}$~GeV while all particles remain weakly interacting.  The
hypothesis that interactions between $W/Z$ bosons and Higgs particles
remain weak up to the GUT scale, plays a key r\^ole in explaining the
experimental value of the electroweak mixing parameter $\sin^2
\theta_w$.  Based on the SM particle spectrum, the electroweak mixing
parameter evolves in this scenario from the symmetry value 3/8 at the
GUT scale down to $\sim $~0.2 at ${\cal O} (100 $~GeV).  Even though
additional degrees of freedom are needed to account for the difference
from the experimentally observed value 0.23, the hypothesis that the
particle interactions remain weak up to the GUT scale is nevertheless
strongly supported by this result.  From the additional requirement of
vacuum stability, lower bounds on the Higgs mass can be derived.
Negative loop corrections to the Higgs potential due to heavy top
quarks can only be balanced if the Higgs mass is sufficiently large.
Based on these arguments, the SM Higgs mass would be expected in the
window $130 \lessim M_H \lessim 180$ GeV for a top mass value of about
175 GeV (c.f. Fig.\ref{13}).  This mass range agrees nicely with the
most probable estimate of the Higgs mass from the high-precision
electroweak data \cite{503B}: $M_H = 159 \stackrel{+153}{\scriptstyle
  -86} $~GeV.

\vspace*{-1mm}
\GS The SM Higgs boson  can be discovered at
the LHC in the region above LEP2, including a firm overlap of the two
machines, up to the canonical upper limit of $M_H \sim 800 $~GeV.  In
the theoretically preferred intermediate mass range below the $ZZ$
decay threshold, the experimental search is difficult.  

\vspace*{-1mm}
\GS A large variety of channels can be exploited to search for Higgs
particles in the Higgs-strahlung \cite{504,504A} and fusion processes
\citer{505,507} at \ee colliders. The signature is very clear and
the background almost negligible so that the properties of the Higgs
boson can easily be reconstructed, in particular in the preferred
intermediate mass region.   In the Higgs-strahlung process \ee
$\ra ZH$, recoil-mass techniques can be used in final states with
leptonic $Z$ decays, or the Higgs particle may be reconstructed in $H
\ra b \bar{b}, WW$ directly.  The $WW$ fusion process \ee $\ra
\bar{\nu}_e \nu_e H$ requires the reconstruction of the Higgs
particle.

\STS Once the Higgs boson is found, it will be very important to
explore the properties which reveal the physical nature of the
particle.   The zero-spin of the Higgs particle is
reflected in the angular distribution of the Higgs-strahlung process
which asymptotically must approach the $\sin^2\theta$ law.  Of
paramount importance is the measurement of the couplings to gauge
bosons and matter particles.  The strength of the couplings to $Z$ and
$W$ bosons is reflected in the size of the \ee production cross
sections.  The strength of the couplings to fermions can be measured
through the decay branching ratios and Higgs bremsstrahlung off top
quarks.  These measurements are important instrumentaria to establish
the Higgs mechanism experimentally.  Finally, the Higgs potential
itself, which provides the physical basis of the Higgs phenomenon,
must be reconstructed by measuring the triple and quartic Higgs
self-couplings \cite{508}.  This appears possible only by exploiting
multi-Higgs production in the fusion mechanism at TeV energies and
maximum possible luminosity.

\GS 
\subsection[Decays of the Higgs Boson]{Decays of the Higgs Boson}

The profile of the Higgs particle is uniquely determined if the Higgs
mass is fixed. For Higgs particles in the lower part of the
intermediate mass range $M_Z \le M_H \le 2M_Z$, the main decay modes
\cite{509} are fermion decays, in particular $b\bar{b}$ final states,
\begin{equation}
\Gamma (H \to f\bar{f} ) = \frac{G_F N_C}{4 \sqrt{2}\pi}
m_f^2 (M_H^2)M_H
\end{equation}
and in the upper part \WW and $ZZ$ pairs with one of the two gauge
bosons being virtual below the threshold,
\begin{equation}
\Gamma (H \to VV ) = \frac{3 G_F^2 M_Z^4}{16 \pi ^3}
M_H R_V (M_V^2/M_H^2) \to 2(1) \frac{\sqrt{2} G_F}{32 \pi}M_H^3
\; \; \hspace{2mm} [V= W (Z)]
\end{equation}

\begin{figure}[hb]
\begin{center}
\epsfig{file=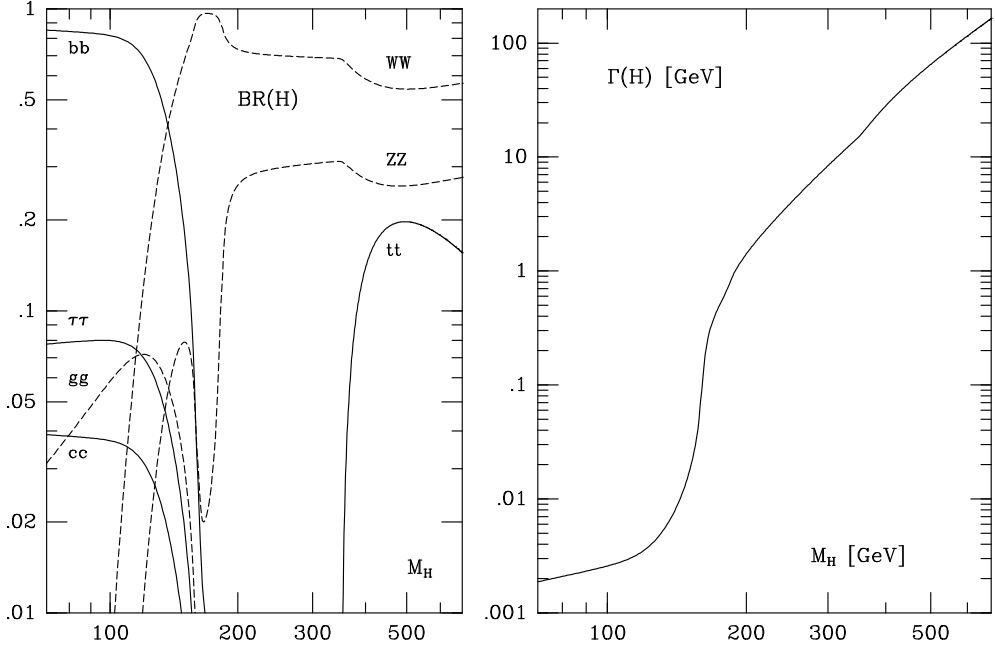,width=9cm,angle=-90}
\end{center}
\vspace*{-0.5cm}
\caption[]{\it 
  Branching ratios of the main decay modes of the SM Higgs boson and
  total decay width; Ref.\protect\cite{509}.
  \protect\label{HBRG}\label{HSMdecay}}
\end{figure}

Above the \WW threshold, the Higgs particles decay almost exclusively
into the $W/Z$ channels, except in the mass range near the $t\bar{t}$
decay threshold.  Below 140~GeV, the decays $H \to \tau ^+ \tau ^-,
c\bar{c}$ and $gg$ are also important besides the dominating
$b\bar{b}$ channel.  By adding up all possible decay channels, we
obtain the total Higgs decay width shown in Fig.\ref{HBRG} for $m_t=
$~175~GeV.  Up to masses of 140~GeV, the Higgs particle is very
narrow, $\Gamma (H) \le $~10~MeV.  After opening the mixed
real/virtual gauge boson channels, the state becomes rapidly wider,
reaching $\sim $~1~GeV at the $ZZ$ threshold.  The width cannot be
measured directly in the intermediate mass range.  Only above $M_H \ge
$~200~GeV it becomes wide enough to be resolved experimentally.

\STS
\subsection[The Production of Higgs Bosons]
{The Production of Higgs Bosons}

The main production mechanism for Higgs particles of moderate mass and
at moderate energies in \ee collisions is the Higgs-strahlung off the
$Z$ boson line \cite{504,504A}
\[
e^+e^- \stackrel{Z}{\longrightarrow} Z H
\]
The cross section is given by
\begin{equation}
\sigma(e^+e^- \to ZH)= \frac{G_F^2 M_Z^4}{96 \pi s}( v_e^2 + a_e^2 )
  \lambda ^{\frac{1}{2}} \frac{\lambda + 12 M_Z^2 /s}
  {(1-M_Z^2 /s)^2}
\end{equation}
where $\lambda$ is the usual 2--body phase space coefficient.  For a
given Higgs mass $M_H$, the Higgs-strahlung cross section is maximal
for the c.m.~energy $\sqrt{s} \sim M_Z +2 M_H$.  Beyond the threshold
region, the cross section for Higgs-strahlung scales as $s^{-1}$ and
vanishes asymptotically.  With rising energy the fusion mechanisms, in
particular \WW fusion, become increasingly important \citer{505,507},
\begin{eqnarray*}
e^+e^- &\stackrel{WW}{\longrightarrow}&  \bar{\nu}_e \nu_e H\\
e^+e^- &\stackrel{ZZ}{\longrightarrow}&  e^+e^- H
\end{eqnarray*}
The corresponding cross sections rise logarithmically with energy; 
for \WW fusion:
\begin{equation}
\sigma(e^+e^- \to \bar{\nu}_e \nu _e H) \to \frac{
G_F^3 M_W^4}
{4 \sqrt{2} \pi ^3}
\left[
\left( 1+\frac{M_H^2}{s} \right) \log \frac{s}{M_H^2} -2
\left( 1-\frac{M_H^2}{s} \right)
\right]
\end{equation}
Due to the reduced $Z$ charges, the \cs for $ZZ$ fusion is about one
order of magnitude smaller; the same applies for $ZZ$ fusion in $e^-e^-$
collisions.  However, with two leptons in the final states, recoil
mass techniques can be applied which allow a more effective background
rejection compared to the neutrino channel.

\begin{figure}[ht]
\begin{center}
\epsfig{file=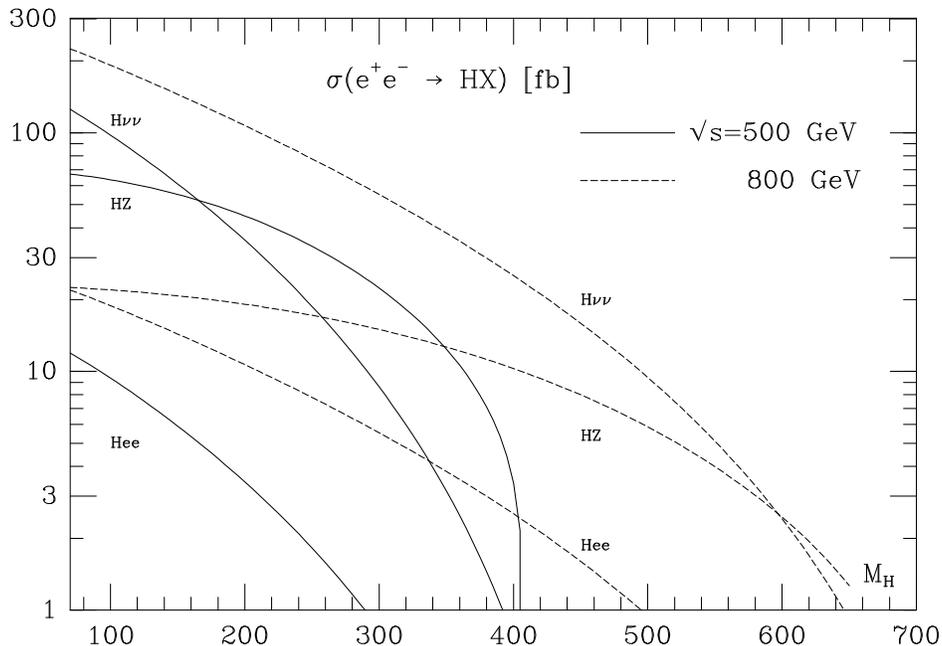,width=9cm,angle=-90}
\end{center}
\vspace*{-.5cm}
\caption[]{\it 
  The cross section for the production of SM Higgs bosons in
  Higgs-strahlung $e^+e^- \rightarrow ZH$ and $WW/ZZ$ fusion $ e^+e^-
  \rightarrow \overline{\nu} \nu / e^+e^- H$; solid curves: $\sqrt{s}
  = 500$ GeV, dashed curves: $\sqrt{s}= 800$ GeV.
  \protect\label{15}\label{HSMprod}}
\end{figure}

\STS The cross sections for the Higgs-strahlung process and the
fusion processes are shown in Fig.\ref{15} for $\sqrt{s}=500 $ and
800~GeV.  Several thousand events will be produced for the envisaged
luminosities.

\begin{figure}[ht]
\begin{center}
\hspace*{-5mm}
\epsfig{file= 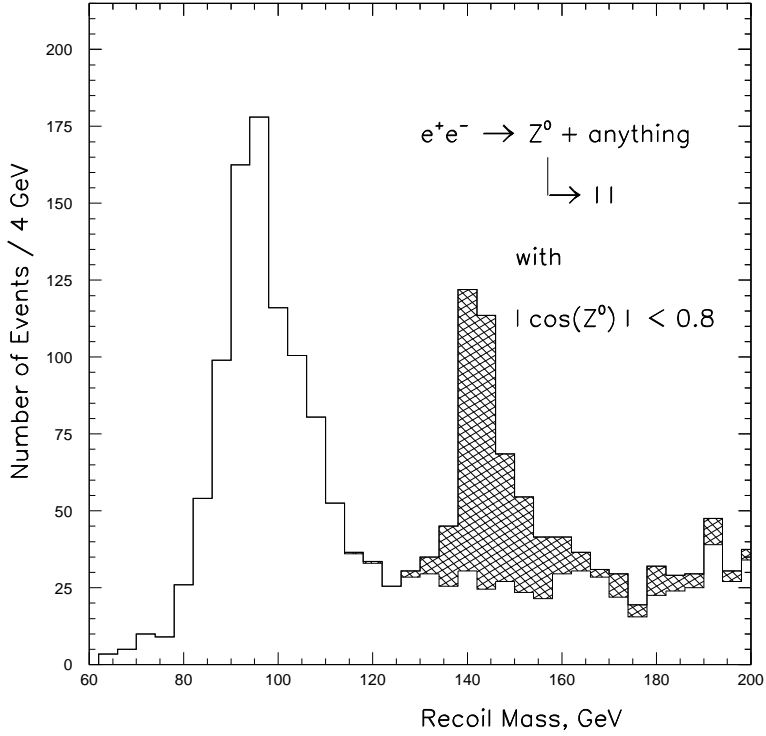,width=11cm}
\end{center}
\caption[]{\it 
  Dilepton recoil mass analysis of Higgs-strahlung $e^+e^- \rightarrow
  ZH \rightarrow l^+ l^- + \rm{anything}$ in the intermediate Higgs
  mass range for $M_H = 140$ GeV. The c.m. energy is
  $\sqrt{s}$=360~GeV and the integrated luminosity $\int {\cal
    L}=$50~fb$^{-1}$.  Ref.\protect\cite{511A}.
  \protect\label{16}\label{expHSMpeak}}
\end{figure}

\GS The recoiling $Z$~boson in the two-body reaction $\epem \to Z H$
is mono-energetic and the Higgs mass can be derived from the energy
of the $Z$~boson, $M_H^2 = s -2\sqrt{s} E_Z + M_Z^2$.  The detection
of the Higgs boson in this channel is independent of the Higgs decay
properties, thus providing a very powerful tool for the search of this
particle.  Initial-state bremsstrahlung and beamstrahlung smear out
the peak slightly \cite{510}.  $ZZ$~production does not pose a serious
background problem; if efficient $b$~tagging devices are used, the
Higgs signal can be extracted even for masses close to the $Z$~mass
\cite{511}. Signal and background are shown in Fig.\ref{16}. A
similar clear peak can be observed in the fusion process $\epem \to
\bar{\nu_e} \nu_e H$ by collecting the decay products of the Higgs
boson.  The dominant background process in this case is the reaction
$\epem \to (e^+) \nu _e W^-$, with the final state positron traveling
undetected along the beam pipe; this background is negligible also
with experimental resolution effects taken into account.

\subsection[Higgs Production in $\gamma \gamma$ Collisions]
{Higgs Production in $\gamma \gamma$ Collisions}

The production of Higgs bosons in $\gamma \gamma$ collisions
\cite{f103A} can be exploited to determine important properties of
these particles, in particular the two-photon width.  The $H\gamma
\gamma$ \cp is built up by loops of charged particles.  If the mass of
the loop \p is generated through the Higgs mechanism, the decoupling
of the heavy \ps is lifted and the $\gamma \gamma$ width reflects the
spectrum of these states with masses possibly far above the Higgs
mass.

\STS Together with the measurement of the branching ratio $BR_{\gamma
  \gamma}$ at the LHC, or if enough events can be generated at \ee
linear colliders in Higgs-strahlung, the measurement of the $\gamma
\gamma$ partial width can be used to determine the total width of the
Higgs boson in a range where it cannot be resolved experimentally.

\GS The two-photon width is related to the $\gamma \gamma$ production
\cs by
\begin{equation}
\sigma(\gamma \gamma_{J_z=0} \rightarrow H) = 
\frac{16 \pi^2\Gamma(H\rightarrow \gamma \gamma)}{M_H}
\times BW
\end{equation}
where $BW$ denotes the Breit--Wigner resonance factor in terms of the
energy squared.  For narrow Higgs bosons the observed \cs is found by
folding the parton \cs with the invariant $\gamma \gamma$ energy flux
$\tau d{\cal L}^{\gamma \gamma}/d\tau$ for $J_z^{\gamma \gamma}=0$ at
$\tau=M_H^2/s_{ee}$.

\begin{figure}[ht]
\begin{center}
\epsfig{file=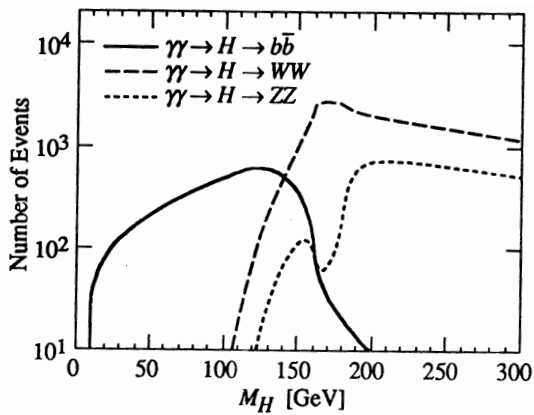,width=7.6cm,angle=-2.0} 
\end{center}
\vspace{-0.5cm}
\caption[]{\it 
  Production rate of Standard Model Higgs bosons into the three
  exclusive final states relevant for the intermediate-- and heavy
  mass regions in ${\gamma \gamma}$ collisions. A value of $4 \cdot
  10^{-2} \rm{fb}^{-1}/\rm{GeV}$ is assumed for $d{\cal L}^{\gamma
    \gamma}/dW_{\gamma \gamma}$. Ref.\protect\cite{f103A}.
  \protect\label{f0.23}}
\end{figure}

\STS The event rate for the production of Higgs bosons in $\gamma
\gamma$ collisions of Weizs\"acker--Williams photons is too small to
play a r\^ole in practice. However, the rate is sufficiently large if
the photon spectra are generated by Compton back-scattering of laser
light, Fig.\ref{f0.23}; the $\gamma \gamma$ luminosity in such a
Compton collider is expected to be only slightly smaller than the
luminosity in \ee collisions. In the Higgs mass range between 100 and
150 GeV, the final state consists primarily of $b\overline{b}$ pairs.
The large $\gamma \gamma$ continuum background is suppressed in the
$J_z^{\gamma \gamma}=0$ polarization state.  For Higgs masses above
150~GeV, $WW$ final states become dominant, supplemented in the ratio
1:2 by $ZZ$ final states above the $ZZ$ decay threshold.  While the
continuum $WW$ background in $\gamma \gamma$ collisions is very large,
the $ZZ$ background appears under control for masses up to order
300~GeV \cite{f103B}.  The error on the partial $\gamma \gamma$ decay
width of the Higgs boson is expected in the range of 10\%
\cite{f103A,f103C}.

\GS Additional sources of Higgs bosons are provided by $e\gamma$
collisions \cite{f103D}.  The process $e\gamma \rightarrow \nu W H$ is
generated at the tree level while Higgs production in $e\gamma
\rightarrow eH$ proceeds through the fusion of the real and virtual
photon.  For $\sqrt{s} = 500$~GeV the \css are larger than 10~fb for
Higgs masses below 250~GeV.  For $\sqrt{s} = 800$~GeV, this limit is
raised to 450~GeV in the $\nu W H$ process, with an initial \cs of
100~fb for a Higgs mass of 100~GeV.

\subsection[The Profile of the Higgs Particle]
{The Profile of the Higgs Particle}

To establish the Higgs mechanism experimentally, the nature of this
particle must be explored by measuring {\it all} its characteristics,
the mass and lifetime, the external quantum numbers spin-parity, the
couplings to gauge bosons and fermions, and last but not least the
self-couplings.  This program can be realized at \ee~colliders in
consecutive steps.

\GS The {\it mass} of the Higgs particle can be determined at \ee
linear colliders very precisely.  This can be achieved by exploiting
the kinematical constraints in the four-jet topology, the $\tau
\bar{\tau} q \bar{q}$ final state and the leptonic channels in
Higgs-strahlung events \cite{512}.  For an integrated luminosity
$\int \LUM =$~50~fb$^{-1}$ at $\sqrt{s}= 500 $~GeV, a precision of
$\pm 180 $~MeV can be reached; at this level, systematic errors due to
the measurement of the beam energy can still be neglected.

\GS The width of the state, i.e. the {\it lifetime} of the particle,
can be measured directly above the $ZZ$ decay threshold where the
width grows rapidly.  In the lower part of the intermediate mass range
the width can be measured indirectly by combining the branching ratio
for $H \to \gamma \gamma$, accessible at the LHC, with the measurement
of the partial $\gamma \gamma$ width, accessible through $\gamma
\gamma$ production at a Compton collider.  In the upper part of the
intermediate mass range, the combination of the branching ratios for
$H \to WW, ZZ$ decays with the production \css for \WW~fusion and
Higgs-strahlung, which can be expressed both through the partial
Higgs-decay widths to \WW and $ZZ$ pairs, will allow us to extract
the width of the Higgs particle.  Thus, the width of the Higgs \p will
be determined throughout the entire possible mass range if the
experimental results from LHC, \ee {\it and optional} $\gamma \gamma$
colliders can be combined.

\begin{figure}[ht]
\begin{center}
\epsfig{file=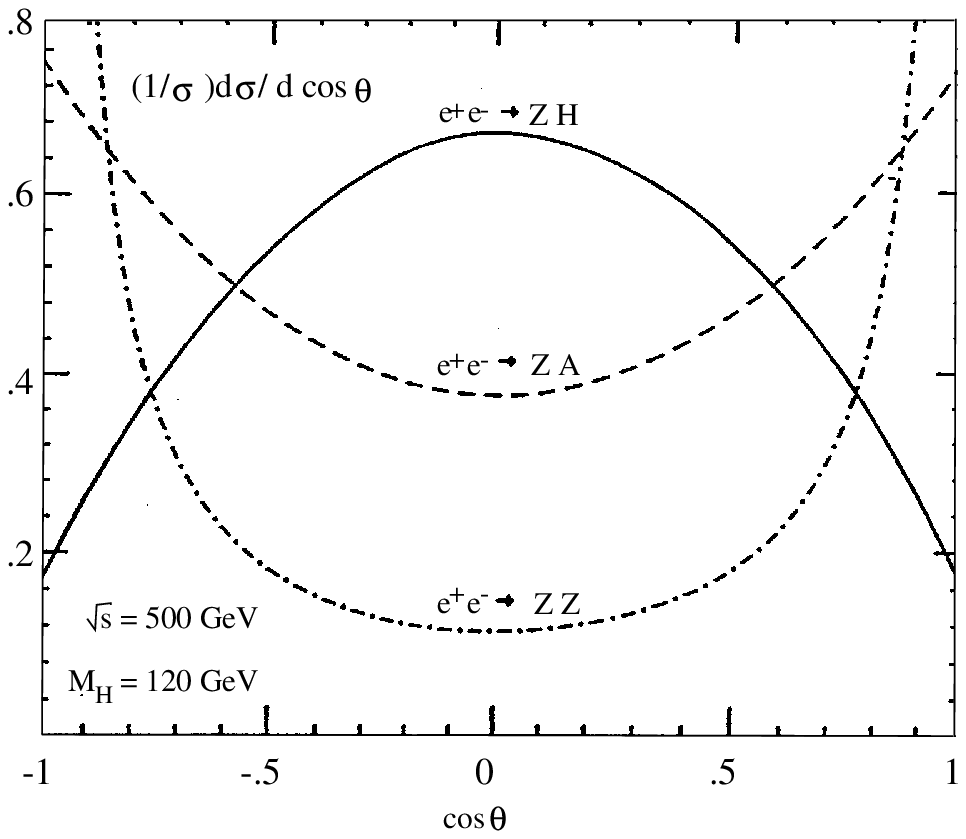,width=8.2cm}
\raisebox{1mm}{\epsfig{file=higgs_sm_spin_exp.eps,width=6.4cm}}
\end{center}\vspace{-0.5cm}
\caption[]{\it 
  Left: Angular distribution of $Z/H$ bosons in Higgs-strahlung,
  compared with the production of pseudoscalar particles and the $ZZ$
  background final states; Ref.\protect\cite{504A}.  Right: The same
  for the signal plus background in the experimental simulation of
  Ref.\protect\cite{512}.  \protect\label{17cp}\label{dsigtheta}}
\end{figure}

\GS The angular distribution of the $Z / H$ bosons in the \Hs process
is sensitive to the {\it spin and parity} of the Higgs \p \cite{504A}.
Since the production amplitude is given by ${\cal A} (0^+) \sim
\vec{\varepsilon_{Z^*}} \cdot \vec{\varepsilon_Z}$, the $Z$~boson is
produced in a state of longitudinal polarization at high energies --
in accord with the equivalence theorem.  As a result, the angular
distribution
\begin{equation}
d\sigma / d \cos \theta \sim \sin^2\theta
+8M_Z^2 / (\lambda s)
\end{equation}
approaches the spin-zero $\sin^2\theta$ law asymptotically.  This may
be contrasted with the distribution $\sim 1 + \cos^2\theta$ for
negative parity states which follows from the transverse polarization
amplitude ${\cal A} (0^-) \sim \vec{\varepsilon_{Z^*}} \times
\vec{\varepsilon_Z} \cdot \vec{k_Z}$.  It is also characteristically
different from the distribution of the background process $\epem \to
ZZ$ which, as a result of $t/u$--channel $e$ exchange, is strongly
peaked in the forward/backward direction, Fig.\ref{17cp}.

\begin{figure}[ht]
\begin{center}
\epsfig{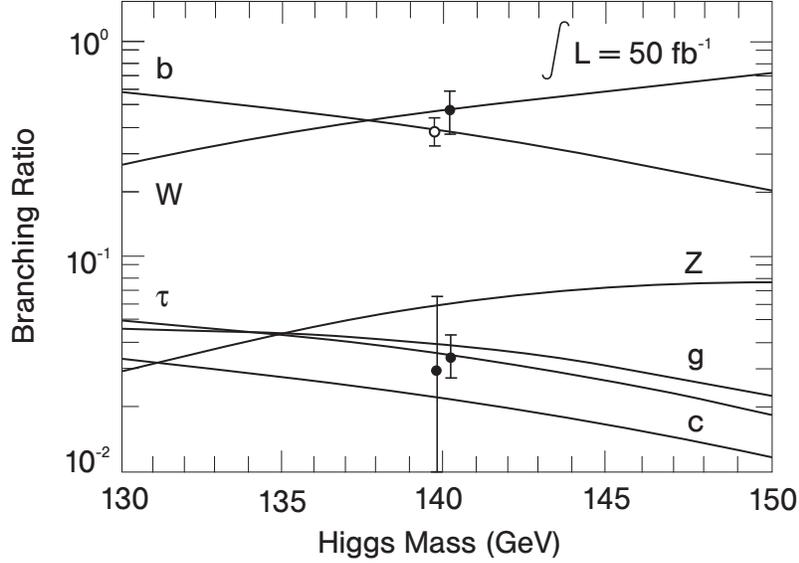}
\end{center}
\caption[]{\it 
  The measurement of decay branching ratios of the SM Higgs boson for
  $M_H = 140$ GeV.  In the bottom part of the figure the small error
  bar belongs to the $\tau$ branching ratio, the large bar to the sum
  of the charm and gluon branching ratios which were not separated in
  the simulation of Ref.\protect\cite{513}. In the upper part of the
  figure the open circle denotes the $b$ branching ratio, the full
  circle the $W$ branching ratio. \protect\label{18}\label{BRs}}
\end{figure}

\GS Since the fundamental particles acquire masses through the
interaction with the Higgs field, the strength of the {\it Higgs
  couplings} to fermions and gauge bosons is set by the masses of
these particles.  It will therefore be a very important task to
measure the Higgs couplings to the fundamental particles, which are
uniquely predicted by the very nature of the Higgs mechanism.  The
Higgs couplings to massive gauge bosons can be determined from the
production \css with an accuracy of $\pm 3$~\%, the $HZZ$ coupling in
the \Hs and the $HWW$~coupling in the fusion process.  For Higgs
couplings to fermions, either loop effects in $H \rightleftharpoons
gg, \gamma \gamma$ [mediated by top quarks] must be exploited, or the
branching ratios $H \to b\bar{b}, c\bar{c}, \tau ^+\tau^-$ in the
lower part of the intermediate mass range; they provide a {\it direct}
determination of the Higgs Yukawa \cps to these fermions.  This is
exemplified for a Higgs mass of 140~GeV in Fig.\ref{18}.

\begin{figure}[ht]
\begin{center}
\hspace*{6mm}
\epsfig{file=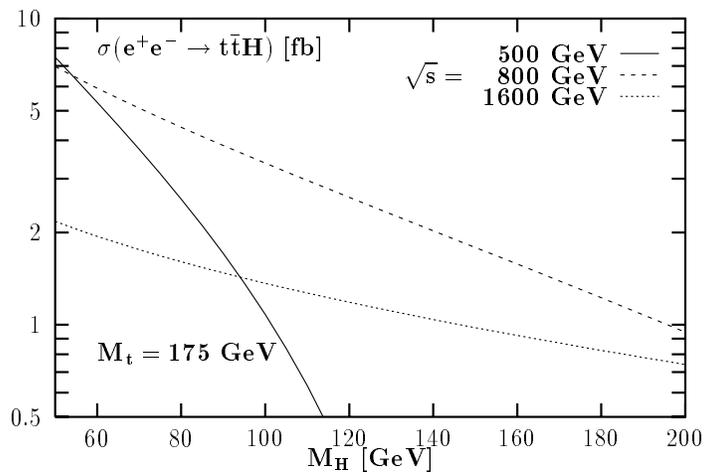,width=9.7cm}
\end{center}
\vspace{-1cm}
\caption[]{\it 
  The \cs for bremsstrahlung of SM Higgs bosons off top quarks in the
  Yukawa process $\epem \ra t\bar{t}H$.  [The amplitude for radiation
  off the intermediate $Z$--boson line is small];
  Ref.\protect\cite{514}.  \protect\label{higgssm}}
\end{figure}

\STS The Yukawa coupling of the intermediate Higgs boson to the top
quark in the range $M_H \le $~120~GeV can be measured directly in the
bremsstrahlung process $\epem \to t\bar{t} H$ in which primarily the
top quarks radiate the Higgs boson in high energy \ee~collisions
\cite{514}.  Since the top quark is very heavy, the $ttH$ \cp may
eventually provide essential clues to the nature of the mechanism
breaking the electroweak symmetries.  Even though the experiment is
difficult due to the small cross section, Fig.\ref{higgssm}, and the
complex topology of the $bbbbWW$ final state, this analysis is an
important experimental task to explore the electroweak symmetry
breaking.  For large Higgs masses above the $t\bar{t}$ threshold, the
decay channel $H \to t\bar{t}$ increases the \cs of $\epem \to
t\bar{t} Z$ through the reaction $\epem \to Z H (\to t\bar{t})$
\cite{515}.  Higgs exchange between $t\bar{t}$ quarks also affects the
excitation curve near the threshold at a level of a few percent.

\vspace*{-1.5cm}

\clearpage

\GS The Higgs mechanism, based on a non-zero value of the Higgs field
in the vacuum, must finally be made manifest experimentally by
reconstructing the interaction 
potential which generates the non-zero
Higgs field in the vacuum.  This program can be carried out by
measuring the strength of the {\it trilinear and quartic
  self-couplings} of the Higgs particles.  This is a very difficult
task since the processes to be exploited are suppressed by small
couplings and phase space.  Nevertheless, the problem can be solved in
the high energy phase of the \ee~linear colliders for sufficiently
high luminosities \cite{508}.  The best suited reaction for the
measurement of the trilinear coupling for Higgs masses in the
theoretically preferred ${\cal O}(100 $~GeV) mass range, is the $WW$
fusion process
\[
\epem \to \bar{\nu}_e \nu _e H H
\]
in which, among other mechanisms, the two-Higgs final state is
generated by the $s$--channel exchange of a virtual Higgs \p so that
this process is sensitive to the trilinear $HHH$ coupling in the Higgs
potential.  Since the \cs is only a fraction of 1~fb at an energy of
$\sim $~1.6~TeV, an integrated luminosity of $\sim 1,000 $~fb$^{-1}$
is needed to isolate the events.  The quartic coupling $H^4$ seems to
be accessible only through loop effects in the foreseeable future.

\GS To sum up, we conclude from the preceding discussion that \ee
linear colliders with energies in the range of 300 to 500~GeV are the
ideal instruments to search for Higgs particles in the intermediate
mass range, {\it a~priori} the theoretically most attractive range,
and to establish its characteristic properties experimentally.  In the
high energy phase of the colliders, important parameters of the Higgs
potential can be reconstructed which are necessary for generating the
spontaneous breaking of the electroweak symmetries.  \STS

\section[Supersymmetry]{Supersymmetry}

Even though no {\it direct} experimental evidence has emerged yet for
the existence of super\-symmetry \cite{N4} in Nature, the concept has
so many attractive features that it may be considered as a prime
target of present and future experimental particle research.
Arguments in favor of supersymmetry are deeply rooted in particle
physics.  Supersymmetry unifies matter and forces, and if realized
locally, it plays a crucial r$\hat{\rm o}$le in a quantum theory of
gravity.  In relating particles of different spins to each other,
fermions and bosons, low-energy supersymmetry stabilizes the masses
of fundamental Higgs scalars in the context of very high energy scales
associated with grand unification \cite{601}.  Besides solving this
part of the hierarchy problem, supersymmetry may even be closely
related to the physical origin of the Higgs mechanism itself
\cite{602}: In supergravity inspired realizations of \ssy theories
\cite{602A}, incorporating universal scalar masses at the scale of the
grand unification, one of the scalar masses squared can evolve down to
negative values and thus induce spontaneous symmetry breaking in the
electroweak sector.  This is possible if the top mass has a value
between 150 and 200~GeV; all other masses squared of squarks and
sleptons remain positive so that ${\rm U}(1)_{EM}$ and ${\rm SU}(3)_C$
are unbroken.

\GS The minimal supersymmetric extension of the Standard Model
\cite{603}, MSSM, is based on the symmetry group $\rm SU(3)\times
SU(2) \times U(1)$ of the Standard Model.  The gauginos are the
supersymmetric spin--$\frac{1}{2}$ partners of the gauge bosons.  The
matter particles, quarks and leptons, are associated with scalar
supersymmetric particles, squarks and sleptons.  To preserve
supersymmetry and to keep the theory free of anomalies, two Higgs
doublets are needed, the supersymmetric partners of which are
spin--$\frac{1}{2}$ higgsinos.  Charged/neutral higgsinos mix in
general with the non-colored gauginos, forming charginos and
neutralinos.  Supersymmetric partners carry a multiplicative quantum
number $R=-1$ [$R=+1$ for ordinary particles] which is conserved in
this model.  Supersymmetric particles are therefore generated in pairs
and the lightest supersymmetric particle ($LSP$) is stable.  This
particle is in general identified with the lightest neutralino, but it
could also be the sneutrino.

\GS Strong support for supersymmetry and the particle spectrum of the
minimal supersymmetric standard model in the mass range of several
hundred GeV follows from the high-precision measurement of the
electroweak mixing angle $\sin^2\theta_w$ \cite{604,604A}.  The value
predicted by the MSSM, $\sin^2\theta_w = 0.2336 \pm 0.0017$, is
matched surprisingly well by the value determined by the LEP and other
experiments, $\sin^2\theta_w = 0.2315 \pm 0.0003$, the theoretical
uncertainty being less than 2 permille.

\GS In the simplest realization of supersymmetric grand unified
theories, with the supersymmetry breaking parameters taken to be
universal at the GUT scale, five parameters specify the supersymmetric
particle sector.  They can be chosen as the (universal) scalar mass
parameter $m_0$; the (universal) gaugino mass $M_{\demi}$; $\tb$, the
ratio of the vacuum expectation values $v_2/v_1$ associated with the
two neutral Higgs fields; the (universal) trilinear scalar coupling
$A_0$; and the sign of $\mu$, the Higgs mass parameter in the
superpotential.  Evolving the universal mass parameters from the GUT
scale down to the electroweak scale, the entire spectrum of the Higgs
particles and the supersymmetric \ps can be generated, see, for
example, the analysis in Ref.\cite{605}.  It is well-known that the
mass of the lightest Higgs particle is less than about 150~GeV in the
MSSM; this bound follows from the fact that the quartic couplings are
given by the gauge couplings.  The non-colored particles,
charginos/neutralinos and sleptons, are in general significantly
lighter than colored particles in this scenario.  The lightest of
these \ps can have masses in the range of 100 to 200~GeV.

\begin{figure}[ht]
\begin{center}
\epsfig{file=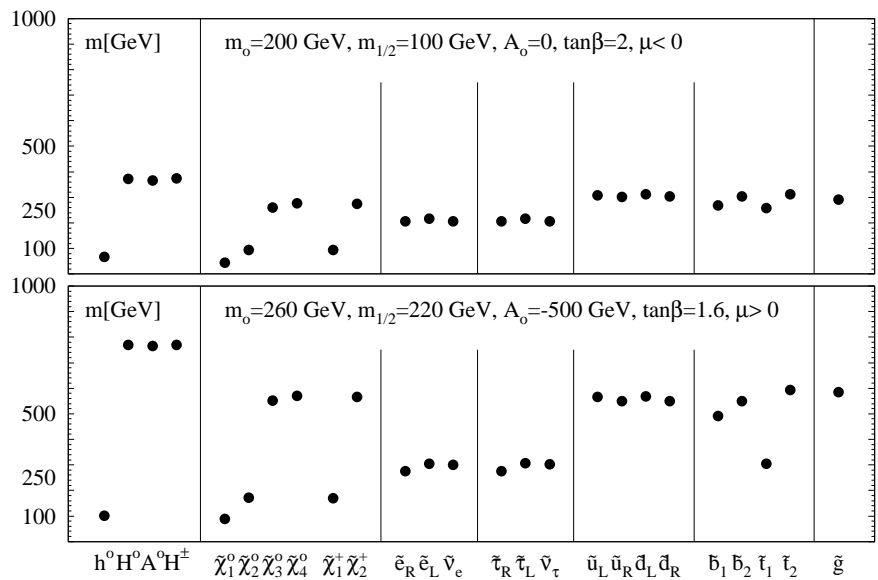,width=12cm}
\end{center}
\vspace*{-.6cm}
\caption[]{\it 
  Illustration of \ssy \p masses in two typical points of the
  parameter space of minimal supergravity.  \protect\label{f6xx}}
\end{figure}

\STS This general discussion is quantified in Table~\ref{t3} for a few
illustrative examples.  The input parameters [$m_0, M_{\demi}, A_0,
\tb, {\rm sgn} \ \mu$] have been chosen such that they are compatible
with constraints on the low-energy MSSM from the $b \to s \gamma$
decays \cite{606}, demanding $1 \NT 10^{-4} < {\rm BR}(b \to s \gamma
) < 4 \NT 10^{-4}$.  Moreover, cosmological constraints are taken into
account by requiring the matter density in the universe, primarily
composed of relic neutralinos, to be bounded as predicted in the mixed
hot/cold dark matter scenario, $\Omega h^2 \sim 0.15$ to 0.4, or more
generally, by $\Omega h^2 < 1$ as required by the age of the universe
\cite{607}.  The mass parameters have been computed within the
approximate solutions of Refs.\cite{608,609}.  The particles of
Table~\ref{t3} which are accessible at a c.m.~energy of $\sqrt{s} =
500 $~GeV are marked by one asterisk, the particles which are
accessible at 1~TeV, by two asterisks.  All other particles in the table
can be produced at a c.m.~energy of about 2~TeV.  The range of the
masses is illustrated for the two points B and G in Fig.\ref{f6xx}.
  
\vspace*{-2mm}
\STS
\subsection[SUSY Higgs Particles]{SUSY Higgs Particles}

One of the prime arguments for introducing supersymmetry is the
solution of the hierarchy problem.  By assigning fermions and bosons
to common multiplets, large 
radiative corrections can be canceled in a
natural way \cite{601} by adding up bosonic and \linebreak

\begin{table}[ht!]
\begin{center}
\begin{tabular}{|l||lll|lll||l|}
\hline
Point \rule{0mm}{5mm}  & A & B & C & D & E & F & G \\
\hline
\hline
\multicolumn{8}{|l|}{\rule[-3mm]{0mm}{8mm}SGUT Parameters}      \\
\hline
\hline
$m_0$         
\rule{0mm}{5mm}     & 125 & 200 & 125 & 200 & 100 & 200 & 260\\
$M_{\demi}$          & 175 & 100 & 175 & 400 & 200 & 400 & 220\\
$A_0$              &   0 &   0 &   0 &   0 &   0 &   0 & -500\\
$\tb$              &   2 &   2 &   2 &  10 &  10 &  10 & 1.6\\
${\rm sgn}(\mu )$  & --  & --  & +   & --  & +   & +  & + \\
\hline
\multicolumn{8}{|l|}{\rule[-3mm]{0mm}{8mm}Mass Parameters}     \\
\hline
\hline
$h^0$          
\rule{0mm}{5mm}      
&  74\A &  68\A &  86\A & 114\A & 110\A & 116\A & 101\A \\
$A^0$                & 433\B & 367\B & 434\B & 660 & 321\B & 660 & 765\\
$H^0$                & 438\B & 373\B & 439\B & 660 & 321\B & 661 & 769\\
$H^{\pm}$            & 440\B & 374\B & 440\B & 665 & 330\B & 665 & 769\\
\hline
$\tilde{\chi_1^0}$ \rule{0mm}{5mm}  &  75\A &  44\A &  64\A & 164\A 
&  78\A & 163\A & 88\A \\
$\tilde{\chi_2^0}$   & 151\A &  95\A & 122\A & 321\A & 144\A & 315\A 
& 171\A \\
$\tilde{\chi_3^0}$   & 352\A & 259\A & 350\A & 579\B & 289\A & 577\B 
& 552\B \\
$\tilde{\chi_4^0}$   & 364\A & 276\A & 378\A & 585\B & 313\A & 591\B 
& 571\B \\
$\tilde{\chi_1^{\pm}}$&151\A &  94\A & 119\A & 321\B & 142\A & 315\B 
& 170\A \\
$\tilde{\chi_2^{\pm}}$&363\A & 275\A & 375\A & 588\B & 314\A & 591\B 
& 567\B \\
\hline
$\tilde{l_R}$  \rule{0mm}{5mm}      & 146\A & 207\A & 146\A & 257\C 
& 134\A & 257\C & 275\B \\
$\tilde{l_L}$        & 181\A & 216\A & 181\A & 354\B & 182\A & 354\B 
& 305\B \\
$\tilde{\nu_L}$      & 170\A & 207\A & 170\A & 345\B & 163\A & 345\B 
& 300\B \\
\hline
$\tilde{u_R}$   \rule{0mm}{5mm}     & 403\B & 303\B & 403\B & 817 
& 443\B & 817 & 550\\
$\tilde{u_L}$        & 415\B & 307\B & 415\B & 850 & 457\B & 850 & 566\\
$\tilde{d_R}$        & 402\B & 304\B & 402\B & 813 & 442\B & 813 & 549\\
$\tilde{d_L}$        & 419\B & 313\B & 419\B & 853 & 463\B & 853 & 569\\
$\tilde{b_1}$        & 376\B & 268\B & 376\B & 766 & 420\B & 766 
& 492\B \\
$\tilde{b_2}$        & 402\B & 304\B & 402\B & 811 & 445\B & 811 & 550\\
$\tilde{t_1}$        & 339\B & 258\C & 252\C & 611 & 325\B & 598 
& 305\B \\
$\tilde{t_2}$        & 415\B & 311\B & 473\B & 823 & 505\D & 832 & 594\\
\hline
$\tilde{g}$    \rule{0mm}{5mm}      & 459\B & 291\B & 460\B & 953 
& 515 & 953 & 584\\
\hline
\end{tabular}
\parbox{14cm}{
\caption[]{{\it 
    Typical mass spectra of \ssy \ps derived from various sets of
    supergravity parameters.  Particles with one asterisk can be
    produced at collider energies of $\sqrt{s}=500$~GeV, with two
    asterisks at 1~TeV; all the \ps can be produced at about 2~TeV.
    [Particles which are located just at the borderline between two
    collider energies, are characterized by oblique strokes.]
    \protect\label{t3} }}}
\end{center}
\end{table}
\clearpage

\noindent
opposite-sign
fermionic loops.  As a result of the bosonic-fermionic supersymmetry,
Higgs bosons can be retained as elementary spin-zero particles with
masses close to the scale of the electroweak symmetry breaking even in
the context of very high grand unification scales.  The minimal
supersymmetric extension of the \SM serves as a useful guideline into
this area.  Only a few phenomena are specific to this minimal version,
many of the characteristic patterns are realized also in more general
extensions.  High-energy \ee colliders can easily cope with the
experimental problems in such general scenarios since methods of
analysis appropriate to these machines are quite robust, and do not
rely upon specific favorable circumstances.

\begin{figure}[ht]
\begin{center}
\epsfig{file=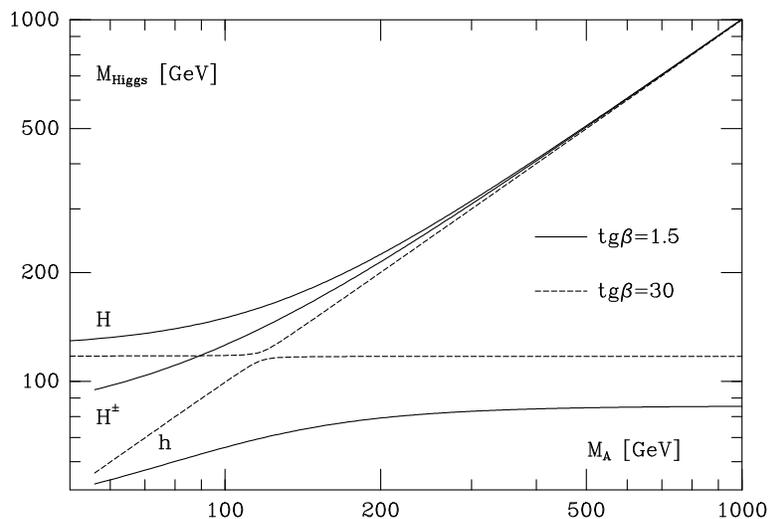,width=7cm,angle=-90}
\end{center}
\vspace{-.5cm}
\caption[]{\it 
  The masses of the Higgs bosons in the Minimal Supersymmetric
  Standard Model MSSM for two representative values of $\tb = 1.5$ and
  30.  \protect\label{f601}\label{SUSYmH}}
\end{figure}

\STS The Higgs spectrum in the MSSM consists of five \ps
\cite{611,612}, $h^0, H^0, A^0$ and $H^\pm$, the states $h^0, H^0$ and
$A^0$ being ${\cal CP}$ even and odd, respectively.  Besides the
masses, two mixing angles define the properties of the scalar
particles and their interactions with gauge bosons and fermions: the
ratio of the two vacuum expectation values $\tb = v_2/v_1$ and a
mixing angle $\alpha$ in the neutral ${\cal CP}$--even sector.
Supersymmetry leads to several relations among these parameters and,
in fact, only two of them are independent.  These relations impose, at
the tree-level, a strong hierarchical structure on the mass spectrum $
[ M_h<M_Z , M_A < M_H$ and $M_W <M_{H^\pm}]$ which however is broken
by radiative corrections $\sim G_F m_t^4 \log{\tilde{m}_t^2/m_t^2}$
for the large top quark mass \cite{613} (c.f. Fig.\ref{f601}).

\GS The couplings of the neutral Higgs bosons to fermions and gauge
bosons will in general depend on the angles $\alpha$ and $\beta$.  The
pseudoscalar boson $A^0$ does not have tree-level couplings to gauge
bosons, and its couplings to (up) down type fermions are (inversely)
proportional to $\tb$.  The couplings are in general strongly
dependent on the input parameter $\tb$ and the masses.  The couplings
to down (up) type fermions are enhanced (suppressed) compared to the
SM Higgs couplings.  If $M_h$ is very close to its upper limit for a
given value of $\tb$, the couplings to fermions and gauge bosons are
SM like; this decoupling limit \cite{613B} is realized if the
pseudoscalar mass $M_A$ exceeds 300~GeV.

\GS
\noindent
a) \underline{Decays} \hfill  \\

\begin{figure}[ht]
\begin{center}
\epsfig{file=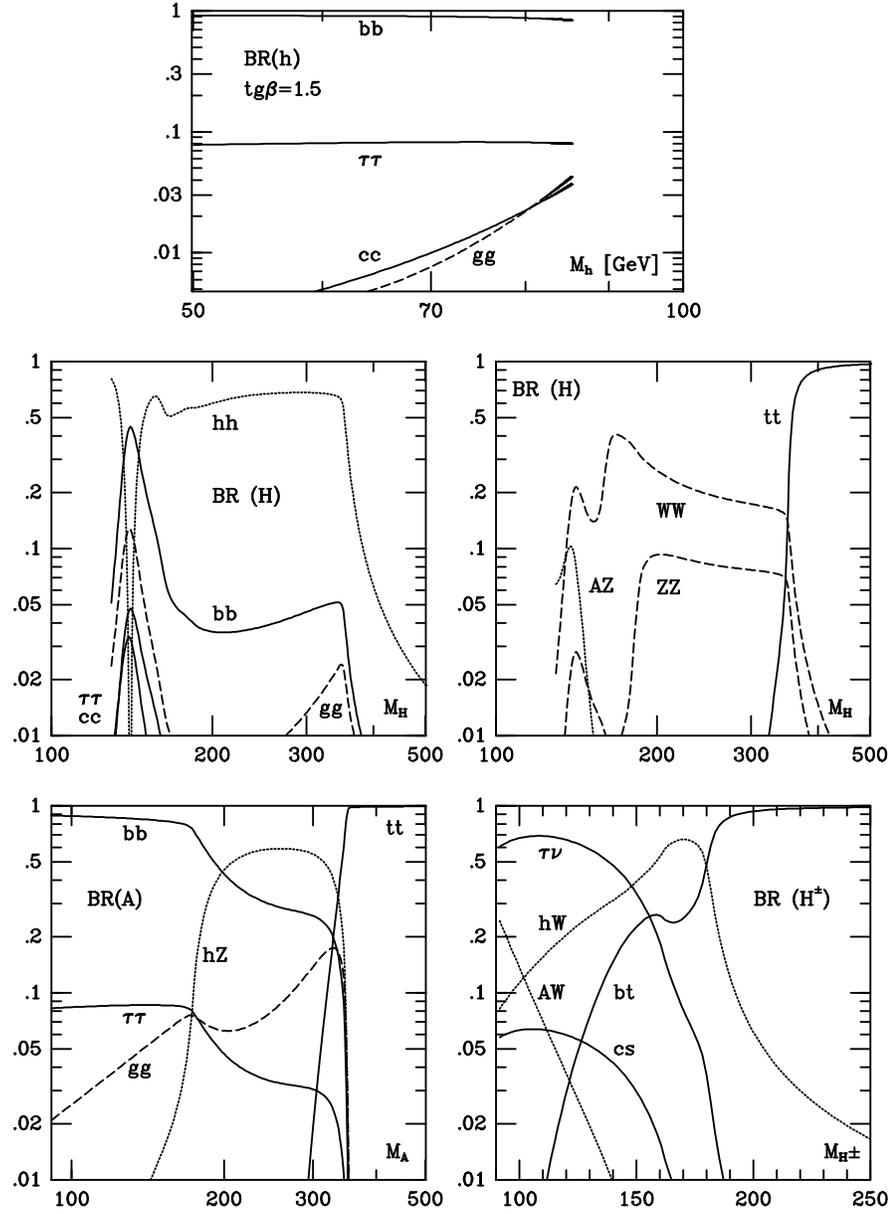,width=14cm}
\end{center}
\vspace{-2.5cm}
\caption[]{\it 
  Branching ratios of the main decay modes of the five MSSM Higgs
  bosons to SM particles and in cascade decays for $\tb = 1.5$;
  Refs.\protect\cite{613A,615}.
  \protect\label{f602}\label{alldecays}}
\end{figure}

\begin{figure}[ht]
\begin{center}
\vspace*{-1.cm}
\epsfig{file=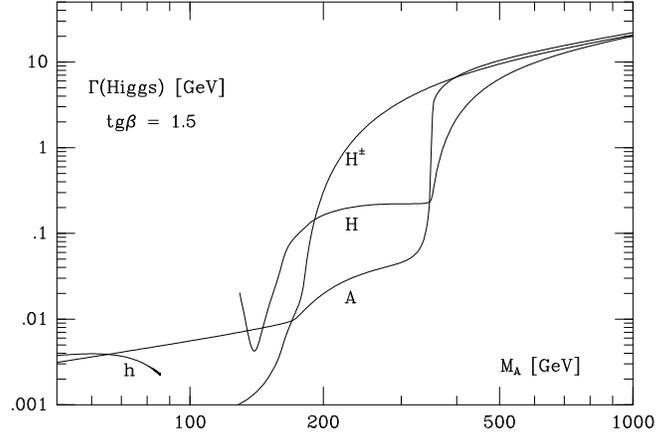,width=6.0cm,angle=-90}
\end{center}
\vspace{-.5cm}
\caption[]{\it 
  Total SM plus cascade decay widths of the five MSSM Higgs bosons for
  $\tb = 1.5$.  \protect\label{f603}\label{Gtot}}
\end{figure}

\STS
\noindent
The lightest {\it neutral Higgs boson} $h^0$ will decay mainly into
fermion pairs since its mass is smaller than $\sim$~130~GeV,
Fig.\ref{f602} (c.f. Ref.\cite{613A} for a comprehensive summary).
This is, in general, also the dominant decay mode of the pseudoscalar
boson $A^0$.  For values of $\tb$ larger than unity and for masses
less than $\sim$~140~GeV, the main decay modes of the neutral Higgs
bosons are decays into $b \bar{b}$ and $\tau^+ \tau^-$ pairs; the
branching ratios are of order $ \sim 90\%$ and $8\%$, respectively.
The decays into $c \bar{c}$ pairs and gluons are suppressed especially
for large $\tb$.  For large masses, the top decay channels $H^0, A^0
\rightarrow t\bar{t}$ open up; yet for large $\tb$ this mode remains
suppressed and the neutral Higgs bosons decay almost exclusively into
$b\bar{b}$ and $\tau^+\tau^-$ pairs.  If the mass is high enough, the
heavy ${\cal CP}$--even Higgs boson $H^0$ can in principle decay into
weak gauge bosons, $H^0 \to WW, ZZ$.  Since the partial widths are
proportional to $\cos^2 (\beta-\alpha)$, they are strongly suppressed
in general, and the gold-plated $ZZ$ signal of the heavy Higgs boson
in the \SM is lost in the supersymmetric extension.  As a result, the
total widths of the Higgs bosons are much smaller in \ssy theories
than in the Standard Model, c.f. Fig.\ref{Gtot}

\begin{figure}[ht]
\begin{center}
\vspace*{-1.2cm}
\hspace*{-.5cm}
\epsfig{file=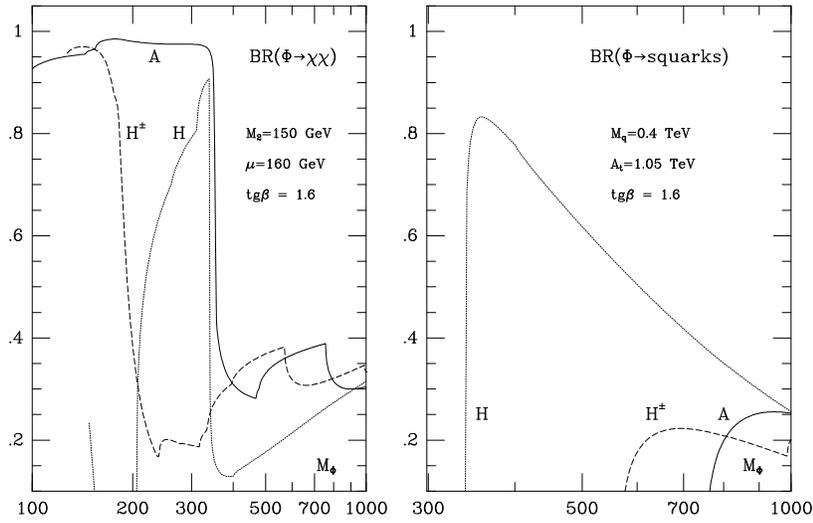,width=7.4cm,angle=-90}
\end{center}
\caption[]{\it 
  Decays of MSSM Higgs bosons to charginos/neutralinos and sfermions;
  Ref.\protect\cite{613A,615}.  \protect\label{susydec} }
\end{figure}

\STS The heavy neutral Higgs boson $H^0$ can also decay into two
lighter Higgs bosons.  Other possible channels are Higgs cascade
decays and decays into supersymmetric particles \citer{614,616},
Fig.\ref{susydec}.  In addition to light sfermions, Higgs boson decays
into charginos and neutralinos could eventually be important.  These
new channels are kinematically accessible at least for the heavy Higgs
bosons $H^0, A^0$ and $H^\pm$; in fact, the branching fractions can be
very large and they can become dominant in some regions of the MSSM
parameter space.  Decays of $h^0$ into the lightest neutralinos
$(LSP)$ are also important, exceeding 50\% in some parts of the SUSY
parameter space.  These decays strongly affect experimental search
techniques.  In particular, neutral Higgs decays into the $LSP$ which
would be invisible, could jeopardize the search for the Higgs
particles at hadron colliders where these decay modes are very
difficult to detect.  At \ee colliders however, missing mass
techniques allow us to find these events easily; this is most obvious
for the ${\cal CP}$--even Higgs bosons which can be produced in
association with the $Z$ boson.

\clearpage

\vspace*{0.5cm}

\vspace{2.5cm}

\clearpage
  
The {\it charged Higgs particles} decay into fermions but also, if
allowed kinematically, into the lightest neutral Higgs and a $W$
boson.  Below the $tb$ and $Wh$ thresholds, the charged Higgs
particles will decay mostly into $\tau \nu_{\tau}$ and $c\bar{s}$
pairs, the former being dominant for $\tb >1$.  For large $M_{H^\pm}$
values, the top-bottom decay mode $H^+ \to t\bar{b}$ becomes
dominant.  In some parts of the SUSY parameter space, decays into \ssy
\ps may exceed 50 percent.

\GS Adding up the various decay modes, the width of all five Higgs
bosons remains very narrow, being of order 10~GeV even for large
masses.
  
\GS
\noindent
b) \underline{Production} \hfill \\

\STS
\noindent
The search for the neutral SUSY Higgs bosons at \ee colliders will be
a straight\-forward ex\-tension of the search performed at LEP2, which
is expected to cover the mass range up to $\sim 90 \ \makebox{to} \ 
100$~GeV for neutral Higgs bosons, depending on $\tb$.  Higher
energies, $\sqrt{s}$ in excess of $250$~GeV, are required to sweep the
entire parameter space of the MSSM.

\GS The main production mechanisms of {\it neutral Higgs bosons} at
\ee colliders \cite{612, 615, 617} are the \Hs process and associated
pair production, as well as the fusion processes:
\begin{eqnarray}
(a) \ \ {\rm Higgs-strahlung} \hspace{1cm} \epem & 
\stackrel{Z}{\longrightarrow} & Z+h/H \hspace{5cm}
\nonumber  \\
(b) \ \ {\rm Pair \ Production} \hspace{13.6mm} \epem & 
\stackrel{Z}{\longrightarrow} & A+h/H 
\nonumber \\
(c) \ \ {\rm Fusion \ Processes} \hspace{10.7mm} \ \epem & 
\stackrel{WW}{\longrightarrow} & \overline{\nu_e} \ \nu_e \ + h/H 
\hspace{3.3cm} \nonumber  \\
\epem & 
\stackrel{ZZ}{\longrightarrow} &  \epem + h/H  \nonumber
\end{eqnarray}
The ${\cal CP}$--odd Higgs boson $A^0$ cannot be produced in fusion
processes to leading order.  The cross sections for the four \Hs and
pair production processes can be expressed as
\begin{eqnarray}
\sigma(\epem \ra Z + h/H) & =& \sin^2/\cos^2(\beta-\alpha) \ \sigma_{SM}
\nonumber \\
\sigma(\epem \ra A + h/H) & =& \cos^2/\sin^2(\beta-\alpha) \
\bar{\lambda} \  \sigma_{SM}
\end{eqnarray}
where $\sigma_{SM}$ is the SM cross section for \Hs and the coefficient
$\bar{\lambda} \sim \lambda^{3/2}_{Aj} / \lambda^{\demi}_{Zj}$ accounts 
for the suppression of the $P$--wave 
$Ah/H$ cross sections near the threshold.

\STS The cross sections for the Higgs-strahlung and for the pair
production, likewise the cross sections for the production of the
light and the heavy neutral Higgs bosons $h^0$ and $H^0$, are mutually
complementary to each other, coming either with coefficients
$\sin^2(\beta-\alpha)$ or $\cos^2(\beta-\alpha)$.  As a result, since
$\sigma_{SM}$ is large, at least the lightest ${\cal CP}$--even Higgs
boson must be detected.

\begin{figure}[ht]
\begin{center}
\hspace*{5mm}
\epsfig{file=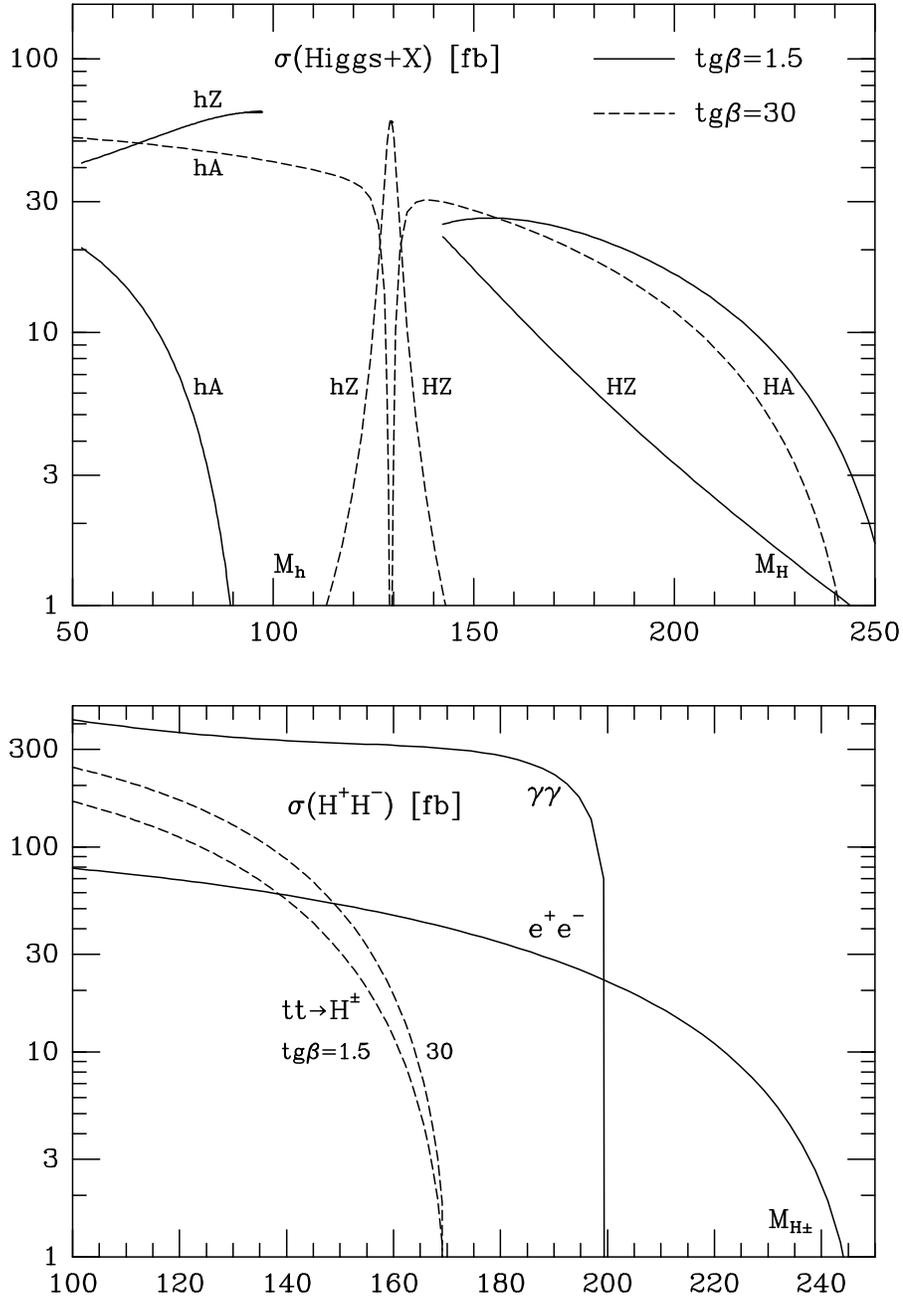,width= 15cm}
\end{center}
\vspace{-2.5cm}
\caption[]{\it 
  Production cross sections of MSSM Higgs bosons at $\sqrt{s} =
  500$~GeV: Higgs-strahlung and pair production; upper part: neutral
  Higgs bosons, lower part: charged Higgs bosons.
  Ref.\protect\cite{613A}.  \protect\label{f604}\label{prodcs}}
\end{figure}
\clearpage 

\STS Representative examples of the cross sections for the production
mechanisms of the neutral Higgs bosons are shown as a function of the
Higgs masses in Fig.\ref{f604} for $\tb= 1.5$ and 30.  The cross
section for $hZ$ is large for $M_h$ near the maximum value (allowed
for $\tb$), where it is of order 50~fb, corresponding to $\sim$ 2,500
events for an integrated luminosity of 50 fb$^{-1}$.  By contrast, the
cross section for $HZ$ is large if $M_h$ is sufficiently below the
maximum value associated with $\tb$ [implying small $M_H$].  For $h^0$
and for light $H^0$, the signals consist of a $Z$ boson accompanied by
a $b\bar{b}$ or $\tau^+ \tau^-$ pair.  The signal is easy to separate
from the background which comes mainly from $ZZ$ production if the
Higgs mass is close to $M_Z$.  For the associated channels $\epem \to
Ah$ and $AH$, the situation is opposite to the previous case: The
cross section for $Ah$ is large for light $h^0$ whereas $AH$ pair
production is the dominant mechanism in the complementary region for
heavy $H^0$ and $A^0$ bosons.  The sum of the two cross sections
decreases from $\sim 50$ to 10~fb if $M_A$ increases from $\sim 50$ to
200 GeV at $\sqrt{s} = 500 $~GeV.  In major parts of the parameter
space, the signals consist of four $b$ quarks in the final state,
requiring facilities for efficient $b$ quark tagging.  Mass
constraints will help to eliminate the backgrounds from QCD jets and
$ZZ$ final states.  For the $WW$ fusion mechanism, the cross sections
are larger than for \Hs if the Higgs mass is moderately small -- less
than 160~GeV at $\sqrt{s} = 500$ GeV.  However, since the final state
cannot be fully reconstructed, the signal is more difficult to
extract.  As in the case of the \Hs processes, the production of light
$h^0$ and heavy $H^0$ Higgs bosons complement each other in $WW$
fusion too.

\STS Once the heavy Higgs \ps $H^0$ and $A^0$ are discovered, the {\it
  negative parity of the pseudoscalar Higgs boson} $A^0$ must be
established.  For large $A^0, H^0$ masses the decays to $t\bar{t}$
final states can be used to discriminate between the different parity
assignments \cite{618}.  For example, the $W^+$ and $W^-$ bosons of
the $t$ and $\bar{t}$ decays tend to be emitted parallel and
anti-parallel for $A^0$ and $H^0$ decays, respectively, in the plane
perpendicular to the $t\bar{t}$ axis.  For light $A^0, H^0$ masses,
$\gamma \gamma$ collisions appear to provide a viable solution
\cite{618}.  The fusion of Higgs particles by linearly polarized
photon beams depends on the angle between the polarization vectors.
For scalar $0^+$ particles the production amplitude $\sim
\vec{\varepsilon}_1 \cdot \vec{\varepsilon}_2$ is non-zero only for
parallel polarization vectors while pseudoscalar particles $0^-$ with
amplitudes $\sim \vec{\varepsilon}_1 \times \vec{\varepsilon}_2$
require perpendicular polarization vectors.  The experimental set-up
for Compton back-scattering of laser light can be tuned in such a way
that the linear polarization of the generated hard photon beams
approaches values close to $100\%$.  This method requires high
luminosities.

\GS The {\it charged Higgs bosons}, if lighter than the top quark, can
be produced in top decays, $t \ra b + H^+$, with a branching ratio
varying between $2\%$ and $20\%$ in the kinematically allowed region.
Since the cross section for top pair production is of order 0.5 pb at
$\sqrt{s} = 500$~GeV, this corresponds to 1,000 to 10,000 charged
Higgs bosons at a luminosity of 50~fb$^{-1}$.  Since for $\tb$ larger
than unity, the charged Higgs bosons will decay mainly into $\tau
\nu_\tau$, this results in a surplus of $\tau$ final states over $e,
\mu$ final states in $t$ decays, an apparent breaking of lepton
universality.  For large Higgs masses the dominant decay mode is the
top decay $H^+ \to t \overline{b}$.  In this case the charged Higgs
particles must be pair produced in \ee colliders:
\[
              \epem \to H^+H^-
\]
The cross section depends only on the charged Higgs mass.  It is of
order 100 fb for small Higgs masses at $\sqrt{s} = 500$~GeV, but it
drops very quickly due to the $P$--wave suppression $\sim \beta^3$
near the threshold.  For $M_{H^{\pm}} = 230$~GeV, the cross section
falls to a level of $\simeq 5\,$~fb, which for an integrated
luminosity of $50\,{\rm fb}^{-1}$ corresponds to 250 events.  The \cs
is considerably larger for $\gamma \gamma$ collisions.

\GS The reconstruction of the {\it Higgs potential} is much more
complicated in \ssy theories than in the \SM since a large ensemble of
trilinear and quartic couplings between the Higgs \ps are predicted in
two-doublet scenarios.  Nevertheless, it has been demonstrated that
from Higgs cascade decays like $H \to hh$, and also from Higgs pair
production in the continuum, trilinear couplings can be reconstructed
in part of the Higgs parameter space \cite{619}.

\begin{figure}[ht]
\begin{center}
\hspace*{-1cm}
\epsfig{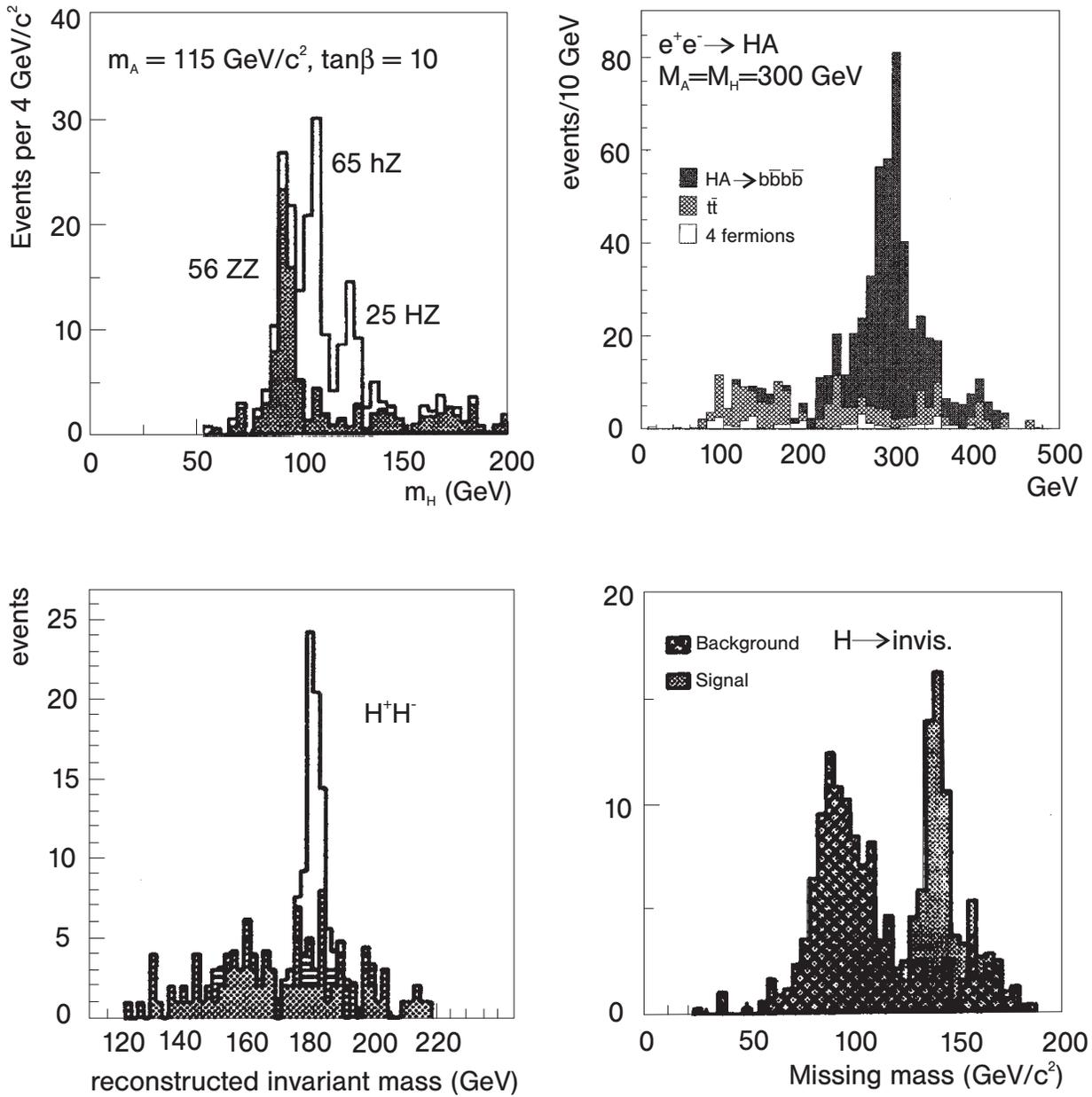}
\end{center}
\vspace{-.5cm}
\caption[]{\it 
  Experimental simulations of the search for MSSM Higgs bosons in
  Higgs-strahlung $hZ / HZ$, heavy pair production $HA$, charged
  Higgs production $H^+ H^-$, and neutral invisible Higgs decays in
  Higgs-strahlung.  Refs.\protect\citer{620,620A}.
  \protect\label{f605}}
\end{figure}

\GS Experimental search strategies have been summarized for neutral
Higgs bosons in Refs.\cite{620,620AA} and charged Higgs bosons in
Ref.\cite{620A}.  Examples of the results for \Hs $Zh, ZH$ and pair
production $Ah$, $AH$ and $H^+H^-$ are given in Fig.\ref{f605}.
Visible as well as invisible decays are under experimental control
already for an integrated luminosity of 10~fb$^{-1}$.

\GS
\noindent
c) \underline{Experimental Summary} \hfill \\

\noindent
The preceding discussion of the MSSM Higgs sector at \ee linear
colliders can be summarized in the following two points:

\STS
\noindent
{$(i)$} The lightest ${\cal CP}$--even Higgs particle $h^0$ can be
detected in the entire range of the MSSM parameter space, either
via the Higgs-strahlung process $\epem \to hZ$ or via pair
production $\epem \rightarrow hA$.  This conclusion holds true even at
a c.m. energy of 250 GeV, independently of the squark mass values; it
is also valid if decays to invisible neutralino and other SUSY \ps
will be realized.

\STS
\noindent
{$(ii)$} The area in the parameter space where {\it all} SUSY Higgs
bosons can be discovered at \ee colliders is characterized by $M_H,
M_A \lessim \frac{1}{2} \sqrt{s}$, independently of $\tb$.  The $h^0,
H^0$ Higgs bosons can be produced either using \Hs or using $Ah, AH$
associated production; charged Higgs bosons will be produced in
$H^+H^-$ pairs.

\STS The properties of the SUSY Higgs bosons can be explored in the
same way as the 

\clearpage

\noindent
 Higgs \p of the \SM so that the profile of the Higgs
\ps can be reconstructed in \ssy theories.

\GS It has been established  that at least one of the Higgs
\ps in the SUSY spectrum could be discovered at the LHC. It has been
assumed, however, in these analyses that the Higgs \ps decay only into
SM particles.  The general consequences of decays to SUSY particles,
partly invisible, have not yet been studied experimentally.  A problem
arises from the difficulties of establishing the heavy neutral Higgs
bosons, above the top threshold, in the interesting parameter range of
small to moderate $\tb$. This is a
problem for most of the Higgs mass estimates in supergravity inspired
parametrizations.  The detection of charged Higgs bosons is
guaranteed, at this point, only in top decays, restricting the mass
range accessible for this \p to rather low values.  Thus, there are
large areas in the SUSY Higgs parameter space where the ensemble of
individual Higgs particles are not accessible {\it in toto} at the
same time.

\GS
\noindent
d) \underline{Non--Minimal Supersymmetric Extensions} \hfill \\

\noindent
The minimal supersymmetric extension of the \SM (MSSM) may appear very
restrictive for supersymmetric theories in general, in particular in
the Higgs sector where the quartic couplings are identified with the
gauge couplings.  However, it turns out that the mass pattern of the
MSSM is quite typical if the theory is assumed to be valid up to the
GUT scale -- the motivation for supersymmetry {\it per se}.  This
general pattern has been studied thoroughly within the
next-to-minimal extension: The MSSM, incorporating two Higgs
isodoublets, is augmented by introducing an isosinglet field $N$.
This extension leads to a model \cite{621,622} which is generally
referred to as the (M+1)SSM.

\STS The additional Higgs singlet can solve the so-called
$\mu$--problem [i.e. to explain why $\mu = {\cal O}(M_W)$] by
eliminating the $\mu$ higgsino parameter in the potential and
replacing its effect by the vacuum expectation value of the $N$ field,
which can be naturally related to the usual vacuum expectation values
of the Higgs isodoublet fields.  In this scenario the superpotential
involves the two trilinear couplings $H_1 H_2 N$ and $N^3$.  The
consequences of this extended Higgs sector will be outlined below in
the context of (s)grand unification including universal soft breaking
terms of the supersymmetry \cite{622}.

\GS The Higgs spectrum of the (M+1)SSM includes, besides the minimal
set of Higgs particles, one additional scalar and pseudoscalar Higgs
particle.  The neutral Higgs \ps are in general mixtures of the
iso-scalar doublets, which couple to $W, Z$ bosons and fermions, and
the iso-scalar singlet, decoupled from the non-Higgs sector.
The\pagebreak\ 
trilinear self-interactions contribute to the masses of the Higgs
particles.  In contrast to the minimal model, the mass of the charged
Higgs \p could be smaller than the \W mass.  Since the trilinear \cps
increase with energy, upper bounds on the mass of the lightest neutral
Higgs boson $h_1^0$ can be derived, in analogy to the Standard Model,
from the assumption that the theory be valid up to the GUT scale:
$m(h_1^0) \lessim 140 $~GeV.  Thus despite the additional
interactions, the distinct pattern of the minimal extension remains
valid also in more complex \ssy \sces \cite{623}.  In fact, the mass
bound of 140~GeV for the lightest Higgs particle is realized in almost
all \ssy theories.  If $h_1^0$ is (nearly) pure iso-scalar, it
decouples from the gauge boson and fermion system and its r\^ole is
taken by the next Higgs \p with a large isodoublet component, implying
the validity of the mass bound again.

\begin{figure}[ht]
\begin{center}
\epsfig{file=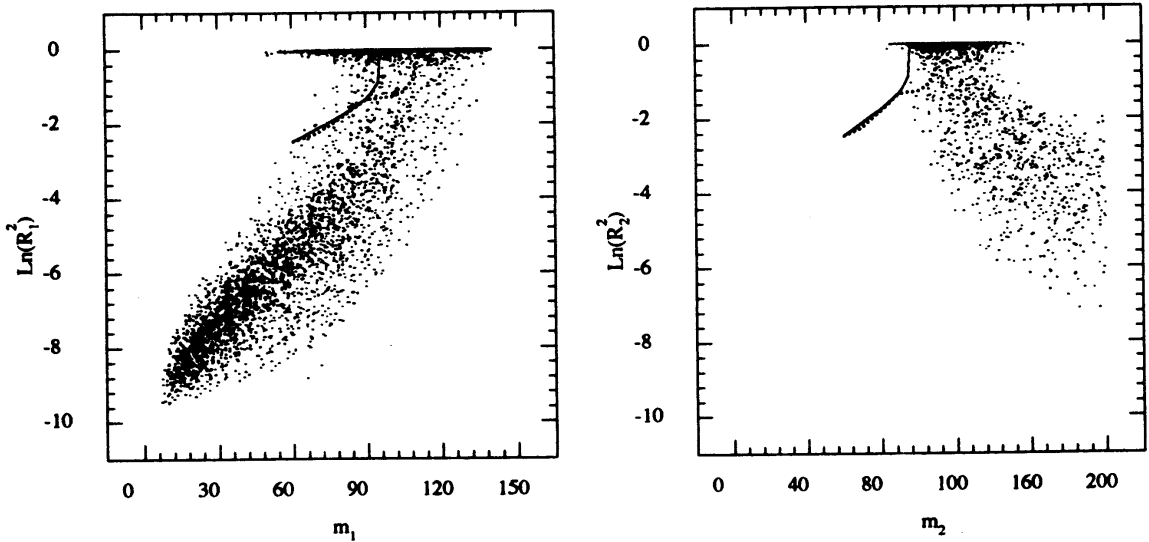,width=15.5cm}
\end{center}
\vspace{-.5cm}
\caption[]{\it 
  The couplings $ZZh_1$ and $ZZh_2$ of the two lightest $CP$--even
  Higgs bosons in the next-to-minimal supersymmetric extension of
  the Standard Model, $(M + 1) SSM$. The solid lines indicate the
  accessible range at LEP2, the dotted lines for an energy of 205~GeV.
  The scatter plots are solutions for an ensemble of possible SUSY
  parameters defined at the scale of grand unification.
  Ref.\protect\cite{622}.  \protect\label{f621}\label{Ellw}}
\end{figure}

\STS The \cps $R_i$ of the ${\cal CP}$--even neutral Higgs \ps $h_i^0$
to the $Z$ boson, $ZZh_i^0$, are defined relative to the usual SM
coupling.  If the Higgs \p $h_1^0$ is primarily isosinglet, the \cp
$R_1$ is small and the \p cannot be produced by Higgs-strahlung.
However, in this case $h_2^0$ is generally light and couples with
sufficient strength to the $Z$ boson; if not, $h_3^0$ plays this r\^ole.
This \sce is quantified in Fig.\ref{f621} where the \cps $R_1$ and
$R_2$ are shown for the ensemble of allowed Higgs masses $m(h_1^0)$

\noindent 

and $m(h_2^0)$ [adopted from Ref.\cite{624}; see also
Ref.\cite{622,L141A}].  Two different regions exist within the GUT
(M+1)SSM: A densely populated region with $R_1 \sim 1$ and $m_1 > 50
$~GeV, and a tail with $R_1 < 1$ to $<< 1$ and small $m_1$.  Within
this tail, the lightest Higgs boson is essentially a gauge singlet
state so that it can escape detection at LEP [full/solid lines].  If
the lightest Higgs boson is essentially a gauge singlet, the second
lightest Higgs \p cannot be heavy.  In the tail of diagram~\ref{f621}a
the mass of the second Higgs boson $h_2^0$ varies between 80~GeV and,
essentially, the general upper limit of $\sim 140 $~GeV.  $h_2^0$
couples with full strength to $Z$ bosons, $R_2 \sim 1$.  If in the
tail of diagram~\ref{f621}b this \cp becomes weak, the third Higgs
boson will finally take the r\^ole of the leading light particle.

\GS To summarize: Experiments at \ee colliders are in a no-lose
situation \cite{L141A} for detecting the Higgs \ps in general \ssy
theories even for c.m. energies as low as $\sqrt{s} \sim 300 $~GeV.

\STS
\subsection[Supersymmetric Particles]{Supersymmetric Particles}

The only guidelines for estimating the mass spectra of \ssy particles,
follow from the embedding of low-energy supersymmetry into grand
unified theories and the requirement of avoiding the fine tuning of
parameters.  The second principle is hard to quantify, yet for
plausible \sces such as those presented in Table~\ref{t3}, the spectra
do conform to this principle.  The embedding of low-energy
supersymmetry into supergravity \sces with universal soft
SUSY-breaking parameters, reduces the number of free parameters,
generally of order one hundred, to a few.  In addition, problems of
${\cal CP}$ violation are removed etc.  Given that the large ensemble
of masses and mixing angles is reduced to a small number, many
relations can be found among the observables which can be scrutinized
with high accuracy.

\STS It is evident from the table of masses derived for various SGUT
\sces that a large number of particles can be expected which are
accessible in the first phase of the \ee linear colliders.  In
particular, the color-neutral charginos/neutralinos and sleptons,
lighter than the colored squarks and gluinos, can be produced in \ee
collisions and studied thoroughly in the clean environment of these
machines.  Moreover, stop \ps could be light as well, partly a result
of mixing effects induced by the large Yukawa \cp between L and R
states in this sector.  The experiments at \ee colliders will not only
allow high-precision measurements of masses and couplings, but also
of such subtle effects as mixings.

\begin{figure}[ht]
\begin{center}
\epsfig{file=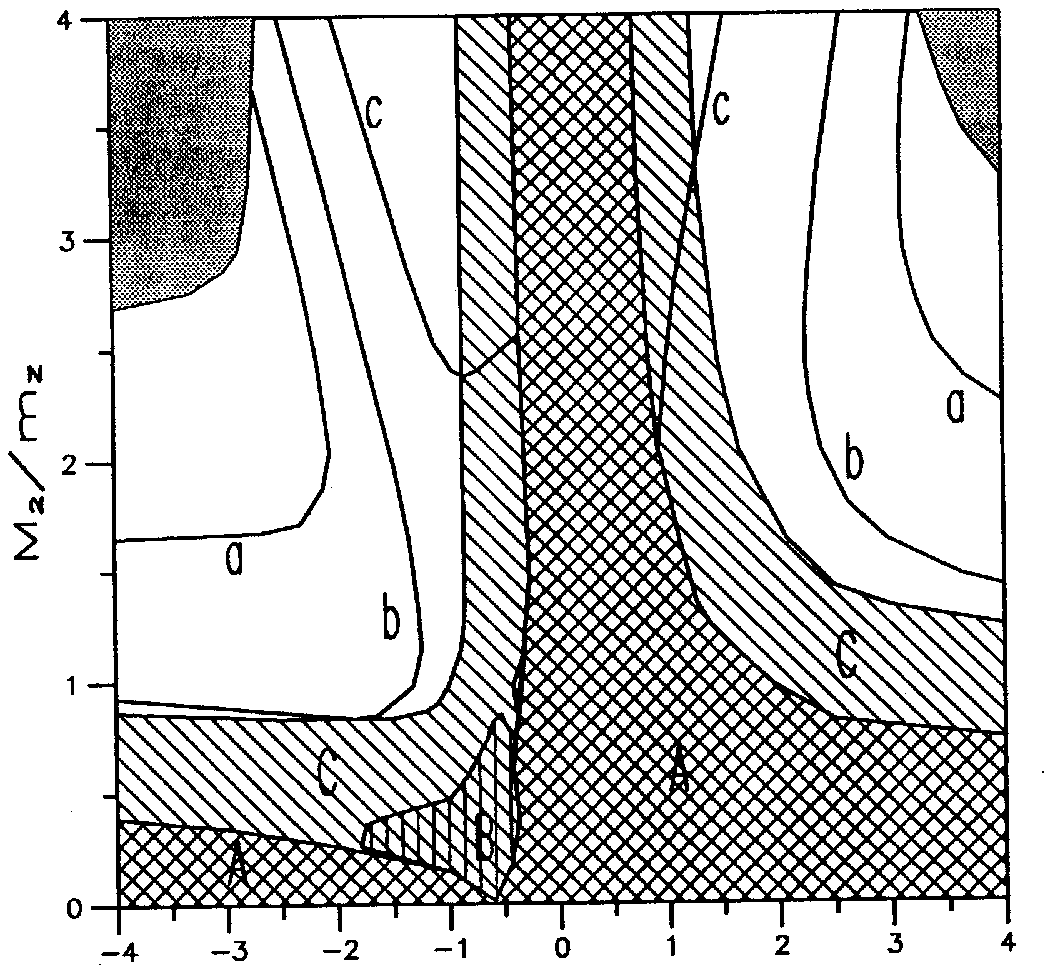,width=8.3cm}
\raisebox{-2mm}{\epsfig{file=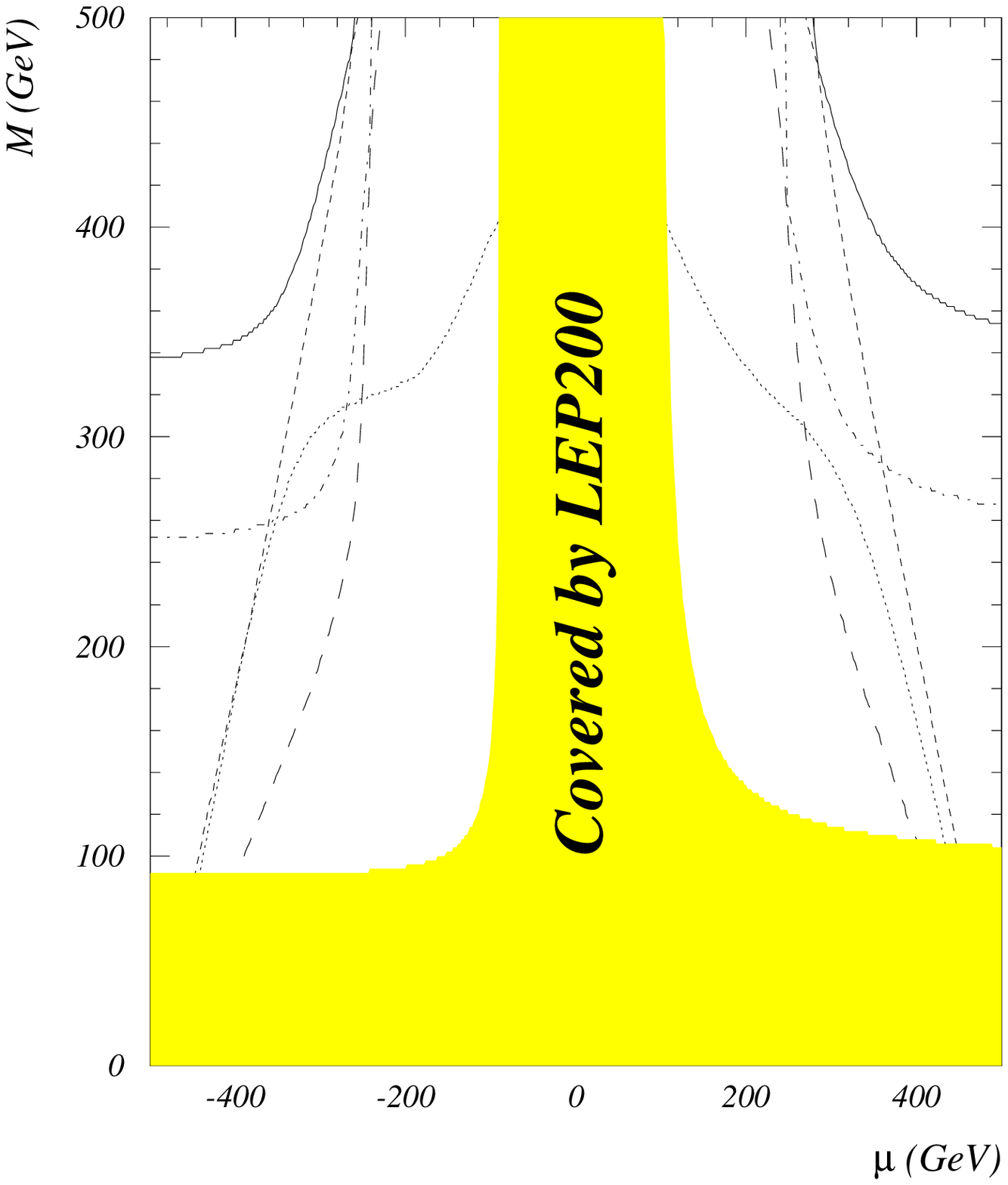,width=6.9cm}}
\end{center}
\vspace{-0.5cm}
\caption[]{\it 
  Left: The impact of chargino searches in the [$\mu, M_2$] plane of
  the MSSM for $\tan\beta=2$ at $\sqrt{s} = 500$ GeV.  The hatched
  region is covered by LEP2; in the grey region the chargino mass
  exceeds the limit of $m_{\tilde{\chi}_1^{\pm}} = 250$ GeV.  The
  curves a,b,c correspond to mass differences
  $m_{\tilde{\chi}^{\pm}_1} - m_{\tilde{\chi}^0_1}$ of 80, 50 and 20
  GeV, respectively.  Ref.\protect\cite{632}.  Right: Supersymmetry
  parameter space [$\mu, M_2$] for neutralino production for
  $\tan\beta=4$ at $\sqrt{s} = 500$~GeV.  Shown are the limits for
  $e^+e^- \rightarrow \tilde{\chi}^0_i \tilde{\chi}^0_j$ in the
  combinations: 12 (solid); 13 (short dashes); 14 (dots); 22
  (dot-dashes); 23 (long dashes).  Ref.\protect\cite{631}.
  \protect\label{f641}\label{inos}}
\end{figure}

\GS
\noindent
a) \underline{Charginos and Neutralinos} \hfill \\

\STS
\noindent
The ensemble of the two charginos $\tilde{\chi}_i^+$ and the four
neutralinos $\tilde{\chi}_i^0$, mixtures of the [non-colored]
gauginos and higgsinos, include the lightest \ssy particle ($LSP$) in
a large part of parameter space.  In the MSSM with conserved
$R$--parity, the neutralino $\tilde{\chi}_1^0$ with the smallest mass
is in general the lightest \ssy particle and stable. Only in
exceptional cases sneutrinos are the lightest SUSY particles.   The
heavier neutralinos and the charginos decay into (possibly virtual)
gauge and Higgs bosons plus the $LSP$, $\tilde{\chi}_i^0 \to
\tilde{\chi}_1^0 + Z/W$ and $\tilde{\chi}_1^0 + H$, or if they are
heavy enough, into neutralino/chargino cascades, and sleptons plus
leptons \cite{631}.  At the end of the cascades, the events will
consist of jet pairs, leptons and $LSP$'s which escape undetected.

\STS Neutralinos and charginos are easy to detect and to study with
high accuracy at \ee colliders.  They are produced in pairs
\begin{eqnarray*}
\epem & \to & \tilde{\chi}_i^+ \tilde{\chi}_j^-
\hspace{1.5cm} [i,j = 1,2] \\
\epem & \to & \tilde{\chi}_i^0 \tilde{\chi}_j^0
\hspace{1.5cm} \hspace{2mm} [i,j = 1,..,4] 
\end{eqnarray*}
through $s$--channel $\gamma, Z$ exchange and $t$--channel sneutrino
or selectron exchange. \linebreak\ [Experimental details are presented
in the second appendix.]  \hfill\ The accessible SUSY

\begin{figure}[ht]
\vspace*{-1.2cm}
\begin{center}
\hspace{-3mm}
\epsfig{file= 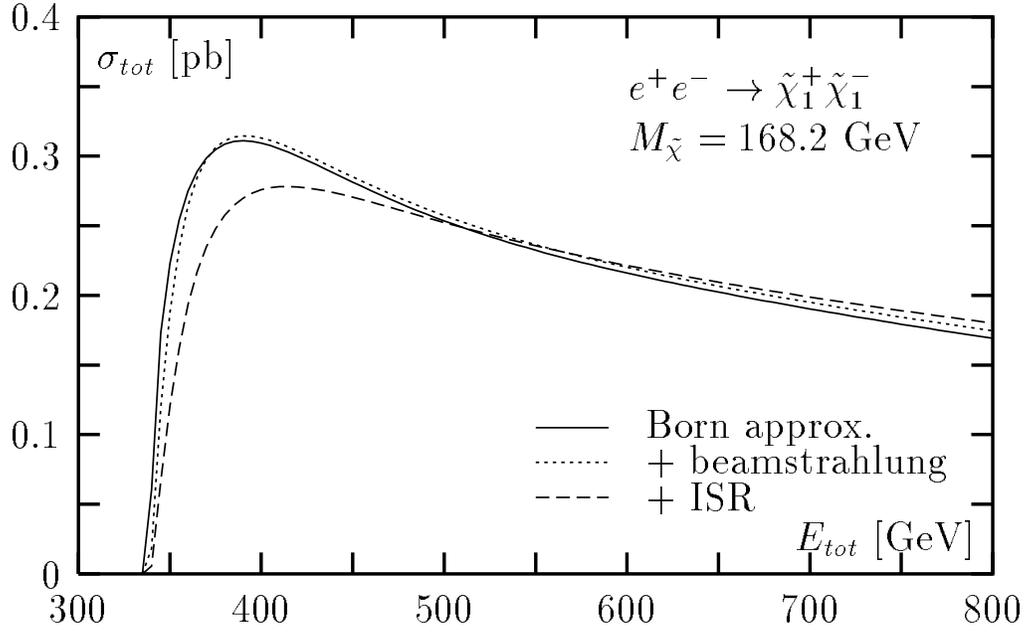,width=13.8cm} \\
\vspace*{-1.3cm}
\hspace*{-7mm}
\epsfig{file= 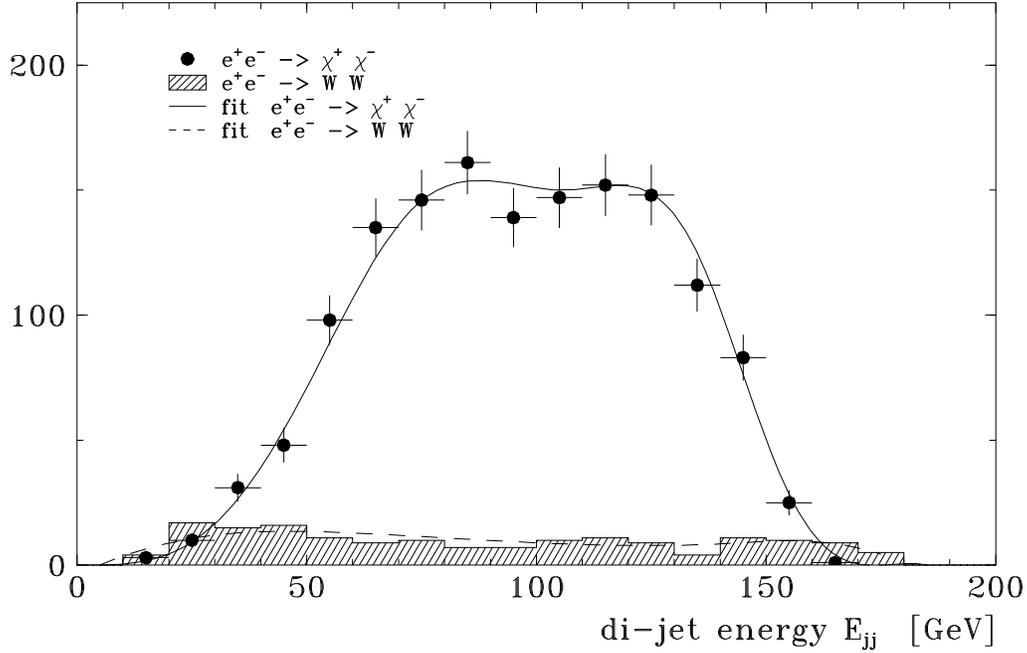,width=11.0cm,angle=90}
\end{center}
\vspace{-0.7cm}
\caption[]{\it 
  Upper part: Threshold behavior of the cross section for $e^+e^-
  \rightarrow \tilde{\chi}^+_1 \tilde{\chi}^-_1$ including
  initial-state radiation and beamstrahlung.
  Lower part: Simulation of the energy spectrum in the decay
  $\tilde{\chi}^+_1 \rightarrow \tilde{\chi}^0_1 + jj$ based on the
  input values $m_{\tilde{\chi}^+_1} = 168.2$~GeV and
  $m_{\tilde{\chi}^0_1} = 88.1$~GeV at $\sqrt{s}$ = 500 GeV;
  Ref.\protect\cite{635}.  \protect\label{f642}\label{chitdecay}}
\end{figure}
\clearpage 
\noindent
 parameter range in the $[\mu,
M_2]$ plane is shown in Fig.\ref{f641} for the production of various
chargino and neutralino pairs at 500~GeV colliders.  [$\mu$ is the
higgsino mass parameter while $M_2$ is the SU(2) gaugino mass; the
U(1) gaugino mass $M_1$ is generally assumed to be related to $M_2$,
with a coefficient $5/3 \tan^2\theta_w$, motivated by supergravity
models.]  Compared to the region which can be explored by LEP2, a
substantial extension can be anticipated.  Since the \css are as large
as ${\cal O}(100 \ {\rm fb})$, enough events will be produced to
discover these \ps for masses nearly up to the kinematical limit.  In
fact, it has been demonstrated by detailed experimental simulations
that charginos can be detected with masses up to the beam energy if
the mass difference $m(\tilde{\chi}_1^+) - m(\tilde{\chi}_1^0)$ is
sufficiently large \cite{632}.

\begin{figure}[b]
\begin{center}
\epsfig{file= 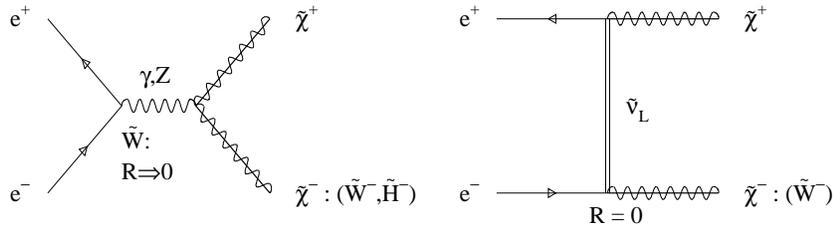,width=12cm}
\end{center}
\vspace{-.5cm}
\caption[]{\it 
  Mechanisms contributing to the production of charginos in \ee
  collisions.  Given are the gaugino/higgsino components that can be
  excited in $s$-- and $t$--channel diagrams; also indicated is the
  impact of right-handed electron polarization.
  \protect\label{fnew}\label{expsearch}}
\end{figure}

\STS The properties of the neutralinos and charginos can be studied in
great detail at \ee colliders.  The decay energy spectrum in
$\tilde{\chi}_1^+ \to \tilde{\chi}_1^0 + W^+$ allows us to measure the
mass of the $ \tilde{\chi}_1^+$ \p within $\delta m_{\tilde{\chi}_1^+}
= 1.1$~GeV, and even better for the neutralino, c.f. Fig.\ref{f642}.
From the fast onset $\sim \beta$ of the spin--$\frac{1}{2}$ excitation
curve near the threshold, the masses can be measured very accurately.
Performing a threshold scan, the error on the $\tilde{\chi}_1^{\pm}$
mass can be reduced to a very small value.  Accuracies on the $
\tilde{\chi}_1^{\pm}$ and $ \tilde{\chi}_1^0$ masses of
\begin{eqnarray*}
\delta m_{\tilde{\chi}_1^{\pm}} & \approx 100 & \ {\rm MeV}\\
\delta m_{\tilde{\chi}_1^0}     & \approx 600 & \ {\rm MeV}
\end{eqnarray*}
can finally be achieved.

\STS
Using polarized $e^{\pm}$
 beams, the decomposition of the states,
$\tilde{\chi}_1^- = \alpha \tilde{W}^- + \beta \tilde{H}^-$ into Wino
and higgsino components can be determined \citer{134A,134C}.  In
general, both the Wino and higgsino components of the charginos are
produced through $s$--channel $\gamma, Z$ exchange, Fig.\ref{fnew}.
However, at high energies $\gamma$ and $Z$ are demixed to $B^0/W^3$
bosons so that only the higgsino component is generated for
right-handedly polarized electrons.
  In the second diagram,
describing the $t$--channel $\tilde{\nu}_L$ exchange, only the Wino
component couples which can be switched off/on by operating right/left
polarized beams.  Since the energy and angular dependence are
different for the two states of L/R polarization, the mixing
parameters $\alpha$ and $\beta$ can be determined directly.  Similar
techniques can be applied for neutralinos.

\GS In some areas of the parameter space of supergravity models,
spectacular single--$\gamma$ and $\gamma\gamma$ final states are
predicted in \ee collisions \cite{108A}.  In these models the lightest
neutralino may decay into a photon $\gamma$ plus a gravitino
$\tilde{G}$ which escapes undetected, or the $\tilde{\chi}^0_1
\tilde{G}$ final state is produced directly:
\begin{eqnarray*}
\epem  \to & \tilde{\chi}^0_1 \tilde{G} & \to \gamma \tilde{G}
\tilde{G} \to \hspace{0mm} 
\gamma +\notE \\
\epem  \to & \hspace{0.1mm} 
\tilde{\chi}_1^0 \tilde{\chi}_1^0 &\hspace{0mm}
 \to \hspace{0mm}
\gamma \gamma \tilde{G} \tilde{G}  \to \hspace{0mm} 
\gamma \gamma +\notE 
\end{eqnarray*}

The most important background events to these processes are generated
by the radiative return to the $Z$, $\epem \rightarrow n \gamma + Z$
with the $Z$ decaying into pairs of neutrinos.  This background can be
eliminated by requiring the invariant missing mass, represented by
$\notE$, to be below the $Z$ mass.  [Pure photonic events can also be
generated in $\epem \to \tilde{\chi}_1^0 \tilde{\chi}_2^0$ and
$\tilde{\chi}_2^0\tilde{\chi}_2^0$ when the branching ratio for decays
$\tilde{\chi}_2^0 \to \gamma \tilde{\chi}_1^0$ is non-negligible.]

\GS
\noindent
b) \underline{Sleptons} \hfill \\

\STS
\noindent
The superpartners of the right-handed leptons decay into the
associated SM partners
and neutralinos/charginos.   In major parts of
the SUSY
  parameter space the dominant decay mode is $\tilde{\mu}_R \to
\mu + \tilde{\chi}_1^0$ \cite{631}.  For the superpartners of the
left-chiral sleptons, the decay pattern is more complicated since,
besides the $\tilde{\chi}_1^0$ channels, decays into leptons and
charginos can also occur.  In \ee collisions, sleptons are produced in
pairs:
\begin{eqnarray*}
\epem   & \to  &  \tilde{\mu}_L^+  \tilde{\mu}_L^- \; , \;
\tilde{\mu}_R^+  \tilde{\mu}_R^- \; , \;
\tilde{\tau}_L^+  \tilde{\tau}_L^- \; , \;
\tilde{\tau}_R^+  \tilde{\tau}_R^-                  \\
\epem   & \to  &  \tilde{e}_L^+  \tilde{e}_L^- \; , \;
\tilde{e}_R^+  \tilde{e}_R^- \; , \;
\tilde{e}_L^+  \tilde{e}_R^- \; , \;
\tilde{e}_R^+  \tilde{e}_L^-                  \\
\epem   & \to  &  \tilde{\nu}_{e L}  \bar{\tilde{\nu}}_{e L} \; , \;
\tilde{\nu}_{\mu L}  \bar{\tilde{\nu}}_{\mu L} \; , \;
\tilde{\nu}_{\tau L}  \bar{\tilde{\nu}}_{\tau L}
\end{eqnarray*}
For charged sleptons, the production proceeds via $\gamma, Z$ exchange
in the $s$--channel. In the case of selectrons, additional
$t$--channel neutralino exchange is present, which is also responsible
for the production of the mixed left and right-chiral selectron
states.  For sneutrinos, the production process is mediated by
$s$--channel $Z$--exchange and, in the case of electron-sneutrinos, by
$t$--channel exchange of charginos in addition.

\begin{figure}[ht]
\begin{center}
\vspace*{-1.5cm}
\epsfig{file=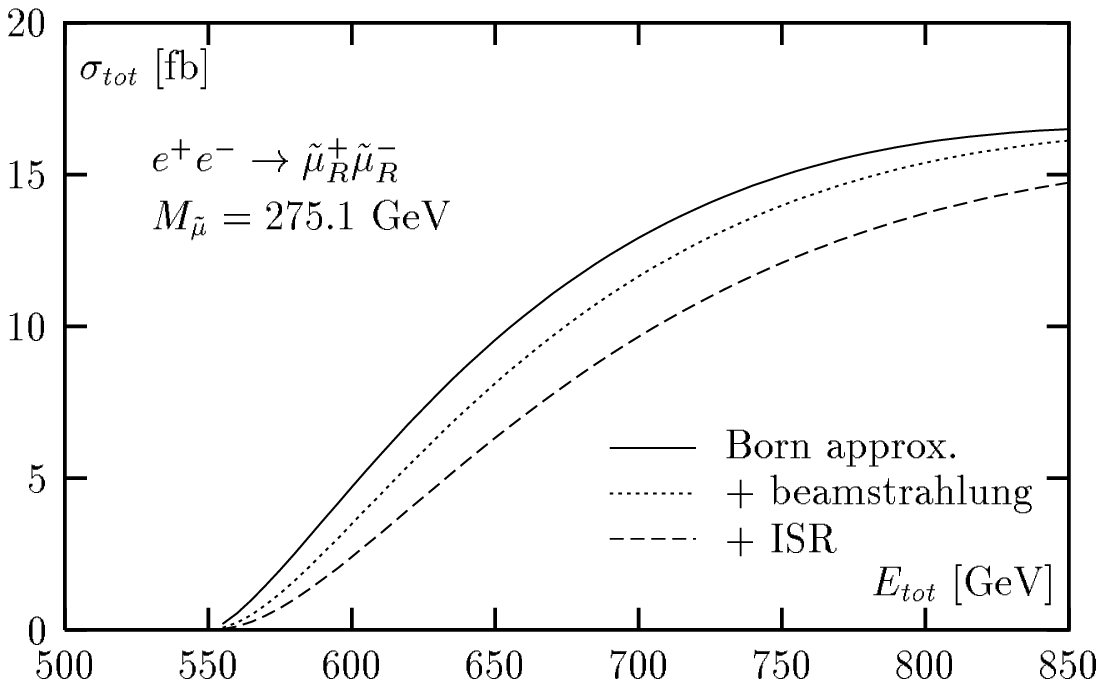,width=13.3cm} \\
\vspace*{-1.0cm}
\hspace*{-8mm}
\epsfig{file=susy_smu_dec.ps,width=11.0cm,angle=90} \\
\end{center}
\vspace*{-1cm}
\caption[]{\it 
  Upper part: The cross section of the production of $\tilde{\mu}
  \tilde{\mu}$ pairs in $\epem$ collisions for different electron-beam
  polarizations.  Lower part: Decay lepton energy spectrum in
  $\tilde{\mu}_R \to\mu \tilde{\chi}^0_1$ for $m_{\tilde{\mu}_R}$ =
  275.1 GeV and $m_{\tilde{\chi}^0_1}$ = 88.1 GeV at the energy
  $\sqrt{s}$ = 800 GeV; also shown is the background from $WW$ pair
  production.  Ref.\protect\cite{635}.
  \protect\label{f645}\label{mut}\label{f644}\label{eetoll}}
\end{figure}

\STS The \css for the pair production of sleptons are of the order of
5 to 15 fb ,c.f. Fig.\ref{f644} (upper part), so that their discovery
is very easy up to the kinematical limit \cite{634}.  Enough events
will be produced to study their detailed properties.  From the sharp
threshold behavior of the excitation curve and/or decay spectra the
masses

\clearpage

\noindent
of the $\tilde{\mu}$ and $\tilde{\chi}_1^0$ can be determined
\cite{635} up to a level of
\[
\delta m_{\tilde{\mu}}=1.8 \hspace{2mm} {\rm GeV} 
\] 
c.f.\ Fig.\ref{f645} (lower part).  By analysing the angular
distribution in the production process, the spin of the sleptons can
be checked to be zero.  The polarization of the $e^{\pm}$ beams will
help to identify the \cps of these particles.

\STS The opportunity to analyze the $\tau$ helicity \cite{636} in the
decay $\tilde{\tau}_L^- \to \tau_{\lambda}^- \tilde{\chi}_1^0$
($\lambda = $~L/R) can be exploited to discriminate between the
gaugino and higgsino components of $\tilde{\chi}_1^0$.  The gaugino
component of $\tilde{\chi}_1^0$, coupled in a chirality-conserving
vertex, gives rise to 
left-handed $\tau_L^-$ states, the higgsino
component, coupled in a chirality-flip vertex, to right-handed
$\tau_R^-$ states.  The analysis can readily be extended to the more
complicated \sce of L/R mixed $\tilde{\tau}$ states.

\GS Selectrons can be produced in association with gauginos in $e
\gamma$ collisions: $e \gamma \to \tilde{e} \tilde{\chi}_1^0 \to e
\tilde{\chi}_1^0 \tilde{\chi}_1^0$ etc.  For small $\tilde{\chi}_1^0$
masses, the kinematic range of the selectron mass extends beyond the
$e^{\pm}$ beam energy \cite{637}.  However, it seems difficult to
exploit this window in practice, since for masses beyond the $e^{\pm}$
beam energies the rates are quite low.

\GS
\noindent
c) \underline{Stop particles} \hfill \\

\STS
\noindent
The stop particles $\tilde{t}_1$ may have small masses \cite{638}
compared to the other squarks for two reasons.  First, due to the
large Yukawa terms, the mass term of the top squark may evolve to much
lower values than the mass terms for first and second generation
squarks [see e.g. Ref.\cite{615}].  Second, mixing due to large Yukawa
terms between $\tilde{t}_L$ and $\tilde{t}_R$ leads to a large
splitting of the mass values associated with the mass eigenstates
$\tilde{t}_1$ and $\tilde{t}_2$.  Stop experiments can therefore be
exploited to measure the soft SUSY-breaking trilinear scalar \cp $A$.
[Similar phenomena may also be observed in the $\tilde{b}$ and
$\tilde{\tau}$ sectors, less pronounced though \cite{639}.]  In \ee
collisions the following final states can be generated \cite{639}
\[
\epem \to \tilde{t}_1 \bar{\tilde{t}}_1,
\tilde{t}_2 \bar{\tilde{t}}_2 \; \; \hspace{+2mm} {\rm and } \;\; 
\hspace{+2mm}
\tilde{t}_1 \bar{\tilde{t}}_2 \;  + \;  c.c
\]
either by $\gamma, Z$ exchange for diagonal or $Z$ exchange for mixed
final states.  The production rates are determined by the masses and
the mixing angle $\Theta_{\tilde{t}}$ \cite{640}.  Depending on the
mass ratios, stop \ps can decay into many final states.  For heavy
states, $\tilde{t}_2$ in particular \cite{641}: $\tilde{t}_i \to t
\tilde{\chi}_j^0, b \tilde{\chi}_j^+, t \tilde{g}$ or $\tilde{t}_i \to
\tilde{b} W^+, \tilde{b} H^+$, among which the first two modes are in
general dominant.

\STS Depending on the masses and energies, the production \css are
generally in the range between 10 and 100 fb so that the search can be
performed very efficiently and the properties of the \ps can be
determined in detail.  \hfill\ Using polarized $e^-$ 

\begin{figure}[ht]
\begin{center}
\epsfig{file=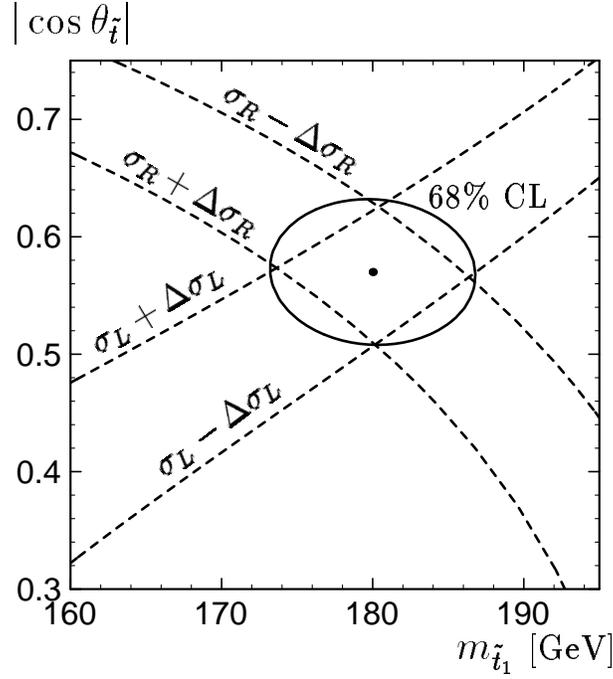,width=9cm}
\end{center}
\vspace{-.5cm}
\caption[]{\it 
  The measurement of the light stop mass $m_{\tilde{t}_1}$ and the
  mixing angle $\Theta_{\tilde{t}}$ of the stop sector in $ e^+e^-
  \rightarrow \tilde{t}_1 \overline{\tilde{t}}_1$ for 90\% left-- and
  right-handedly polarized electron beams at $\sqrt{s}$ = 500 GeV.
  The central values correspond to $m_{\tilde{t}_1}$ = 180 GeV and $|
  \cos \Theta_{\tilde{t}} | = 0.57$.  Ref.\protect\cite{639}.
  \protect\label{f647}\label{ttplot}}
\end{figure}

\noindent
beams, both the
$\tilde{t}_1$ mass and the mixing angle $\Theta_{\tilde{t}}$ can be
measured in diagonal pair production $\epem \to \tilde{t}_1
\bar{\tilde{t}_1}$ of the lightest stop state.  A study of this
reaction has been performed at $\sqrt{s}=500$~GeV,
$m_{\tilde{t}_1}=180$~GeV for the left-right stop mixing angle $| \cos
\Theta_{\tilde{t}} | =0.57$ which corresponds to the minimum of the
cross section \cite{639}.  The \css at tree level for these parameters
are $\sigma_L= 48.6$~fb and $\sigma_R=46.1$~fb for 90\% left-- and
right-polarized $e^-$ beams, respectively.  Based on detailed
simulations, the experimental errors on these \css are estimated to be
$\Delta \sigma_L= \pm 6$~fb and $\Delta \sigma_R= \pm 4.9$~fb.
Figure~\ref{f647} shows the resulting error bands and the
corresponding error ellipse in the $[m_{\tilde{t}_1}, \cos
\Theta_{\tilde{t}}]$ plane.  The experimental accuracy for the
$\tilde{t}_1$ mass and the stop mixing angle are $m_{\tilde{t}_1}=180
\pm 7$~GeV and $| \cos \Theta_{\tilde{t}} | =0.57 \pm 0.06$.

\STS The sbottom system can be treated analogously.  If $\tb$ is
moderate, mixing effects can be neglected.  Taking
$m_{\tilde{b}_1}=200$~GeV, $m_{\tilde{b}_2}=220$~GeV, the \css and the
expected experimental errors are $\sigma_L(\epem \ra \tilde{b}_1
\bar{\tilde{b}}_1)= 61.1 \pm 6.4$~fb, $\sigma_R(\epem \ra \tilde{b}_2
\bar{\tilde{b}}_2)= 6 \pm 2.6$~fb for 90\% left-- and right-polarized
$e^-$ beams.  The resulting experimental errors are
$m_{\tilde{b}_1}=200 \pm 4$~GeV, $m_{\tilde{b}_2}=220 \pm 10$~GeV.
With these results, the mass of the heavier stop particle can be
predicted: $m_{\tilde{t}_2}=289 \pm 15$~GeV.

\STS This prediction allows us to fix the MSSM parameters
in a subtle domain.  Assuming that $\mu$ and $\tb$ are known from
other neutralino/chargino experiments [e.g. $\mu = -200$~GeV and $\tb
= 2$], the soft breaking parameters of the stop and sbottom systems
can be derived: $m_{\tilde{Q}}= 195 \pm 4$~GeV, $m_{\tilde{U}}= 138
\pm 26$~GeV, $m_{\tilde{D}}= 219 \pm 10$~GeV, $A_t=-236 \pm 38$~GeV if
$\cos \Theta_{\tilde{t}} >0$, and $A_t=36 \pm 38$~GeV if $\cos
\Theta_{\tilde{t}} <0$.

\GS
\noindent
d) \underline{SUSY at LC and LHC: La Cohabitation}

\STS
\noindent 
 Since the mass 
scale of low-energy supersymmetry is restricted to order 1~TeV, the
LHC will either find signals of supersymmetry or rule out low-energy
supersymmetry.  In this collider, squarks and gluinos can be detected
with masses up to about 1.5 to 2~TeV \cite{804}.  \ee linear
colliders can provide the necessary complementary information in the
sector of non-colored \ssy particles.

\STS The analysis of \ssy \ps at the LHC can be divided into two
categories.  First, the gross features of \ssy phenomena, such as
missing transverse energy, dileptons etc., will be measured.  These
measurements will allow us to determine the typical mass scale of
colored \ssy particles.  Second, if in specific scenarios cascade
decays with favorable branching ratios can be exploited, a remarkably
high precision can be reached in determining mass differences between
\ssy \ps at the LHC \cite{804}.  For example, from the analysis of the
decay chain $\tilde{g} \to \tilde{b} \to \tilde{\chi}_2^0 \to
\tilde{\chi}_1^0$ in ``LHC Point 3'' of the study Ref.\cite{804}, the
mass difference $m(\tilde{\chi}_2^0) - m(\tilde{\chi}_1^0)$ can be
measured within $\pm 50 $~MeV, and $m(\tilde{g}) - m(\tilde{b})$
within 2~GeV accuracy.  In a similar way, the masses of other squarks
can be determined within a few percent.  The corresponding observation
and high-precision analyses of the heavier charginos and neutralinos
have not yet been demonstrated.  The situation in the scalar slepton
sector is similar.  If the rates are large enough to generate signals
of sleptons, their properties can be determined in some points of the
SUSY parameter space. Assuming that the soft SUSY-breaking parameters
are universal, the entire set of the basic parameters can be generated
at the LHC with an accuracy at the percent level.

\STS This experimental scenario at the LHC is easy to compare with the
\sce at \ee colliders. While the discovery limits
at the LHC can be extended to values above 1 TeV, the experiments at
the \ee machines provide high-precision information on all the \ssy
\ps which are kinematically accessible, in the high energy range for
masses up to $\sim$1 TeV. In particular, the analysis of the
color-neutral states, charginos/neutralinos and sleptons which are
generally (much) lighter than the colored states, can be carried out
with high accuracy, independently of specific assumptions and solely
based on kinematics.  Masses, for instance, can be measured by
exploiting threshold effects or simple decay kinematics.  Even such
subtle properties as the mixing of states can be analyzed thoroughly
in polarization experiments.  Thus, a systematic and high-precision
study can be performed which will resolve the complexities of the \ssy
phenomena.  In this way, the complete supersymmetry \sce can be
reconstructed up to the kinematical limit in a comprehensive form and
the structure of the underlying microscopic theory can be uncovered.

\STS
\noindent
\subsection[Testing SUSY--GUT]{Testing SUSY--GUT} 

\STS
\noindent
The high precision with which masses, \cps and mixing parameters will
be determined at \ee colliders, can be exploited to test the structure
of the underlying theories \cite{134B}.  If minimal supergravity is
the fundamental theory, the observable properties of the
superparticles can be expressed by a small set of parameters defined
at the GUT scale.  As a result, many relations can be found among the
masses of the superparticles and other observables which can
stringently be tested at \ee colliders.

\STS An overconstrained set of observables can be collected, slepton
and gaugino/neutralino masses and production cross sections, for
instance, which can be expressed by five basic parameters, as shown
for a few examples in Table~\ref{teq}, c.f.\ Refs.\ \citer{134A,134C}.

\STS
\begin{table}[ht!]
\begin{center}
\begin{tabular}{|llllllll|}
\hline
\rule{0mm}{6mm}
Masses \hfill :
&$m(\tilde{l}_L)$           & $\Leftarrow$  & $m_0$&& $M_{\demi}$&& 
$\tb$
 \hspace{3mm} \\
&$m(\tilde{l}_R)$           & $\Leftarrow$  & $m_0$&& $M_{\demi}$&& 
$\tb$
 \\
&$m(\tilde{\chi}_1^{\pm})$  & $\Leftarrow$  &&& $M_2$& $\mu$ &      
$\tb$
 \\
\rule[-3mm]{0mm}{5mm}
&$m(\tilde{\chi}_{1,2}^0)$  & $\Leftarrow$  && $M_1$& $M_2$& $\mu$& 
$\tb$
 \\
\hline
\rule{0mm}{6mm}
Cross Sections \rule[0mm]{0mm}{9mm} \hfill :
&$\sigma(e_L^-e^+ \to \tilde{e}_L^-\tilde{e}_L^+)$ 
                  & $\Leftarrow$  & $m_0$ & $M_1$ & $M_2$ & $\mu$ & 
$\tb$ \\
&$\sigma(e_R^-e^+ \to \tilde{e}_R^-\tilde{e}_R^+)$ 
                  & $\Leftarrow$  & $m_0$ & $M_1$ & $M_2$ & $\mu$ & 
$\tb$ \\
&$\sigma(e_L^-e^+ \to \tilde{\chi}_1^-\tilde{\chi}_1^+)$ 
                  & $\Leftarrow$  & $m_0$ &       & $M_2$ & $\mu$ & 
$\tb$ \\
\rule[-3mm]{0mm}{5mm}
&$\sigma(e_R^-e^+ \to \tilde{\chi}_1^-\tilde{\chi}_1^+)$ 
                  & $\Leftarrow$  &       &       & $M_2$ & $\mu$ & 
$\tb$ \\
\hline
\end{tabular}
\end{center}
\caption[]{\it 
The dependence of a representative set of observables
on the underlying SUSY and SUGRA parameters.
[The parameters are defined below.]
\protect\label{teq}}
\end{table}

\begin{figure}[ht]
\begin{center}
\hspace*{-1cm}
\epsfig{file= 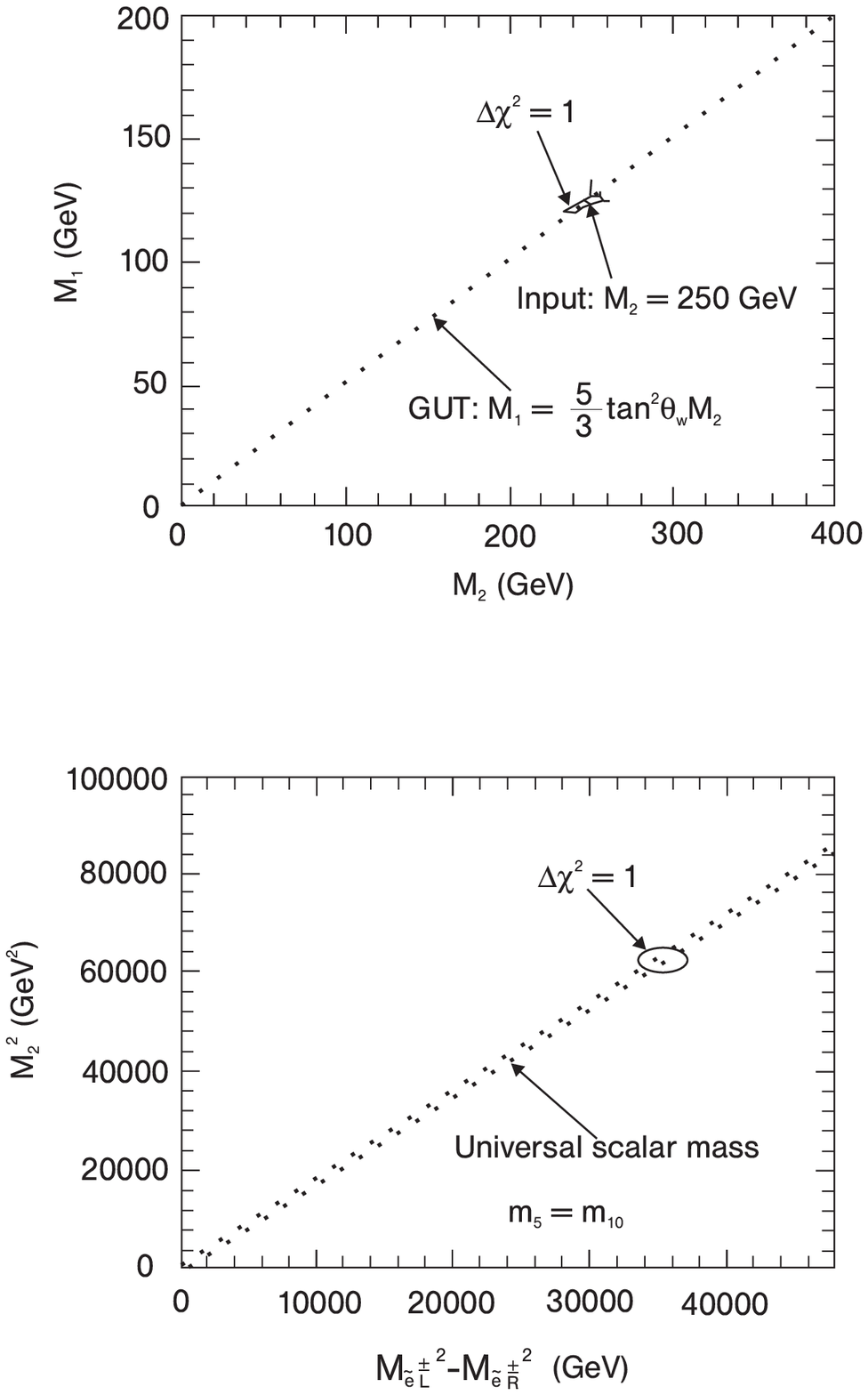, width= 11cm}
\end{center}
\vspace{-.5cm}
\caption[]{\it 
  Testing mass relations at \ee colliders between gauginos, and
  between selectrons (lower part), as predicted in minimal
  supergravity; Refs.\protect\cite{134A,134B}.
  \protect\label{f648}\label{2SGUTt}}
\end{figure}

\GS Two characteristic examples \cite{134B} should illustrate the
great potential of \ee colliders in this context:

\STS {$(i)$} The {\it gaugino masses} at the scale of $\rm SU(2)
\times U(1)$ symmetry breaking are related to the common gaugino mass
$M_{\demi}$ at the GUT scale by the running gauge couplings:
\begin{equation}
M_i = \frac{\alpha_i}{\alpha_{\rm GUT}} M_{\demi} \hspace{8mm}
[ \; i = 1,2,3 \; {\rm for \; U(1), SU(2), SU(3)} \; ]
\end{equation}
with $\alpha_{\rm GUT}$ being the gauge \cp at the unification scale.
The mass relation in the non-color sector
\begin{equation}
\frac{M_1}{M_2} = \frac{5}{3} \tan^2 \theta_W \approx \frac{1}{2}
\end{equation}
can be tested stringently to $\sim 0.5 $~\% by measuring the masses
and production \css of charginos/neutralinos and sleptons as shown in
Fig.\ref{f648} (upper part).

\STS {$(ii)$} In a similar way the {\it slepton masses} can be
expressed in terms of the common scalar mass parameter $m_0$ at the
GUT scale, contributions $\sim M_{\demi}^2$ due to the evolution from
the GUT scale down to low energies, and the D terms related to the
electroweak symmetry breaking.  The masses of the charged R/L sleptons
and the sneutrino can be written as
\begin{eqnarray}
m^2(\tilde{l}_R)   & =  & m_0^2 + \kappa_R M_{\demi}^2
- \sin^2 \theta_w \cos 2\beta \, M_Z^2           \nonumber        \\
m^2(\tilde{l}_L)   & =  & m_0^2 + \kappa_L M_{\demi}^2
- {\textstyle \frac{1}{2}} ( 1 - 2 \sin^2 \theta_W ) 
\cos 2\beta \, M_Z^2
\nonumber  \\
m^2(\tilde{\nu}_L) & =  & m_0^2 + \kappa_L M_{\demi}^2
+ {\textstyle \frac{1}{2}} \cos 2\beta \, M_Z^2
\end{eqnarray}
with $\kappa_R = 0.15$ and $\kappa_L = 0.52$ determined by the
solution of the evolution equations.  These expressions give rise to
simple relations among slepton masses after eliminating the common
scalar mass parameter:
\begin{eqnarray}
m^2(\tilde{l}_L) - m^2(\tilde{\nu_L})   &  =  &
- \cos^2\theta_w \cos 2\beta \, M_Z^2        \nonumber     \\
m^2(\tilde{l}_L) - m^2(\tilde{l}_R)   &  =  &
(\kappa_L - \kappa_R) M_{\demi}^2
- {\textstyle \frac{1}{2}} ( 1 - 4 \sin^2 \theta_W ) 
\cos 2\beta \, M_Z^2
\end{eqnarray}
The second relation follows from assuming the universality of the
scalar masses, in particular $m_0({\bf 5^*}) = m_0({\bf 10})$ within
$\rm SU(5)$.  This assumption can be tested by relating the mass
difference between $\tilde{e}_L$ and $\tilde{e}_R$ to the $\rm SU(2)$
gaugino mass $M_{\demi}$, as shown in Fig.\ref{f648} (lower part).

\subsection[Supersymmetry with R--Parity Violation]
{Supersymmetry with R--Parity Violation}

The most general gauge and supersymmetry invariant Lagrangian with
minimal particle content admits also the following Yukawa interactions
\cite{f153A}
\begin{equation}
W=\lambda_{ijk}L_iL_j\overline{E}_k
+\lambda_{ijk}'L_iQ_j\overline{D}_k
+\lambda_{ijk}''\overline{U}_i\overline{D}_j\overline{D}_k
\label{star1}
\end{equation}
where $L$, $Q$ are the left-handed lepton and quark superfields while
$\overline{E}$, $\overline{D}$, $\overline{U}$ are the corresponding
right-handed fields.  If both lepton-number violating
($\lambda_{ijk}$ and $\lambda_{ijk}'$) and baryon-number violating
($\lambda_{ijk}''$) \cps were present, they would give rise to very
fast proton decay.  In the MSSM all terms in Eq.(\ref{star1}) are
eliminated by imposing a multiplicative symmetry, $R$--parity: all SM
\ps are assigned $R=+1$, their superpartners have $R=-1$ \cite{f153B}.
However, since the $R$--parity has no {\it a~priori} justification,

\clearpage

\noindent
the possibilities of having either $\Delta L \neq 0$ or $\Delta B \neq
0$ (but not both) should be investigated as well.  If $R$--parity is
broken, the model differs from the MSSM in two important features:

a) The lightest SUSY \p (LSP) is not protected any more, and it can
decay into conventional particles.  As a result, the signature of
missing energy for MSSM processes disappears.

b) Supersymmetric \ps can be produced singly in collisions of
conventional particles, thus extending the mass range that can be
probed for a given energy.

\begin{figure}[hb]
\begin{center}
\hspace*{-1cm}
\epsfig{file=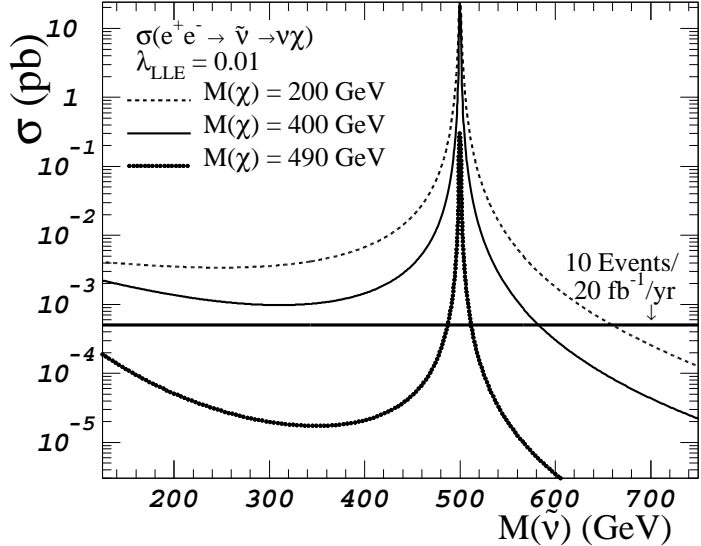,width=9.5cm,angle=0}
\end{center}
\vspace{-.5cm}
\caption[]{\it 
  Typical behavior of \css for $R$--parity breaking supersymmetry in
  the lepton sector; Ref.\protect\cite{f153C}.  \protect\label{fXR}}
\end{figure}

\STS To illustrate a typical case, the production of a single
superparticle in \ee collisions may be outlined, Fig.\ref{fXR},
arising from the lepton-number violating term
$\lambda_{LLE}L_eL_i\overline{E_e}$, $i=\mu$, $\tau$ \cite{f153C},
\[
\epem \rightarrow \tilde{\nu_i} \rightarrow
l_i^{\pm}\tilde{\chi}_k^{\mp},  \ \nu_i \tilde{\chi}_k^0
\]
with the subsequent decays
\begin{eqnarray*}
\tilde{\chi}_1^-
 &\rightarrow&
\tilde{\chi}_1^0l^-\nu, \ e^-l_i^-e^+, \ or \ e^-\nu_e\nu_i
\\
\tilde{\chi}_1^0
 &\rightarrow&
e^{\pm}l_i^{\mp}\nu_e, \ e^{\pm}e^{\mp}\nu_i
\end{eqnarray*}
The large number of charged leptons in the final state can be
exploited as a characteristic signature in this scenario.  These
events should be clearly visible at \ee \linebreak

\clearpage

\noindent
colliders provided the \cps $\lambda_i$ are sufficiently large.  For
example, at $\sqrt{s}=500$~GeV single chargino production with
$m_{\chi^+}=400$~GeV should be observed for $\lambda_{LLE}=0.01$ and a
sneutrino mass in the range of 300 to 600 GeV. Depending on the
couplings $\lambda_{LLE}$, \linebreak $\tilde{\nu}$ decays to lepton
pairs \cite{f153D}

\[
\epem \rightarrow \tilde{\nu_i} \rightarrow
\epem,\mu^+\mu^-,\tau^+\tau^-
\]
provide signatures that are very easy to detect experimentally.

\GS If the $LQ\bar{D}$ interaction term is non-zero, leptoquark-like
phenomena are predicted to occur. This case will be discussed in
detail in the following section.

\section[The Alternative: Compositeness]{The Alternative: Compositeness}

In \ssy extensions of the Standard Model, the fundamental \ps are
pointlike down to distances close to the Planck length.  However, if
light Higgs \ps do not exist, the electroweak bosons become strongly
interacting at energies of order 1~TeV.  As one among other physical
scenarios, this could be interpreted as a signal of composite
substructures of these \ps at a scale of $10^{-17} $~cm.  Moreover,
the proliferation of quarks and leptons could be taken as evidence for
possible substructures of the matter particles \cite{701}.  In this
picture, masses and mixing angles are a consequence of the
interactions between a small number of elementary constituents -- in
perfect analogy to the quark/gluon picture of hadrons.  No theoretical
formalism has been set up so far which would reconcile, in a
satisfactory manner, the small masses in the \SM with the tiny radii
of these particles which imply very large kinetic energies of these
constituents.  However, the lack of theoretical formalism does not
invalidate the physical picture or its motivation.

\STS
\subsection[Bounds on the Electron Radius]
{Bounds on the Electron Radius}

In this agnostic approach, stringent bounds have been derived from
high energy scattering experiments on possible non-zero radii of
leptons, quarks and gauge bosons from $Z$ decay data \cite{702} and
Bhabha scattering \cite{703} in \ee collisions, as well as from
electron-quark and quark-quark scattering at HERA \cite{147A} and
the Tevatron \cite{704}, respectively.  From these analyses the
compositeness scale has been bounded to less than $10^{-17} $~cm.

\begin{figure}[ht]
\begin{center}
\hspace*{-5mm}
\epsfig{file=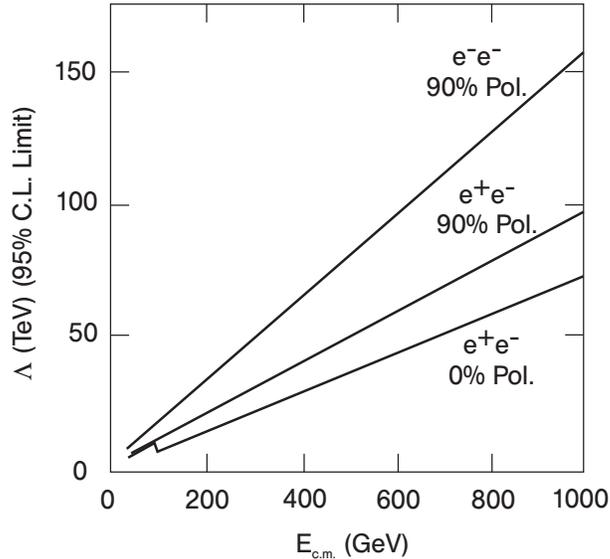,width=8cm,angle=0}
\end{center}
\vspace{-.5cm}
\caption[]{\it 
  Bounds on the compositeness scale of electrons, extracted from large
  angle M{\o}ller scattering $e^-e^- \to e^-e^-$. At $\sqrt{s} =
  1$~TeV the integrated luminosity is assumed to be $\int \LUM $~$= 80
  {\rm fb}^{-1}$; Ref.\protect\cite{705}.  \protect\label{radius}}
\end{figure}

\STS M{\o}ller scattering $e^- e^- \to e^- e^-$ at high energies
provides a very powerful instrument to set limits on electron
compositeness.  This problem has been studied in Ref.\cite{705}, based
on four-electron contact interactions which can be generated by the
exchange of electron constituents \cite{706}:
\begin{equation}
{\cal L}_C = \frac{2 \pi}{\Lambda_c^2} \bar{e}_L \gamma_{\mu}
e_L \cdot \bar{e}_L \gamma_{\mu} e_L
\end{equation}
The strength of the interaction has been set to $g_*^2 / 4\pi = 1$.
The (inverse) contact scale $\Lambda_c$ can be identified, within an
uncertainty of a factor of order 3, with the radius of the electron.
Detailed experimental simulations have shown that M{\o}ller scattering
is superior to Bhabha scattering in this context, a simple consequence
of the bigger \cs in the central rapidity region.  The high
polarization that can be achieved for electron beams, gives M{\o}ller
scattering another advantage.  At c.m. energies of 1~TeV, the bound on
electron compositeness can be set to
\[
\Lambda_c \approx 150 \; \; {\rm TeV} \hspace{+3mm}
\Rightarrow R_e \lessim \hspace{+3mm}
10^{-18} {\rm cm}
\]
for an integrated luminosity of $\int \LUM \sim$~$ 100 $~fb$^{-1}$ if
polarized electrons are used, Fig.\ref{radius}.  These high-energy
electron-electron scattering experiments will provide us with direct
and unambiguous limits on the radius of the electron.  This is in
contrast to high-precision $(g-2)_e$ and $(g-2)_{\mu}$ measurements,
the interpretation of which depends on dynamical assumptions on the
underlying constituent theory.

\STS
\subsection[Excited States]{Excited States}

In compositeness pictures, excited states should be observed with
masses of order of the compositeness scale $\Lambda_c$, i.e. $m_* \sim
1 $~TeV.  Such states can be produced pairwise and singly, generated
by the exchange of constituents \cite{707}: $\epem \to e_*^+ e_*^-$
and $\epem \to e^+ e_*^-$ etc.  Since the constituent exchange
interactions are strong, the masses of the excited states which can be
probed, extend up to the kinematical limit $m(e_*) \sim \sqrt{s}$ for
single $e_*$ production.  The decay modes of the excited states,
besides magnetic dipole decays to gauge bosons $e^* \to e \gamma, e Z$
and $\nu W$, are contact decays $e^* \to e + l \bar{l}$ and $e + jj$
with branching ratios of similar size.

\STS
\subsection[Leptoquarks]{Leptoquarks}

A very exciting prediction of fermion compositeness is the existence
of leptoquarks \cite{708}.  They are novel bound states of
subconstituents which build up leptons and quarks in this scenario.
While the size of the \cps to $\gamma$ and $Z$ bosons follows from the
electroweak symmetries, the Yukawa \cps to leptons and quarks are
bound by experiment.  In the interesting mass range, these Yukawa \cps
are expected to be weak.

\STS These \ps can also occur in grand unified theories.  Moreover, in
\ssy theories in which the $R$ parity is broken, scalar particles,
squarks or sleptons, may be coupled to quarks and leptons, giving rise
to production mechanisms and decay signatures analogous to leptoquarks
\cite{dreiner}.  However, whereas leptoquarks {\it per se}
disintegrate solely to leptons and quarks, a wide variety of decay
modes is in general expected for squarks and sleptons, including the
large ensemble of standard \ssy decay channels, see e.g. \cite{f163A}.
Since leptoquark bound states in the compositeness picture build up a
tower of states with non-zero spins, the phenomenology of the two
scenarios is clearly distinct.

\begin{figure}[hb]
\begin{center}
\epsfig{file=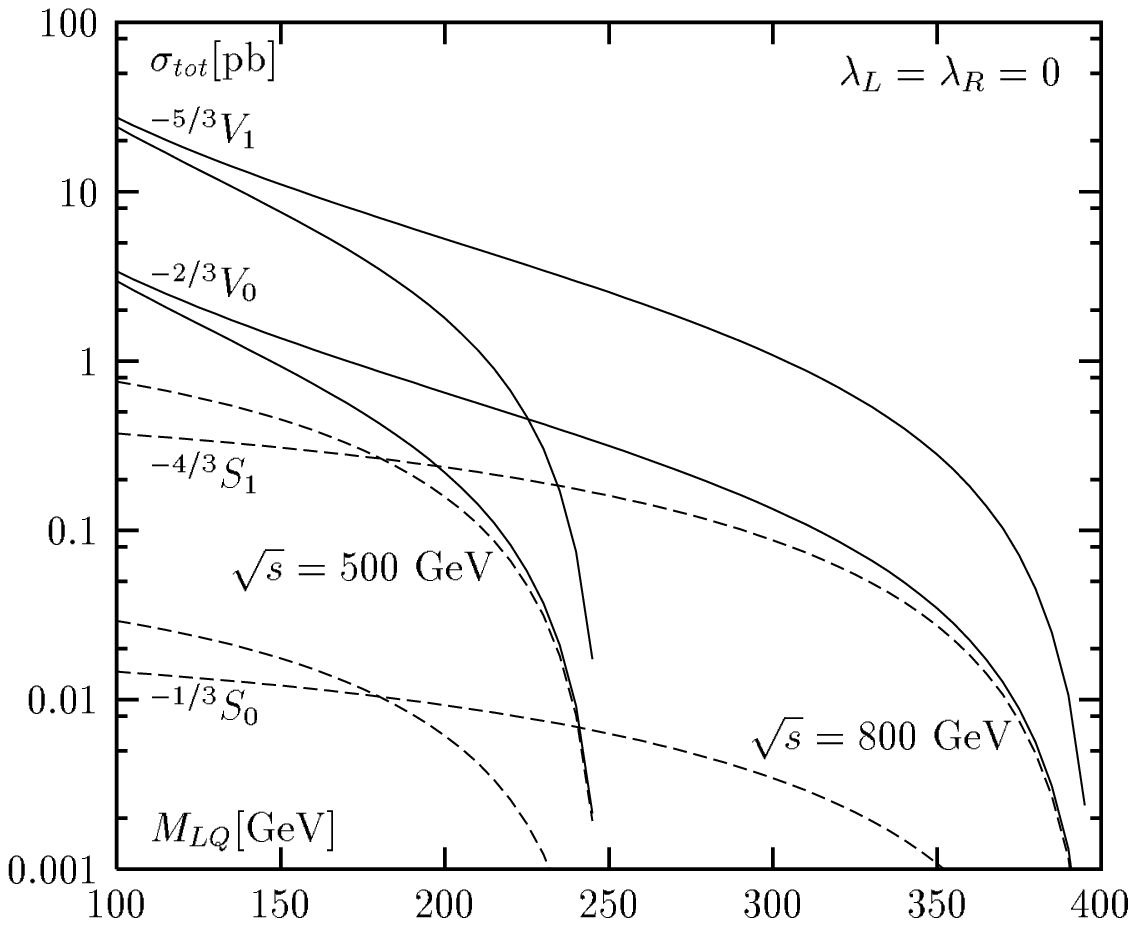 , width=11cm }
\end{center}
\vspace*{-0.5cm}
\caption[]{\it 
  Total cross section for leptoquark pair production at fixed
  center-\-of-\-mass energy as a function of the leptoquark mass
  $M_{LQ}$ assuming vanishing Yukawa couplings; corrections due to
  beamstrahlung and ISR are included. Ref.\protect\cite{712}.
  \protect\label{lepto}} \vspace*{-10cm}
\end{figure}

\GS Leptoquarks can exist in a large variety of states carrying
$[l_iq_j]$ or $[l_i\overline{q}_j]$ quantum numbers $(i, j = L, R)$
and being scalar or vectorial in the simplest representations
\cite{709}, see Table~\ref{t4}.  
They can be produced in \ee collisions pairwise,
\[
\epem \to LQ + \overline{LQ}
\]
through $s$--channel $\gamma, Z$ exchange and partly through
$t$--channel $q$ exchange \cite{710,711}.  The \css for the production
of scalar leptoquarks scale asymptotically as $\log (s/M_{LQ}^2) / s$.
The \css for vector leptoquarks approach non-zero limits for
$s$--channel $\gamma, Z$ exchange, or they grow with $s$ due to the
$t$--channel $q$ exchange until the rise is damped by form factors
\cite{711}.  The typical size of the \css is illustrated in
Fig.~\ref{lepto}.  \STS

\newpage

\begin{table}[ht]
\vspace*{-2cm}
\begin{footnotesize}
\begin{center}
\begin{tabular}{cc}
\begin{tabular}{|c|c|r|}
\hline \rule{0mm}{5mm}
Type ${}^Q\overline{LQ}_T$ & 
Decay & 
$\sigma_{tot}(s) [$fb$]$ \\[1mm]
\hline \rule{0mm}{5mm}
${}^{-1/3}S_0$ & 
$\displaystyle \begin{array}{c} e_L^- u_L \\
                            e_R^- u_R \\ \nu_e d_L \end{array}$ & 
$6~$ \\[1mm]
\hline \rule{0mm}{5mm}
${}^{-4/3}\tilde{S}_0$ & 
$e_R^- d_R$ & 
$98~$\\[1mm]
\hline \rule{0mm}{5mm}
${}^{+2/3}S_1$ & 
$\nu_e u_L$ & 
$110~$ \\ 
${}^{-1/3}S_1$ & 
$\displaystyle \begin{array}{c}
                 \nu_e d_L \\       
                 e_L^- u_L \end{array}$ & 
$6~$ \\
${}^{-4/3}S_1$ & 
$e_L^- d_L$ & 
$158~$ \\[1mm]
\hline \rule{0mm}{5mm}
$\displaystyle \begin{array}{c} {}^{-2/3}S_{1/2} \\ \\
                                {}^{-5/3}S_{1/2} \end{array}$ &
$\displaystyle \begin{array}{c} \nu_e \bar{u}_L \\
                                e_R^- \bar{d}_R \\
                                e_L^- \bar{u}_L \\
                                e_R^- \bar{u}_R \end{array}$ &
$\displaystyle \begin{array}{r} 65 \\ \\ 149 \end{array}$ \\[9mm]
\hline \rule{0mm}{8mm}
$\displaystyle \begin{array}{c} 
                  {}^{+1/3}\tilde{S}_{1/2} \\
                  {}^{-2/3}\tilde{S}_{1/2} \end{array}$ &
$\displaystyle \begin{array}{c} \nu_e \bar{d}_L \\
                                e_L^- \bar{d}_L \end{array}$ & 
$\displaystyle \begin{array}{r} 27 \\ 39 \end{array}$ \\[4.5mm]
\hline
\end{tabular}
\begin{tabular}{|c|c|r|}
\hline \rule{0mm}{5mm}
Type ${}^Q\overline{LQ}_T$ & 
Decay & 
$\sigma_{tot}(s) [$fb$]$ \\[1mm]
\hline \rule{0mm}{5mm}
$\displaystyle \begin{array}{c} {}^{-1/3}V_{1/2} \\ \\
                                {}^{-4/3}V_{1/2} \end{array}$ &
$\displaystyle \begin{array}{c} \nu_e d_R \\
                                e_R^- u_L \\
                                e_L^- d_R \\
                                e_R^- d_L \end{array}$ &
$\displaystyle \begin{array}{r} 365 \\ \\ 895 \end{array}$ \\[6mm]
\hline \rule{0mm}{8mm}
$\displaystyle \begin{array}{c} 
                  {}^{+2/3}\tilde{V}_{1/2} \\
                  {}^{-1/3}\tilde{V}_{1/2} \end{array}$ &
$\displaystyle \begin{array}{c} \nu_e u_R \\
                                e_L^- u_R \end{array}$ & 
$\displaystyle \begin{array}{r} 353 \\ 247 \end{array}$ \\[4mm]
\hline \rule{0mm}{5mm}
  & 
$e_L^- \bar{d}_R$ &
  \\
${}^{-2/3}V_{0}$ & 
$e_R^- \bar{d}_L$ & 
$222~$ \\
  & 
$\nu_e \bar{u}_R$ &
  \\[1mm]
\hline \rule{0mm}{5mm}
${}^{-5/3}\tilde{V}_0$ & 
$e_R^- \bar{u}_L$ & 
$1370~$ \\[1mm]
\hline \rule{0mm}{5mm}
${}^{+1/3}V_1$ & 
$\nu_e \bar{d}_R$ & 
$942~$ \\
${}^{-2/3}V_1$ & 
$\displaystyle \begin{array}{c}
                 e_L^- \bar{d}_R \\
                 \nu_e \bar{u}_R \end{array}$ & 
$222~$ \\
${}^{-5/3}V_1$ & 
$e_L^- \bar{u}_R$ & 
$1790~$ \\[1mm]
\hline
\end{tabular}
\end{tabular}
\end{center}
\end{footnotesize}
\caption[]{\it The total cross section
  $\sigma_{tot}(s)$ are given for $\protect\sqrt{s}=500$GeV and a
  leptoquark mass $M_{LQ}=200$GeV, assuming vanishing Yukawa
  couplings; corrections due to beamstrahlung and initial state
  radiation are included. Ref.\protect\cite{712}.}
\protect\label{t4}
\end{table}

\vspace*{-1cm}

$+{\textstyle \frac{1}{3}}$
$+{\textstyle \frac{2}{3}}$
\clearpage

\GS The \ps decay to a charged lepton, or a neutrino, and a jet,
giving rise to visible (a) $l^+l^-jj$, (b) $l^{\pm}jj$, and (c) $jj$
final states.  Since leptoquarks generate a peak in the invariant
($lj$) mass, they are easy to detect in the cases (a) and (b) up to
mass values close to the kinematical limit \cite{712}: The discovery
range extends up to $m_{LQ} \lessim (0.37 - 0.49)\sqrt{s}$ for scalar
leptoquarks, and up to $m_{LQ} \lessim (0.48 - 0.5)\sqrt{s}$ for
vector leptoquarks.

\GS Since leptoquarks carry color, they are produced copiously
\cite{713} in hadron collisions through the subprocesses $gg,
q\bar{q}, qq \to LQ + LQ'$ and $gq \to LQ + l$.  Leptoquarks can
therefore be generated at the LHC with very high masses.  Experiments
at \ee colliders are nevertheless important to identify the
electroweak properties of these novel states.


\vfill
\section*{Acknowledgements}

\GS We are very grateful to K.~Hagiwara and M.~E.~Peskin for numerous
discussions on the physics with $e^+e^-$ linear colliders and the 
careful reading of the manuscript. We benefited from comments on this
report by
G.~Altarelli, J.~Ellis, D.~Froidevaux, F.~Gianotti, 
E.~Richter--W\c{a}s, and D.~Zerwas.

\STS 
\noindent
Thanks go also to the secretary Frau S.~G\"unther for tireless
efforts in shaping the layout of the manuscript. 


\end{document}